\documentclass[prd,aps,nofootinbib,floatfix,onecolumn,10pt]{revtex4}
\usepackage{amsmath,graphicx,color,epsfig}

\begin{document}
\title{Charmless Two-body $B_{(s)}\to VP$ decays In Soft-Collinear-Effective-Theory}
\author{ Wei Wang$^{a,b}$, Yu-Ming Wang$^{a,b}$, De-Shan Yang$^b$ and Cai-Dian L\"u$^a$ }
\affiliation{
 $^a$ Institute of High Energy Physics, Chinese Academy
of Sciences, Beijing 100049, P.R. China\\
 $^b$ Graduate University of Chinese Academy of
Sciences, Beijing 100049, P.R. China}

\begin{abstract}
We provide the analysis of charmless two-body $B\to VP$ decays under
the framework of the soft-collinear-effective-theory (SCET), where
$V(P)$ denotes a light vector (pseudoscalar) meson. Besides the
leading power contributions, some power corrections (chiraly
enhanced penguins) are also taken into account. Using the current
available $B\to PP$ and $B\to VP$ experimental data on branching
fractions and CP asymmetry variables, we find two kinds of solutions
in $\chi^2$ fit for the 16 non-perturbative inputs which are
essential in the 87 $B\to PP$ and $B\to VP$ decay channels. Chiraly
enhanced penguins can change several charming penguins sizably,
since they share the same topology. However, most of the other
non-perturbative inputs and predictions on branching ratios and CP
asymmetries are not changed too much. With the two sets of inputs,
we predict the branching fractions and CP asymmetries of other modes
especially $B_s\to VP$ decays. The agreements and differences with
results in QCD factorization and perturbative QCD approach are
analyzed. We also study the time-dependent CP asymmetries in
channels with CP eigenstates in the final states and some other
channels such as $\bar B^0/B^0\to\pi^\pm\rho^\mp$ and $\bar
B_s^0/B_s^0\to K^\pm K^{*\mp}$. In the perturbative QCD approach,
the $(S-P)(S+P)$ penguins in annihilation diagrams play an important
role.  Although they have the same topology with charming penguins
in SCET, there are many differences between the two objects in weak
phases, magnitudes, strong phases and factorization properties.
\end{abstract}
\maketitle

%
\section{Introduction}

Studies on $B$ decays are mainly concentrated on the precise test of
the standard model (SM) and the search for possible new physics (NP)
scenarios. To  map out the apex in the unitarity triangle of the
Cabibbo-Kobayashi-Maskawa (CKM) matrix, many precise experimental
data together with reliable theoretical predictions are required. In
charmless two-body non-leptonic $B$ decays, the main experimental
observables are branching ratios and CP asymmetries. To predict
these observables, one has to compute the hadronic decay amplitudes
$\langle M_1 M_2 | O_i |B\rangle$, where $ O_i$ is typically a
four-quark or a magnetic moment type operator. Since three hadronic
states are involved in these decays, the predictions on these
observables are often polluted by our poor knowledge of the
non-perturbative QCD. Fortunately, it has been suggested that in the
$m_b\to\infty$ limit, decay amplitudes can be studied in a
well-organized way: they can be factorized into the convolution of
non-perturbative objects such as $B$ to light form factors and decay
constants of light pseudoscalars/vectors with perturbative hard
kernels. In recent years, great progresses have been made in studies
of charmless two-body $B$ decays. These decays were  investigated in
the so-called naive factorization
approach~\cite{Wirbel:1985ji,Bauer:1986bm} and the generalized
factorization
approach\cite{Ali:1997nh,Kramer:1993yu,Ali:1998eb,Ali:1998gb,Chen:1999nxa}.
At present, there are three commonly-accepted theoretical approaches
to investigate the dynamics of these decays, the QCD factorization
(QCDF)~\cite{Beneke:1999br,Beneke:2000ry,Beneke:2003zv}, the
perturbative QCD (PQCD)~\cite{Keum:2000ph,Keum:2000wi,Lu:2000em},
and the soft-collinear effective theory
(SCET)~\cite{Bauer:2000yr,Bauer:2001cu}. Despite of many
differences, all of them are based on power expansions in
$\Lambda_{QCD}/m_b$, where $m_b$ is the $b$-quark mass and
$\Lambda_{QCD}$ is the typical hadronic scale. Factorization of the
hadronic matrix elements is proved to hold in the leading power in
$\Lambda_{QCD}/m_b$ in a number of decays.

In the present work, we will focus on the SCET. The matching from
QCD onto SCET is always performed in two stages. The fluctuations
with off-shellness ${\cal O}(m_b^2)$ is firstly integrated out and
one results in the intermediate effective theory. At final stage, we
integrate out the hard-collinear modes with off-shellness ${\cal
O}(m_b\Lambda_{QCD})$ to derive SCET$_{II}$. In $B\to M_1M_2$
decays, both of the final state mesons move very fast and are
generated back-to-back in the rest frame of $B$ meson.
Correspondingly, there exist three typical scales: the $b$ quark
mass $m_b$, the soft scale $\Lambda_{QCD}$ set by the typical
momentum of the light degrees of freedom in the heavy $B$ meson, the
intermediate scale $\sqrt{m_b\Lambda_{QCD}}$ which arise from the
interaction between collinear particles and soft modes. SCET
provides an elegant theoretical tool to separate the physics at
different scales and factorization  for $B\to M_1M_2$ was proved to
hold to all orders in $\alpha_s$ at leading power of
$1/m_b$~\cite{Bauer:2001yt,Bauer:2002nz,Chay:2003zp,Chay:2003ju,Bauer:2004tj}.
After integrating out the fluctuations with off-shellness $m_b^2$,
one reaches the intermediate effective theory SCET$_I$, in which the
generic factorization formula for $B\to M_1 M_2$   is written by:
\begin{eqnarray}
 \langle M_1 M_2|O_i|B\rangle= T(u) \otimes \phi_{M_1}(u)\zeta^{B\to M_2}
 +T_J(u,z) \otimes \phi_{M_1}(u)\otimes \zeta_J^{B\to M_2}
 (z),\label{eq:genericfactorization}
\end{eqnarray}
where $T$ and $T_J$ are perturbatively calculable Wilson
coefficients which depend on the Lorentz structure and flavor
structure. Calculations for these hard kernel functions are
approaching next-to-leading order
accuracy~\cite{Beneke:1999br,Beneke:2000ry,Chay:2003ju,Beneke:2005vv,Beneke:2006mk,Jain:2007dy}.
In the second step, the fluctuations with typical off-shellness
$m_b\Lambda_{QCD}$ are integrated out and one reaches SCET$_{II}$.
In SCET$_{II}$, end-point singularities prohibit the factorization
of $\zeta$, while the function $\zeta_J$ can be further factorized
into the convolution of a hard kernel (jet function) with light-cone
distribution amplitudes:
\begin{eqnarray}
 \zeta_J(z)=\phi_{M_2}(x)\otimes J (z,x,k_+)\otimes \phi_B(k_+).
\end{eqnarray}

An essential question is whether power corrections in SCET can be
analyzed in a similar way. It is almost an impossible task to
include all power corrections, but we can include the relatively
important one. Importance of chiraly enhanced penguins has been
noted long time ago, and numerics show that chiraly enhanced
penguins are comparable with the penguin contributions at leading
power. Thus in both of
QCDF~\cite{Beneke:1999br,Beneke:2000ry,Beneke:2003zv} and
PQCD~\cite{Keum:2000ph,Keum:2000wi,Lu:2000em} approaches, it has
been incorporated into the decay amplitudes besides the leading
power penguins. In SCET, the complete operator basis and the
corresponding factorization formulae for this term are recently
derived in Ref.~\cite{Arnesen:2006vb,Jain:2007dy}. A new
factorization formula for chiraly enhanced penguin was proved to
hold to all orders in $\alpha_s$, and more importantly the
factorization formula does not suffer from the endpoint divergence.
In the factorization formula, a new form factor named $\zeta_{\chi}$
and a twist-3 light-cone distribution amplitude $\phi^{pp}$ are
introduced.

In Ref.~\cite{Bauer:2005kd}, one phenomenological framework is
introduced, in which the expansion at the intermediate scale
$\mu_{hc}=\sqrt {m_b\Lambda_{QCD}}$ is not used. Instead the
experimental data are used to fit the non-perturbative inputs. This
method is very useful especially at tree level, since the function
$T(u)$ is a constant and $T_J(u,z)$ is a function of only $u$. Thus
only a few inputs are required in decay amplitudes. In this
framework, an additional term from the intermediate charm quark
loops, which is called charming
penguin~\cite{Bauer:2004tj,Bauer:2005wb,Bauer:2005kd,Colangelo:1989gi,Ciuchini:1997hb,Ciuchini:2001gv},
is also taken into account. Charming penguins are not factorized
into the LCDAs and form factors, since the heavy charm quark pair
can not be viewed as collinear quarks. They are also treated as
non-perturbative inputs.  This method is first applied to $B\to
K\pi$, $B\to KK$ and $B\to\pi\pi$ decays~\cite{Bauer:2005kd}.
Subsequently, it is extended to charmless two-body $B\to PP$ decays
involving the iso-singlet mesons $\eta$ and
$\eta'$~\cite{Williamson:2006hb}.

In the present work, we extend this method to the $B\to VP$ decays.
We will use the wealth of the experimental data to fit the
non-perturbative inputs (in our analysis, we also take the $B\to PP$
decays into account). In doing this, we would assume SU(3) symmetry
for form factors and charming penguins to reduce the number of
independent non-perturbative inputs: there are totally 16
non-perturbative inputs to be determined. Utilizing the meson
matrices, we give the master equations for the hard kernels for
$B\to M_1M_2$ decays. After analyzing the $B\to VP$ decays at
leading power, we take part of chiraly enhanced penguin into
account. With the chiraly enhanced penguins taken into account, we
find most of the 16 inputs are not changed sizably except charming
penguins. Flavor singlet mesons $\eta$ and $\eta'$ receive
additional contributions (gluonic contributions) from higher Fock
state component. In Ref.~\cite{Williamson:2006hb}, the gluonic form
factors and gluonic charming penguins which are responsible for
$B\to PP$ decays are fitted using the related experimental data.
Since there are not enough experimental results, the authors find
two solutions for these inputs. This situation is changed when
considering $B\to VP$ decays since we have more data to give more
stringent constraint. Incorporating the $B\to VP$ experimental
results for branching fractions and CP asymmetries, we find that our
results are consistent with their second solution. We find two
solutions for the inputs only responsible for $B\to VP$ decays. One
of the solutions for $B\to V$ form factors are smaller than those
given in Ref.~\cite{Jain:2007dy}, where the $B\to\rho_L\rho_L$ data
($\rho_L$ denotes a longitudinally polarized meson), $B\to
\rho^0\rho^-$ and $B\to\rho^+\rho^-$ branching ratios and
CP-asymmetries $S_{\rho^+\rho-}$ and $C_{\rho^+\rho^-}$, are used.
Our second solution for $B\to V$ form factors is more consistent
with them. Generally speaking, charming penguins in SCET have the
similar role with $(S-P)(S+P)$ annihilation penguin operators in
PQCD approach. Both of them are essential to give the correct
branching ratios in these two different approaches. But there are
indeed some differences in predictions on other parameters such as
direct CP asymmetries and mixing-induced CP asymmetries. We also
make some comparisons between these two objects.

The paper is organized as follows. $B\to VP$ decay amplitudes at
leading power are briefly given in Sec.~\ref{sec:decayamplitudes}.
What followed is the factorization analysis in which  chiraly
enhanced penguins are taken into account. In
section~\ref{sec:decayamplitudes}, utilizing the rich experimental
data on branching fractions and time-dependent CP asymmetry
observables, we give two kinds of solutions for the $16$
non-perturbative parameters responsible for $B\to PP$ and $B\to VP$
decays at the leading power accuracy. With the inclusion of chiraly
enhanced penguin, most parameters remain unchanged except the
charming penguin parameters. Predictions on branching fractions and
other observables, including direct CP asymmetries, time-dependent
CP asymmetries and ratios of branching fractions, are given
subsequently. A comparison between charming penguins in SCET and
annihilation diagrams in PQCD approach is presented in
Section~\ref{sec:differences}. Sec.~\ref{sec:conclusions} contains
our conclusions.  In appendix~\ref{sec:Hardkernels}, we give the
master equations for the hard kernels in both $b\to d$ and $b\to s$
transitions.

\section{$B\to VP$ decay amplitudes at leading power in
SCET}\label{sec:decayamplitudes}

In this section, we briefly review the factorization analysis at the
leading power and collect the corresponding leading order
short-distance coefficients. The weak effective Hamiltonian which
describes $b\to D$ ($D=d,s$) transitions are \cite{Buchalla:1995vs}:
 \begin{eqnarray}
 {\cal H}_{eff} &=& \frac{G_{F}}{\sqrt{2}}
     \bigg\{ \sum\limits_{q=u,c} V_{qb} V_{qD}^{*} \big[
     C_{1}  O^{q}_{1}
  +  C_{2}  O^{q}_{2}\Big]- V_{tb} V_{tD}^{*} \big[{\sum\limits_{i=3}^{10,7\gamma,8g}} C_{i}  O_{i} \Big]\bigg\}+ \mbox{H.c.} ,
 \label{eq:hamiltonian}
\end{eqnarray}
where $V_{qb(D)}$ are the CKM matrix elements and in the following
we will also use products of the CKM matrix elements
$\lambda^{(f)}_{q}$ (q=u,c,t) defined by $\lambda^{(f)}_{q}= V_{qb}
V_{qf}^*$. Functions $O_{i}$ ($i=1,...,10,7\gamma,8g$) are the local
four-quark operators or the moment type operators:
 \begin{itemize}
 \item  current--current (tree) operators
    \begin{eqnarray}
   O^{q}_{1}=({\bar{q}}_{\alpha}b_{\alpha})_{V-A}
               ({\bar{D}}_{\beta} q_{\beta} )_{V-A},
    \ \ \ \ \ \ \ \ \ O^{q}_{2}=({\bar{q}}_{\alpha}b_{\beta} )_{V-A}
               ({\bar{D}}_{\beta} q_{\alpha})_{V-A},
    \label{eq:operator12}
    \end{eqnarray}
     \item  QCD penguin operators
    \begin{eqnarray}
      O_{3}=({\bar{D}}_{\alpha}b_{\alpha})_{V-A}\sum\limits_{q^{\prime}}
           ({\bar{q}}^{\prime}_{\beta} q^{\prime}_{\beta} )_{V-A},
    \ \ \ \ \ \ \ \ \
    O_{4}=({\bar{D}}_{\beta} b_{\alpha})_{V-A}\sum\limits_{q^{\prime}}
           ({\bar{q}}^{\prime}_{\alpha}q^{\prime}_{\beta} )_{V-A},
    \label{eq:operator34} \\
     \!\!\!\! \!\!\!\! \!\!\!\! \!\!\!\! \!\!\!\! \!\!\!\!
    O_{5}=({\bar{D}}_{\alpha}b_{\alpha})_{V-A}\sum\limits_{q^{\prime}}
           ({\bar{q}}^{\prime}_{\beta} q^{\prime}_{\beta} )_{V+A},
    \ \ \ \ \ \ \ \ \
    O_{6}=({\bar{D}}_{\beta} b_{\alpha})_{V-A}\sum\limits_{q^{\prime}}
           ({\bar{q}}^{\prime}_{\alpha}q^{\prime}_{\beta} )_{V+A},
    \label{eq:operator56}
    \end{eqnarray}
 \item electro-weak penguin operators
    \begin{eqnarray}
     O_{7}=\frac{3}{2}({\bar{D}}_{\alpha}b_{\alpha})_{V-A}
           \sum\limits_{q^{\prime}}e_{q^{\prime}}
           ({\bar{q}}^{\prime}_{\beta} q^{\prime}_{\beta} )_{V+A},
    \ \ \ \
    O_{8}=\frac{3}{2}({\bar{D}}_{\beta} b_{\alpha})_{V-A}
           \sum\limits_{q^{\prime}}e_{q^{\prime}}
           ({\bar{q}}^{\prime}_{\alpha}q^{\prime}_{\beta} )_{V+A},
    \label{eq:operator78} \\
     O_{9}=\frac{3}{2}({\bar{D}}_{\alpha}b_{\alpha})_{V-A}
           \sum\limits_{q^{\prime}}e_{q^{\prime}}
           ({\bar{q}}^{\prime}_{\beta} q^{\prime}_{\beta} )_{V-A},
    \ \ \ \
    O_{10}=\frac{3}{2}({\bar{D}}_{\beta} b_{\alpha})_{V-A}
           \sum\limits_{q^{\prime}}e_{q^{\prime}}
           ({\bar{q}}^{\prime}_{\alpha}q^{\prime}_{\beta} )_{V-A},
    \label{eq:operator9x}
    \end{eqnarray}
     \item magnetic moment operators
    \begin{eqnarray}
     O_{7\gamma}=-\frac{em_b}{4\pi^2}{\bar{D}}_{\alpha}\sigma^{\mu\nu}
     P_R b_{\alpha}F_{\mu\nu},
    \ \ \ \
    O_{8g}=-\frac{gm_b}{4\pi^2}{\bar{D}}_{\alpha}\sigma^{\mu\nu}
     P_RT^a_{\alpha\beta}b_{\beta}G^a_{\mu\nu},\label{eq:operator7gamma8g}
\end{eqnarray}
\end{itemize}
where $\alpha$ and $\beta$ are color indices and $q^\prime$ are the
active quarks at the scale $m_b$, i.e. $q^\prime=(u,d,s,c,b)$. The
$m_b$ is the $b$ quark mass and we use $m_b=4.8$ GeV. The left
handed current is defined as $({\bar{q}}^{\prime}_{\alpha}
q^{\prime}_{\beta} )_{V-A}= {\bar{q}}^{\prime}_{\alpha} \gamma_\nu
(1-\gamma_5) q^{\prime}_{\beta}  $ and the right handed current
$({\bar{q}}^{\prime}_{\alpha} q^{\prime}_{\beta} )_{V+A}=
{\bar{q}}^{\prime}_{\alpha} \gamma_\nu (1+\gamma_5)
q^{\prime}_{\beta}$. The projection operators are defined as
$P_{L}=(1-\gamma_5)/2$ and $P_{R}=(1+\gamma_5)/2$. The electro-weak
penguin operators $O_{9,10}$ can be eliminated using $e_q\bar qq
=\bar uu+\bar cc- \frac{1}{3} \bar qq$. In the following, we will
work to leading order in $\alpha_s(m_b)$.  In the naive dimensional
regularization (NDR) scheme for $\alpha_s(m_Z)=0.119$, $\alpha_{\rm
em}=1/128$, $m_t=174.3$ GeV, the Wilson coefficients $C_i$ at
leading logarithm order for tree and QCD penguin operators are
\begin{eqnarray}
C_{1-6}(m_b)=&\{1.110,-0.253,0.011,-0.026,0.008,-0.032\}
\label{eq:WC1-6},
\end{eqnarray}
while the Wilson coefficients for electroweak penguin (EWP) operators are:
\begin{eqnarray}
C_{7-10}(m_b)=\{0.09, 0.24, -10.3,2.2\}\times 10^{-3},
\end{eqnarray} and for the magnetic operators
$C_{7\gamma}(m_b)=-0.315$, $C_{8g}(m_b)=-0.149$. We have used the
sign convention for the electromagnetic and strong coupling constant
as $D_\mu=\partial_\mu-igT^aA_\mu^a-ieQ_fA_\mu$, so that the Feynman
rule for the vertex is $igT^a\gamma_\mu+ieQ_f\gamma_\mu$.

In the present work, we will adopt the notations as in
Ref.~\cite{Beneke:2002ph} and use $\lambda=\sqrt
{\Lambda_{QCD}/m_b}$. The emitted quark and anti-quark mainly move
along the direction $n_+$ and the recoiling meson is moving on the
direction $n_-$, where $n_\pm$ are two light-cone vectors:
$n_\pm^2=0$ and $n_+\cdot n_-=2$. The matching from QCD onto SCET
are always performed in two stages. We will first integrate out the
fluctuations with off-shellness ${\cal O}(m_b^2)$ to give the
intermediate effective theory. At final stage, we integrate out the
hard-collinear modes with off-shellness ${\cal O}(m_b\Lambda_{QCD})$
to derive SCET$_{II}$

\subsection{Matching onto SCET$_I$}

To study the decay amplitudes of $B\to M_1M_2$ decays in SCET, we
first consider the possible operators   using the building blocks.
The power counting rule for these blocks has been given in
Ref.~\cite{Beneke:2002ph}. Integrating out the hard scales with
typical off-shellness $m_b^2$, the electro-weak operators can match
onto two kinds of operators in SCET where the situation is similar
with that in $B$ to light form factors: the first kind of operators
involve four quark fields while the second one involves an
additional transverse gluon field. For flavor-singlet mesons, one
needs to consider the operators which are composed by two gluon
fields. Then the leading power operators responsible for $b\to s$
transitions are chosen by:
\begin{eqnarray}
  Q_{1s}^{(0)}(t)&=&
  \left[(\bar s W_{c2})(tn_-)\frac{n\!\!\!\slash_-}{2}(1-\gamma_5)(W_{c2}^\dagger
   u)\right]
   \left[ (\bar u W_{c1})n\!\!\!\slash _+(1-\gamma_5) h_v\right],\nonumber\\
     Q_{2s,3s}^{(0)}(t)&=&
  \left[(\bar u W_{c2})(tn_-)\frac{n\!\!\!\slash_-}{2}(1\mp\gamma_5)(W_{c2}^\dagger
   u)\right]
   \left[ (\bar s W_{c1})n\!\!\!\slash _+(1-\gamma_5) h_v\right],\nonumber\\
  Q_{4s}^{(0)}(t)&=&
  \left[(\bar s W_{c2})(tn_-)\frac{n\!\!\!\slash_-}{2}(1-\gamma_5)(W_{c2}^\dagger
   q)\right]
   \left[ (\bar q W_{c1})n\!\!\!\slash _+(1-\gamma_5) h_v\right],\nonumber\\
     Q_{5s,6s}^{(0)}(t)&=&
  \left[(\bar q W_{c2})(tn_-)\frac{n\!\!\!\slash_-}{2}(1\mp\gamma_5)(W_{c2}^\dagger
   q)\right]
   \left[ (\bar s W_{c1})n\!\!\!\slash _+(1-\gamma_5) h_v\right],\nonumber\\
   Q_{gs}^{(0)}(t)&=& m_b i\epsilon_{\perp\mu\nu}{\rm Tr}
   \left[ [W_{c2}^\dagger iD^\mu_{\perp {c2}} W_{c2}](tn_-)  [W_{c2}^\dagger iD^\nu_{\perp {c2}} W_{c2}] \right]
   \left[ (\bar s W_{c1})n\!\!\!\slash _+(1-\gamma_5) h_v\right
   ],\label{eq:leadingpoweroperators}
\end{eqnarray}
with the trace over the color indices. The operators suppressed by
$\lambda$ are given by:
\begin{eqnarray}
  Q_{1s}^{(1)}(t,s)&=&
  -\frac{1}{m_b}\left[(\bar s W_{c2})(tn_-)\frac{n\!\!\!\slash_-}{n_-v}(1-\gamma_5)(W_{c2}^\dagger
   u)\right]
   \left[ (\bar u W_{c1})(W_{c1}^\dagger iD\!\!\!\!\slash_{\perp c1} W_{c1})(sn_+)(1-\gamma_5) h_v\right],\nonumber\\
     Q_{2s,3s}^{(1)}(t,s)&=&
  -\frac{1}{m_b}\left[(\bar u W_{c2})(tn_-)\frac{n\!\!\!\slash_-}{n_-v}(1\mp\gamma_5)(W_{c2}^\dagger
   u)\right]
   \left[ (\bar s W_{c1})(W_{c1}^\dagger iD\!\!\!\!\slash_{\perp c1} W_{c1})(sn_+)(1-\gamma_5) h_v\right],\nonumber\\
  Q_{4s}^{(1)}(t,s)&=&
  -\frac{1}{m_b}\left[(\bar s W_{c2})(tn_-)\frac{n\!\!\!\slash_-}{n_-v}(1-\gamma_5)(W_{c2}^\dagger
   q)\right]
   \left[ (\bar q W_{c1})(W_{c1}^\dagger iD\!\!\!\!\slash_{\perp c1} W_{c1})(sn_+)(1-\gamma_5) h_v\right],\nonumber\\
     Q_{5s,6s}^{(1)}(t,s)&=&
  -\frac{1}{m_b}\left[(\bar q W_{c2})(tn_-)\frac{n\!\!\!\slash_-}{n_-v}(1\mp\gamma_5)(W_{c2}^\dagger
   q)\right]
   \left[ (\bar s W_{c1})(W_{c1}^\dagger iD\!\!\!\!\slash_{\perp c1} W_{c1})(sn_+)(1-\gamma_5) h_v\right],\nonumber\\
     Q_{7s}^{(1)}(t,s)&=&
  -\frac{1}{m_b}\left[(\bar s W_{c2})(tn_-)n\!\!\!\slash_-\gamma^\perp_\mu (1+\gamma_5)(W_{c2}^\dagger
   u)\right]
   \left[ (\bar u W_{c1})(W_{c1}^\dagger iD^{\mu}_{\perp c1} W_{c1})(sn_+)(1-\gamma_5)(sn_+) h_v\right],\nonumber\\
     Q_{8s}^{(1)}(t,s)&=&
  -\frac{1}{m_b}\left[(\bar s W_{c2})(tn_-)n\!\!\!\slash_-\gamma^\perp_\mu (1+\gamma_5)(W_{c2}^\dagger
   q)\right]
   \left[ (\bar q W_{c1})(W_{c1}^\dagger iD^{\mu}_{\perp c1} W_{c1})(sn_+)(1-\gamma_5) h_v\right],\nonumber\\
   Q_{gs}^{(1)}(t,s)&=& -2m_b i\epsilon_{\perp\mu\nu}{\rm Tr}
   \left[ [W_{c2}^\dagger iD^\mu_{\perp {c2}} W_{c2}](tn_-)  [W_{c2}^\dagger iD^\nu_{\perp {c2}} W_{c2}]
   \right]\nonumber\\
   &&\;\;\;\times
   \left[ (\bar s W_{c1})(W_{c1}^\dagger iD\!\!\!\!\slash_{\perp c1} W_{c1})(sn_+)(1-\gamma_5)
   h_v\right] ,\label{eq:nextleadingpoweroperators}
\end{eqnarray}
where the fields without position argument are at $x=0$. The field
products within the square brackets are color-singlet and we will
neglect the colour-octet operators since they give vanishing matrix
elements at leading order. The operators responsible for $b\to d$
transitions could be directly obtained by replacing $s$ quark fields
by the corresponding $d$ quark fields. Although the operators given
in Eq.~\eqref{eq:nextleadingpoweroperators} are suppressed by
$\lambda$ compared with those in
Eq.~\eqref{eq:leadingpoweroperators}, all of the operators in
Eq.~\eqref{eq:leadingpoweroperators} and
Eq.~\eqref{eq:nextleadingpoweroperators} contribute to $\langle
M_1M_2|O|B\rangle$ at the same power when matching onto SCET$_{II}$.
Hence the effective Hamiltonian are matched onto SCET$_I$ by the
following equation:
\begin{eqnarray}
 {\cal H}_{eff}&=& \frac{G_F}{\sqrt 2} \left\{ \int d\hat t \; \hat c_i  (\hat t) O_i^{(0)}(t)+
  \int d\hat t d\hat s\;\hat b_i (\hat t,\hat s)
  O_i^{(1)}(t,s)\right\},
\end{eqnarray}
with $\hat s=n_+\cdot p's=m_Bs$, $\hat t=n_-\cdot qt=m_Bt$ ($p'$ and
$q$ are the momentum of the recoiling and emitted meson,
respectively). We usually evaluate the Wilson coefficients $c_i(u)$
and $b_i(u,z)$ in momentum space which is related to the ones in
coordinated space by:
\begin{eqnarray}
 c_i(u)= \int d\hat t e^{-i um_B\hat t }\hat c_i (\hat t),\;\;\;
 b_i(u,z)=\int d\hat t e^{-i m_B (u\hat t+z\hat s) }\hat b_i (\hat t, \hat
  s).
\end{eqnarray}
The tree level matching coefficients for the four-body operators in
eq.~\eqref{eq:leadingpoweroperators} are given by:
 \begin{eqnarray}
    c_{1,2}^{(f)}&=&\lambda_u^{(f)}\Big[C_{1,2}+\frac{1}{N_c}C_{2,1}\Big]
    -\lambda_t^{(f)}\frac{3}{2}\Big[\frac{1}{N_c}C_{9,10}
    +C_{10,9}\Big],\nonumber\\
    c_3^{(f)}&=& -\frac{3}{2}\lambda_t^{(f)}\Big[C_7+\frac{1}{N_c}C_8\Big],\nonumber\\
    c_{4,5}^{(f)}&=&-\lambda_t^{(f)}\Big[\frac{1}{N_c}C_{3,4}+C_{4,3}-\frac{1}{2N_c}C_{9,10}-\frac{1}{2}C_{10,9}\Big],
    \nonumber\\
    c_6^{(f)}&=&-\lambda_t^{(f)}\Big[C_5+\frac{1}{N_c}C_6-\frac{1}{2}C_7-\frac{1}{2N_c}C_8\Big],\nonumber\\
    c_{g}^{(f)}&=&0. \label{eq:wcleadingpower}
\end{eqnarray}
The tree level matching of five-body operators leads to:
\begin{eqnarray}
   b_{1,2}^{(f)}  &=&\lambda_u^{(f)}\Big[C_{1,2}+\frac{1}{N_c}\Big(1-\frac{m_b}{\omega_3}\Big)C_{2,1}\Big]
        -\lambda_t^{(f)}\frac{3}{2}\Big[C_{10,9}+\frac{1}{N_c}\Big(1-\frac{m_b}{\omega_3}\Big)C_{9,10}\Big],
        \nonumber\\
   b_3^{(f)}      &=&-\lambda_t^{(f)}\frac{3}{2}\Big[C_7+\Big(1-\frac{m_b}{\omega_2}\Big)\frac{1}{N_c}C_8\Big],
   \nonumber\\
   b_{4,5}^{(f)}  &=&-\lambda_t^{(f)}\Big[C_{4,3}+\frac{1}{N_c}\Big(1-\frac{m_b}{\omega_3}\Big)C_{3,4}\Big]
               +\lambda_t^{(f)}\frac{1}{2}\Big[C_{10,9}+\frac{1}{N_c}\Big(1-\frac{m_b}{\omega_3}\Big)C_{9,10}\Big],
               \nonumber\\
    b_6^{(f)}     &=&-\lambda_t^{(f)}\Big[C_5+\frac{1}{N_c}\Big(1-\frac{m_b}{\omega_2}\Big)C_6\Big]
      +\lambda_t^{(f)}\frac{1}{2}\Big[C_{7}+\frac{1}{N_c}\Big(1-\frac{m_b}{\omega_2}\Big)C_{8}\Big],\nonumber\\%
    b_7^{(f)}     &=&-\lambda_t^{(f)}\frac{3}{2}C_7\frac{1}{N_c}
    \Big(\frac{m_b}{\omega_2}-\frac{m_b}{\omega_3}\Big), \nonumber\\
    b_8^{(f)}     &=&-\lambda_t^{(f)}\Big(C_5-\frac{1}{2}C_7\Big)\frac{1}{N_c}\Big(\frac{m_b}{\omega_2}
         - \frac{m_b}{\omega_3}\Big),\nonumber\\
    b_{g}^{(f)}   &=&\lambda_t^{(f)} C_{8g}
     \frac{\alpha_s(m_b)}{16C_F} \Big(\frac{1}{\bar u}-\frac{1}{u}\Big)\Big[ \frac{2+z}{1-z}
      +2\left(1-\frac{1}{N_c^2}\right)\frac{ u\bar u}{(1-z u)(1-z \bar
      u)}
      \Big],\label{eq:wcsubleadingpower}
      \end{eqnarray}
where $\omega_2=u m_B$ and $\omega_3=-\bar um_B$ with $u$ is the
momentum fraction of the positive quark in the emitted meson. $m_B$
is the $B$-meson mass.
$C_F=(N_c^2-1)/2N_c$ and $N_c=3$. The one-loop corrections are given
in
Refs.~\cite{Beneke:1999br,Beneke:2000ry,Chay:2003ju,Beneke:2005vv,Beneke:2006mk,Jain:2007dy}.
The coefficients $c_g^f$ and $b_g^f$ are zero at ${\cal
O}(\alpha_s^0)$, thus they are not relevant for the present study in
which we concentrate on the  leading order analysis.

In SCET$_I$, the matrix elements of $O_i^{(0,1)}$ can be decomposed
into some simple and universal ones defined as follows:
\begin{eqnarray}
 &&\langle M_1|(\bar \chi
 W_{c2})(tn_-)\frac{n\!\!\!\slash_-}{2}(1-\gamma_5)
 (W_{c2}^\dagger\chi)|0\rangle =\frac{if_{M_1}m_B}{2}\int ^1_0du
 e^{iu\hat t}\phi_{M_1}(u),\nonumber\\
 && \langle M_2|T[(\bar \chi W_{c1})n\!\!\!\slash _+(1-\gamma_5)
 h_v]|B\rangle=m_B\zeta,\nonumber\\
 && \langle M_2|T[(\bar u W_{c1})(W_{c1}^\dagger iD\!\!\!\!\slash_{\perp c1} W_{c1})(sn_+)(1-\gamma_5)
 h_v]|B\rangle=-m_B^2\int dz e^{im_Bz\cdot s}\zeta_J(z),
\end{eqnarray}
where $M_2$ is an arbitrary pseudo-scalar meson or vector meson
except $\eta$ and $\eta'$.

\subsection{Matching to SCET$_{\rm II}$}

The matching of SCET$_{\rm I}$ onto SCET$_{\rm II}$ is performed by
integrating out the degrees of freedom with $p^2\sim \Lambda m_b$.
To do so, it is useful to perform a redefintion of collinear fields:
$q\to Y_s q$, where $Y_s$ is a soft Wilson line. The SCET Lagrangian
contains no leading order interactions between the collinear-2 and
collinear-1 fields after decoupling soft-gluons from collinear-2
sector by a field re-definition. Although soft Wilson lines still
appear in the effective electro-weak operators, the Wilson line only
appear in the combination of $Y_s h_v$. Thus the two kinds of
collinear sectors decouple and the decay amplitudes factorize.

In SCET$_{II}$, the end-point singularity prevents the factorization
of $\zeta$ while the form factor $\zeta_J^{BM}(z)$ can be further
factorized into convolution of light-cone-distribution amplitudes
(LCDAs) and jet functions:
\begin{eqnarray}
\zeta_J^{BM}(z) &= &\frac{f_Bf_M}{m_B}\negthickspace
\int\negthickspace dk_+ dx\phi_B^+(k_+)
    J(z, x, k_+)\phi_{M}(x).
\end{eqnarray}
At the lowest order, $J(z,x,k_+)=\delta(z-x)\alpha_s \pi C_F/(N_c
\bar x k_+)$.

\subsection{Decay amplitudes involving flavor-singlet mesons $\eta$ and $\eta'$}

For iso-singlet mesons $\eta$ and $\eta'$, we adopt the
Feldmann-Kroll-Stech (FKS) mixing scheme
\cite{Feldmann:1998vh,Feldmann:1998sh,Feldmann:1999uf}. In this
scheme, an arbitrary iso-singlet biquark operator $O$ can be written
as a linear combination of $O_{q}\sim (u\bar u+d\bar d)/\sqrt{2}$
and $O_s\sim s\bar s$ operators with the well defined flavor
structure. Matrix elements of $O=c_q O_q+c_s O_s$ between $\eta$,
$\eta'$ states and the vacuum state can be parameterized by
\begin{eqnarray}
\langle0|O|\eta\rangle&=c_q \cos \phi_q  \langle O_q\rangle  - c_s \sin \phi_s \langle O_s \rangle,\\
\langle0|O|\eta'\rangle&=c_q \sin \phi_q  \langle O_q \rangle + c_s
\cos \phi_s \langle O_s \rangle,
\end{eqnarray}
where the four matrix elements $\langle 0|O_{q,s}|\eta{(')}\rangle$
are expressed by the two angles $\phi_{q,s}$ and two reduced matrix
elements $\langle O_{q,s}\rangle$. Phenomenologically, one can
neglect the OZI suppressed matrix elements and obtain
$\phi_q=\phi_s=\theta$. Thus, the mass eigenstates $\eta$, $\eta'$
are related to the flavor basis through:
 \begin{eqnarray}
\eta&=\eta_q \cos \theta -\eta_s \sin \theta,\nonumber\\
\eta'&=\eta_q \sin \theta +\eta_s \cos \theta. \label{eq:etamix}
\end{eqnarray}
For these iso-singlet mesons $\eta_q$ and $\eta_s$, we need in
addition more theoretical inputs which arise from the higher Fock
state component:
\begin{eqnarray}
 &&i\epsilon_{\perp\mu\nu} \langle \eta_q(p)|\mbox {Tr}[[W_{c2}^\dagger iD^\mu_{\perp {c2}}
 W_{c2}](tn_-)
 [W_{c2}^\dagger iD^\nu_{\perp {c2}} W_{c2}]]
 |0\rangle =\int^1_0du e^{iu \hat t}\frac{i}{4}\sqrt {C_F}
 \sqrt {\frac{2}{ 3}} f_{\eta_q}\bar\Phi_P^g(u),\nonumber\\
 &&i\epsilon_{\perp\mu\nu} \langle \eta_s(p)|\mbox {Tr}[[W_{c2}^\dagger iD^\mu_{\perp {c2}} W_{c2}](tn_-)
 [W_{c2}^\dagger iD^\nu_{\perp {c2}} W_{c2}]]
 |0\rangle =\int^1_0du e^{iu \hat t}\frac{i}{4}\sqrt {C_F}
  \sqrt {\frac{1}{ 3}} f_{\eta_s}\bar\Phi_P^g(u),\nonumber\\
 && (\langle \eta_q|T[(\bar \chi W_{c1})n\!\!\!\slash _+(1-\gamma_5)
 h_v]|B\rangle)_g= \sqrt2m_B\zeta_g,\nonumber\\
 && (\langle \eta_s|T[(\bar \chi W_{c1})n\!\!\!\slash _+(1-\gamma_5)
 h_v]|B\rangle)_g=  m_B\zeta_g,\nonumber\\
 && (\langle \eta_q|T[(\bar  \chi W_{c1})(W_{c1}^\dagger iD\!\!\!\!\slash_{\perp c1} W_{c1})(sn_+)(1-\gamma_5)
 h_v]|B\rangle)_g=-  \sqrt2m_B^2\int dz e^{im_Bz\cdot s}\zeta_{Jg}(z),\nonumber\\
 && (\langle \eta_s|T[(\bar  \chi W_{c1})(W_{c1}^\dagger iD\!\!\!\!\slash_{\perp c1} W_{c1})(sn_+)(1-\gamma_5)
 h_v]|B\rangle)_g=-   m_B^2\int dz e^{im_Bz\cdot s}\zeta_{Jg}(z),
\end{eqnarray}
where only the gluonic contributions to $B\to\eta_q,\eta_s$ form
factors are shown. Please note that, our convention is different
from the one used in Ref.~\cite{Williamson:2006hb}, where the form
factors $\zeta_g$ and $\zeta_{Jg}$ are incorporated in the
definition of $\zeta^{BM_2}_{(J)}$. Here we have separated them out
and  the two functions $\zeta^{BM_2}_{(J)}$ do not contain
contributions from the gluonic term. This convention is more
convenient when extracting the hard kernels using master equations
given in the appendix.

In SCET$_{II}$, $\zeta_g$ can not be factorized either for the
presence of end-point singularity  but $\zeta_{Jg}^{BM}(z)$ is given
in terms of the jet functions by:
\begin{eqnarray}
 \zeta_{Jg}^{BM}(z) &=&\frac{f_Bf_M}{m_B}\frac{1}{4} \sqrt{\frac{C_F}{3}}
\int  dk_+dx \phi_B^+(k_+)J_g(z, x, k_+) \bar\Phi_{M}^g(x),
\end{eqnarray}
At the lowest order, $J_g(z,x,k_+)=\delta(z-x)\alpha_s 2 \pi/(N_c
k_+)$.

\subsection{A summary of the factorization formulae}

%
%

In summary, the $b\to s(d)$ decay amplitudes at leading power in
SCET can be expressed by:
\begin{eqnarray}
  A(B\to M_1 M_2)&=&\frac{G_F}{\sqrt{2}} m_B^2 \Big\{f_{M_1} \int\negthickspace du
  \phi_{M_1}(u) T_{1}(u)\zeta^{BM_2}+
  f_{M_1}\int\negthickspace du\phi_{M_1}(u) \int\negthickspace dz
  T_{1J}(u,z) \zeta_J^{BM_2}(z)    \nonumber\\
  &&+f_{M_1} \int\negthickspace du
  \phi_{M_1}(u) T_{1g}(u)\zeta_g^{BM_2}+
  f_{M_1}\int\negthickspace du\phi_{M_1}(u) \int\negthickspace dz
  T_{1Jg}(u,z) \zeta_{Jg}^{BM_2}(z)    \nonumber\\
  && + f_{M_1}^1
  \int\negthickspace du  \bar \Phi_{M_1}^g(u)
  T_{1}^g(u)\zeta^{BM_2}+ f_{M_1}^1 \int\negthickspace du  \bar \Phi_{M_1}^g(u)
  \int\negthickspace dz T_{1J}^g(u,z)\zeta_J^{BM_2}(z) \nonumber\\
  && + f_{M_1}^1
  \int\negthickspace du  \bar \Phi_{M_1}^g(u)
  T_{1g}^g(u)\zeta_g^{BM_2}+ f_{M_1}^1 \int\negthickspace du  \bar \Phi_{M_1}^g(u)
  \int\negthickspace dz T_{1Jg}^g(u,z)\zeta_{Jg}^{BM_2}(z) \nonumber\\
  &&+ \lambda_c^{(f)} A^{M_1
   M_2}_{cc} + (1\leftrightarrow 2)\Big\},\label{eq:factorizationformulainperturbation}
\end{eqnarray}
where $A^{M_1 M_2}_{cc}$ denotes the non-perturbative charming
penguins. $T_i$ are hard kernels which can be calculated using
perturbation theory. In the appendix~\ref{sec:Hardkernels}, based on
the flavor structure of the four-body operators and five-body
operators, we give the master equations for hard kernels $T_i$ which
utilize the coefficients given in Eq.~(\ref{eq:wcleadingpower}) and
Eq.~(\ref{eq:wcsubleadingpower}). For distinct decay channels, one
can easily evaluate the equation to obtain the corresponding hard
kernels.

In SCET, the factorization formula for $B\to M_1 M_2$ is easily
proved to hold to all order in $\alpha_s$: the amplitudes given in
Eq.~\eqref{eq:factorizationformulainperturbation} have the form of a
convolution of the universal light-cone distribution amplitudes  and
the perturbative hard kernels. Utilizing the perturbative expansion
in $\alpha_s(\sqrt{m_b\Lambda})$ for the jet functions and in
$\alpha_s(m_b)$ for the Wilson coefficients, one can predict the
branching ratios, CP asymmetries and other observables for $B\to
M_1M_2$ decays. One can also use another parallel method: the
non-perturbative parameters can be fitted by experimental
measurements on the $B\to M_1M_2$ decays. This approach is
especially useful at leading order in $\alpha_s$, since then the
hard kernels $T_{1}(u)$ are constants, while $T_{1J}(u,z)$ are
functions of $u$ only. Furthermore, at this order terms with hard
kernels $T_{1J}^g(u,z)$, $T_{1}^g(u)$, $T_{1Jg}^g(u,z)$,
$T_{1g}^g(u)$ do not contribute at all. Thus the decay amplitudes of
$B\to M_1M_2$ decays at LO in $\alpha_s(m_b)$ are written by:
\begin{eqnarray}
  A(B\to M_1 M_2)&=&\frac{G_F}{\sqrt{2}} m_B^2 \left\{
  f_{M_1}\left[ \zeta_J^{BM_2}\int\negthickspace du\phi_{M_1}(u)  T_{1J}(u)
  +\zeta_{Jg}^{BM_2}\int\negthickspace du\phi_{M_1}(u)   T_{1Jg}(u)\right] \right.   \nonumber\\
  &&\left.+ f_{M_1}  (T_{1}
  \zeta^{BM_2}+T_{1g}\zeta_g^{BM_2})
  +\lambda_c^{(f)} A^{M_1
   M_2}_{cc} + (1\leftrightarrow 2)\right\},\label{eq:leadingpowerfactorization}
\end{eqnarray}
where the four functions $\zeta^{BM_1}$, $\zeta_g$  and
\begin{eqnarray}
 \zeta_J^{BM_2}=\int dz \zeta_J^{BM_2}(z),\;\;\;\zeta_{Jg}^{BM_2}=\int dz \zeta_{Jg}^{BM_2}(z),
\end{eqnarray}
are treated as non-perturbative parameters to be fitted from
experiment measurements.

In order to reduce the independent inputs, one can utilize the SU(3)
symmetry for $B$ to light form factors and charming penguins.  In
the exact SU(3) limit, only two form factors are needed for $B\to
PP$ decays without iso-singlet mesons:
\begin{eqnarray}\label{zetaSU3}
\zeta_{(J)}^{BP} \equiv \zeta_{(J)}^{B\pi}=
\zeta_{(J)}^{BK}=\zeta_{(J)}^{B_sK}.
\end{eqnarray}
Besides these two form factors, there are two additional new
non-perturbative functions $\zeta_{(J)g}$ in decays involving
iso-singlet mesons $\eta_q$ and $\eta_s$. They are contributions
from the intrinsic gluons. The $B\to V$ form factors are rather
simple, since there is no gluonic contribution at all. The flavor
SU(3) symmetry implies the relation for $B\to V$ form factors:
\begin{eqnarray}
 \zeta_{(J)}^{BV} \equiv\zeta^{B\rho}_{(J)}=\zeta^{B K^*}_{(J)}=\zeta^{B\omega}_{(J)}
 =\zeta^{B_s  K^*}_{(J)}=\zeta^{B_s\phi}_{(J)}.
\end{eqnarray}
If the SU(3) symmetry is assumed for charming penguins, there are
totally five complex charming penguins which depends on the spin and
isospin properties of the emitted mesons and recoiling mesons:
$A^{PP}_{cc}, A^{PV}_{cc}, A^{VP}_{cc}, A_{ccg}^{PP}, A^{VP}_{ccg}$.
$A^{M_1M_2}_{cc}$ denotes the charming penguins in which  the $M_1$
meson is emitted and the $M_2$ meson is recoiled. The two charming
penguins $A_{ccg}^{PP}, A^{VP}_{ccg}$ only contributes to decays in
which a iso-singlet meson is recoiled.

With the assumption of flavor SU(3) symmetry for $B$ to light form
factors and charming penguin terms, the non-perturbative, totally 16
real inputs responsible for $B\to PP$ and $B\to VP$ decays are
summarized in the following:
\begin{eqnarray}
   \zeta^{BP},\;\;\;\zeta_J^{BP}, \;\;\;\zeta_g,\;\;\;\zeta_{Jg},\;\;\;
    \zeta^{BV},\;\;\; \zeta^{BV}_{J},\;\;\;A^{PP}_{cc},\;\;\;
    A^{PV}_{cc},\;\;\;A^{VP}_{cc},\;\;\; A_{ccg}^{PP},\;\;\;
    A^{VP}_{ccg}.
\end{eqnarray}

\section{Chiraly enhanced penguins}\label{sec:corrections}

Power corrections are expected to be suppressed by at least the
factor $\Lambda_{QCD}/m_b$, but chiraly enhanced penguins are large
enough to compete with the leading power QCD penguins as the
suppression factor becomes $2\mu_P/m_b$, where $\mu_P\sim 2$ GeV is
the chiral scale parameter. Thus in both of
QCDF~\cite{Beneke:1999br,Beneke:2000ry,Beneke:2003zv} and
PQCD~\cite{Keum:2000ph,Keum:2000wi,Lu:2000em} approaches, it has
been incorporated in the phenomenological analysis. In the framework
of SCET, the complete operator basis and the corresponding
factorization formulae for the chiraly enhanced penguin are recently
derived in Ref.~\cite{Arnesen:2006vb,Jain:2007dy} and the amplitudes
do not suffer from additional endpoint singularities. The
factorization formula will introduce a new form factor
$\zeta_{\chi}$ and a new light-cone distribution amplitude
$\phi^{pp}$.

As discussed in Ref.~\cite{Jain:2007dy}, there are three different
kinds of chiraly enhanced penguin operators in SCET$_I$:
$Q_A^{1\chi}$, $Q_B^{(1\chi)}$ and $Q_B^{(2\chi)}$. The basis for
the $Q_A^{(1\chi)}$-type operators is given by:
\begin{eqnarray}
 Q_{1(qfq)}^{(1\chi)}&=&\frac{1}{m_b}\left[ (\bar q W_{c1})(1-\gamma_5)
   h_v\right]\left[(\bar s
W_{c2})(tn_-)\frac{n\!\!\!\slash_-}{n_-v}{i\partial}\!\!\!\slash_\perp(1+\gamma_5)(W_{c2}^\dagger
   q)\right],\nonumber\\
 Q_{2(qfq)}^{(1\chi)}&=&Q_{1(qfq)}^{(1\chi)}\frac{3}{2}e_q.
\end{eqnarray}
These two operators  $Q_{1,2}^{(1\chi)}$ will contribute to $B\to
PP, VP, V_LV_L$ decays (here $V_L$ denotes a longitudinally
polarized vector meson). There are in addition several operators
omitted here, as they can only contribute to $B\to V_T V_T$ decays
($V_T$ denotes a transversely polarized vector meson). The second
kinds of operators which are responsible for $B\to PP, VP, V_LV_L$
decays are given by:
\begin{eqnarray}
 Q_{1(qfq)}^{(2\chi)} &=& \frac{-1}{m_b}\left[ (\bar q W_{c1}) \frac{1}{n_+\cdot {i\partial}}
 {i\partial}_\perp\cdot (W_{c1}^\dagger iD_{\perp c1} W_{c1})(sn_+)
 (1+\gamma_5)  h_v\right]\nonumber\\
 &&\times
 \left[(\bar s W_{c2})(tn_-)n\!\!\!\slash_-(1-\gamma_5)(W_{c2}^\dagger q)\right],\\
 Q_{2(fuu)}^{(2\chi)} &=& \frac{-1}{m_b}\left[ (\bar s W_{c1}) \frac{1}{n_+\cdot {i\partial}}
 {i\partial}_\perp\cdot (W_{c1}^\dagger iD_{\perp c1} W_{c1})(sn_+)
 (1+\gamma_5)  h_v\right]\nonumber\\
 &&\times
 \left[(\bar u W_{c2})(tn_-)n\!\!\!\slash_-(1+\gamma_5)(W_{c2}^\dagger u)\right],\\
 Q_{3(qfq)}^{(2\chi)} &=& \frac{-1}{m_b^2}\left[ (\bar q W_{c1})(W_{c1}^\dagger iD\!\!\!\!\slash_{\perp c1}
  W_{c1})(sn_+)(1-\gamma_5) h_v\right]\nonumber\\
 &&\times
 \left[(\bar s W_{c2})(tn_-)\frac{n\!\!\!\slash_-}{n_-v}{i\partial}\!\!\!\!\slash_\perp(1+\gamma_5)
 (W_{c2}^\dagger q)\right],\\
 Q_{4(qfq)}^{(2\chi)} &=& \frac{3}{2}e_qQ_{3(qfq)}^{(2\chi)},
\end{eqnarray}
plus operators with the same Dirac structure but different flavors,
$Q^{(2\chi)}_{1(ufu)}$ and $Q^{(2\chi)}_{1(fuu)}$. If
$n_-$-iso-singlet operators are included, we have two additional
operators $Q_{1(fqq)}^{(2\chi)}$ and $Q_{2(fqq)}^{(2\chi)}$.
Operators $Q^{(2\chi)}_{1-4}$ contribute to $B\to PP, VP, V_LV_L$
decays, while operators which only contribute to $B\to V_T V_T$
decays are also given in Ref.~\cite{Jain:2007dy} but omitted here,
since we mainly concentrate on $B\to PP$ and $B\to VP$ decays.

Matching from QCD to SCET$_I$, one obtains the effective Hamiltonian
expressed by the $(1\chi)$ and $(2\chi)$-type operators contributing
to $B\to PP, VP, V_L V_L$ decays:
\begin{eqnarray}
 {\cal H}_{eff}^\chi=\frac{G_F}{\sqrt2}\left[\int d\hat t \hat c^\chi_{i(F)}(\hat t)Q^{(1\chi)}_{i(F)}( t)
  + \int d\hat t d\hat s
 \hat b^\chi_{i(F)}(\hat t, \hat s)Q^{(2\chi)}_{i(F)}(t, s)\right],
\end{eqnarray}
where the indices run over the operator number $i$ and possibilities
for the flavors $F$ for the $Q_{i(F)}$. $\hat c_{i(F)}^\chi$ and
$\hat b_{i(F)}^\chi$ are the short-distance Wilson coefficients in
coordinate space. At tree level, the corresponding coefficients in
momentum space are:
\begin{eqnarray}
 &&c_{1(qfq)}^\chi=\lambda_t^{(f)} (C_6+\frac{C_5}{N_c})\frac{1}{u\bar
 u},\;\;\; c_{2(qfq)}^\chi=\lambda_t^{(f)} (C_8+\frac{C_7}{N_c})\frac{1}{u\bar
 u},\nonumber\\
 &&b_{1(qfq)}^\chi=\lambda_t^{(f)} \left[\frac{1+uz}{uz}(\frac{C_3}{N_c}-\frac{C_9}{2N_c}) +C_4-\frac{C_{10}}{2}\right],\;\;\;
  b_{2(fuu)}^\chi=3\lambda_t^{(f)} [C_7+\frac{C_8}{N_c}-\frac{C_8}{\bar uzN_c}],\nonumber\\
 &&b_{1(ufu)}^\chi=\frac{2(1+uz)}{uz}\left(-\frac{C_2}{N_c}\lambda_u^{(f)} + \frac{3C_9}{2N_c}\lambda_t^{(f)}\right)
 -(2C_1\lambda_u^{(f)}-3C_{10}\lambda_t^{(f)}),\nonumber\\
 &&b_{1(fuu)}^\chi=\frac{2(1+uz)}{uz}\left(-\frac{C_1}{N_c}\lambda_u^{(f)} + \frac{3C_{10}}{2N_c}\lambda_t^{(f)}\right)
 -(2C_2\lambda_u^{(f)}-3C_{9}\lambda_t^{(f)}),\nonumber\\
 &&b_{3(qfq)}^\chi=\lambda_t^{(f)} (C_6+\frac{C_5}{N_c})\frac{1}{u\bar
 u},\;\;\; b_{4(qfq)}^\chi=\lambda_t^{(f)} (C_8+\frac{C_7}{N_c})\frac{1}{u\bar
 u}.
\end{eqnarray}

Matrix elements for these operators can be parametrized into the
following universal distributions:
\begin{eqnarray}
 &&\langle M|\left[ (\bar q W_{c1}) \frac{1}{\bar n\cdot {i\partial}}{i\partial}_\perp\cdot
 (W_{c1}^\dagger iD_{\perp c1} W_{c1})(sn_+)
 (1+\gamma_5)  h_v\right] | \bar B\rangle= -\frac{\mu_Mm_B}{6}\int dz e^{im_Bz\cdot
 s}\zeta^{BM}_\chi(z),\nonumber\\
 &&\langle M(p)|
 \left[(\bar s W_{c2})(tn_-)\frac{n\!\!\!\slash_-}{n_-v}{i\partial}\!\!\!\!\slash_\perp(1+\gamma_5)(W_{c2}^\dagger
 q)\right]|0\rangle =-\frac{if_{M}\mu_{M}}{3}\int ^1_0du
 e^{iu\hat t}\phi^{pp}_{M}(u),
\end{eqnarray}
where $\mu_M$ is the chiral scale parameter which is set to zero for
vector mesons. Using equation of motion, the pseudo-scalar's
light-cone distribution amplitude $\phi_{P}^{pp}(u)$ can be related
to ones defined in QCD~\cite{Arnesen:2006vb,Hardmeier:2003ig}:
\begin{eqnarray}
 \phi_{P}^{pp}(u)=3u\left[\phi_p+\frac{\phi'_\sigma}{6}+\frac{2f_{3P}}{f_P\mu_P}
 \int \frac{dv}{v}\phi_{3P}(u-v,u)\right].
\end{eqnarray}
In the Wandzura-Wilczek approximation, $\phi_{3P}$ vanishes and one
gets $\phi^{pp}_{P}(u)=6u(1-u)$ for the asymptotic form. With the
above matrix elements, generic decay amplitudes from the chiral
enhanced penguin could be written as:
\begin{eqnarray}
  A^\chi(B\to M_1 M_2)&=&\frac{G_F}{\sqrt{2}} m_B^2 \Big\{-\frac{\mu_{M_1}f_{M_1}}{3m_B} \int\negthickspace du
  \phi^{M_1}_{pp}(u) T_{1}^\chi(u)\zeta^{BM_2}\nonumber\\
  && -\frac{\mu_{M_1}f_{M_1}}{3m_B} \int\negthickspace
  du dz
  \phi^{M_1}_{pp}(u) T_{1J}^\chi(u,z)\zeta^{BM_2}_J(z) \nonumber\\
  &&-\frac{\mu_{M_1}f_{M_1}}{3m_B} \int\negthickspace du
  \phi^{M_1}_{pp}(u) T_{1g}^\chi(u)\zeta_g^{BM_2}\nonumber\\
  &&   -\frac{\mu_{M_1}f_{M_1}}{3m_B} \int\negthickspace du dz
  \phi^{M_1}_{pp}(u) T_{1Jg}^\chi(u,z)\zeta^{BM_2}_{Jg}(z) \nonumber\\
  &&-\frac{\mu_{M_2}f_{M_1}}{6m_B} \int\negthickspace du dz
  \phi^{M_1}(u) T_\chi(u,z)\zeta^{BM_2}_\chi(z) +(1\leftrightarrow2)
  \Big\},\label{eq:chiralyenhancedpenguin}
\end{eqnarray}
where $\zeta_\chi(z)$ can be expressed as convolutions of LCDAs and
jet functions:
\begin{eqnarray}
 \zeta_{\chi}^{BM}(z) &= \frac{f_B f_M}{ m_b}\int_0^1   dx
\int_0^\infty   dk^+ \, \frac{J_\perp(z,k^+,x)}{1-z}
  \phi_B^+(k^+)\phi_{pp}^M(x) .
\end{eqnarray}
Here $J_\perp(z,x,k_+) = {\delta(x -z) \pi \alpha_s C_F}/{(N_c\,
\bar x k_+)}$ at lowest order.

As emphasized in section~\ref{sec:decayamplitudes}, the leading
power SCET phenomenological analysis is very useful especially at
tree level. It does simplify the analysis. Even taking into account
the first four terms in Eq.~(\ref{eq:chiralyenhancedpenguin}), the
scheme for phenomenological studies will remain. But considering the
chiraly enhanced penguins, the factorization formulae involves a new
form factor $\zeta_\chi$ which can not be simplified into a
normalization constant even at tree level. As shown in
Ref.~\cite{Jain:2007dy}, the fifth term proportional to $\zeta_\chi$
is small which does not give sizable contributions. Thus in our
analysis, we neglect it and only consider the first four terms:
\begin{eqnarray}
  A^\chi(B\to M_1 M_2)&=&\pm\frac{G_F}{\sqrt{2}} m_B^2 \left(-\frac{2\mu_{M_1}f_{M_1}}{m_B}\right)\Big\{
  T_{1}^\chi\zeta^{BM_2}+
  T_{1J}^\chi\zeta_J^{BM_2}+
  T_{1g}^\chi\zeta_g^{BM_2}+
  T_{1Jg}^\chi\zeta_{Jg}^{BM_2}\nonumber\\
  &&\;\;\;+(1\leftrightarrow 2)\Big\}.
\end{eqnarray}
For $B\to PP$ decays, the chiraly enhanced penguin takes a plus
sign; while in $B\to VP$ decays, when emitting a pseudoscalar meson,
the amplitude take a minus sign; when a vector meson emitted, there
is no contribution from chiraly enhanced penguin since $\mu_V=0$.


\section{Numerical analysis of  $B\to VP$ decays}
\label{sec:numerical}

\subsection{Input parameters}
In the factorization formulae, we will use the following values for
decay constants of the light pseudo-scalars and vector mesons ( in
units of GeV):
\begin{eqnarray}
 &&f_\pi=0.131,\;\;\; f_K=0.160,\;\;\; f_{\eta_q}=1.07f_\pi=0.140,\;\;\;
 f_{\eta_s}=1.34f_\pi=0.176,\nonumber\\
 &&f_\rho=0.209,\;\;\; f_{K^*}=0.217,\;\;\; f_{\omega}=0.195,\;\;\;
 f_{\phi}=0.231.
\end{eqnarray}
The mixing angle between $\eta_q$ and $\eta_s$ is chosen as
$\theta=39.3^\circ$~\cite{Feldmann:1998vh,Feldmann:1998sh,Feldmann:1999uf}.
For the CKM matrix elements and CKM angles, we use the updated
global fit results from CKMfitter group \cite{Charles:2004jd}:
\begin{eqnarray}
 &&V_{ud}=0.97400,\;\;\;\;\; V_{us}=0.22653,\;\;\; |V_{ub}|=(3.57^{+0.17}_{-0.17})\times
 10^{-3},\nonumber\\
 && V_{cd}=-0.22638 ,\;\;\; V_{cs}=0.97316,\;\;\; V_{cb}=(40.5^{+3.2}_{-2.9})\times
 10^{-3},\nonumber\\
 && |V_{td}|=(8.68^{+0.25}_{-0.33})\times 10^{-3},\;\;\; |V_{ts}|=(40.7^{+0.9}_{-0.8})\times
 10^{-3},\;\;\; V_{tb}=0.999135,\nonumber\\
 &&\beta=(21.7^{+0.017}_{-0.017})^\circ,\;\;\; \gamma=(67.6^{+2.8}_{-4.5})^\circ,\;\;\;
 \epsilon=(1.054 _{-0.051}^{ +0.049} )^\circ.
\end{eqnarray}
For the inverse moments of light-cone distribution amplitudes for
pseudo-scalar mesons, we use the same value as in
Ref.~\cite{Williamson:2006hb}:
\begin{eqnarray}
 && \langle x^{-1}\rangle_\pi=\langle x^{-1}\rangle_{\eta_q}=\langle
 x^{-1}\rangle_{\eta_s}=3.3,\;\;\;\langle x^{-1}\rangle_{K}=3.24,\;\;\;\langle x^{-1}\rangle_{\bar K}=3.42,
\end{eqnarray}
where the inverse moment of vector mesons' light-cone distribution
amplitudes are obtained utilizing the Gegenbauer moments evaluated
in QCD sum rules~\cite{Ball:2007rt}:
\begin{eqnarray}
 && \langle x^{-1}\rangle_\rho=\langle x^{-1}\rangle_{\omega}=3.45, \langle x^{-1}\rangle_\phi=3.54,
 \;\;\;\langle x^{-1}\rangle_{K^*}=2.79,\;\;\;\langle x^{-1}\rangle_{\bar
 K^*}=3.81.
\end{eqnarray}
For the chiral scale parameters, we use a universal value
$\mu_P=2.0$ GeV for pseudo-scalars and $\mu_V=0$ for vectors.

The experimental data of $B\to PP$ and $B\to VP$ branching ratios,
the direct CP asymmetries and the parameters in $B^0/\bar
B^0\to\pi^\pm\rho^\mp$ decays (which are defined in
Eq.~\eqref{eq:mixingCS}, Eq.~\eqref{eq:mixingAcp} and
Eq.~\eqref{eq:mixingAfbarf}) are given by
Heavy-Flavor-Averaging-Group (HFAG) \cite{Barberio:2006bi} and
Particle-Data-Group (PDG)~\cite{Yao:2006px}. The following
mixing-induced CP asymmetries in $B\to PP$ and $B\to VP$ decays are
also used in our analysis:
\begin{eqnarray}
 &&-\eta_fS(K_S\eta')=0.61 \pm 0.07,\;\;\;
 -\eta_fS(K_S\pi^0)=0.38 \pm 0.19,\;\;\; S(\pi^+\pi^-)=-0.61\pm0.08,\nonumber\\
 &&-\eta_f S(\phi K_S)=0.39\pm0.17,\;\;
 S(\pi^0\rho^0)=0.12\pm0.38,\nonumber\\
 &&  -\eta_fS(\rho^0 K_S)=0.61 ^{+0.22}_{-0.24}\pm0.09 \pm 0.08=0.61 ^{+0.25}_{-0.27},\;
 -\eta_fS(\omega K_S)=0.48\pm0.24,
\end{eqnarray}
where $\eta_f$ is the CP eigenvalue for the final state $f$.  The
branching ratio of $\bar B^0\to\bar K^{*0}\pi^0$ is not used in this
fitting, since the experimental data could only be viewed as an
upper bound.

\begin{figure}[thb]
\begin{center}
\includegraphics[scale=0.7]{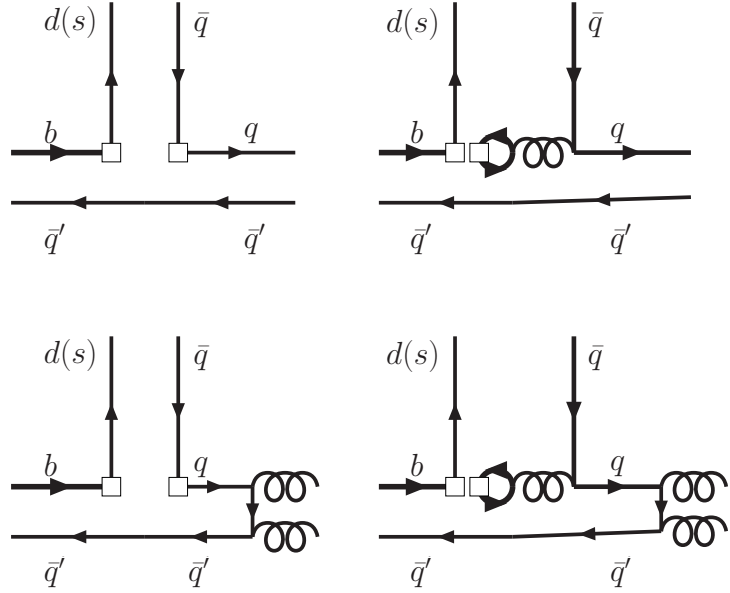}
\caption{Feynman diagrams for chiraly enhanced penguins (left) and
charming penguins (right). The two diagrams in the lower line only
contribute to decays involving $\eta$ or $\eta'$, where $q=q'$. }
\label{diagram:chiralyenhancedpenguin}
\end{center}
\end{figure}


With these data for branching fractions and CP asymmetries, $\chi^2$
fit method is used to determine the non-perturbative inputs: form
factors and charming penguins. Straightforwardly, we obtain the two
solutions for numerical results of the 16 non-perturbative inputs.
At leading order and leading power accuracy, the first solution is
(the charming penguins are given in units of GeV):
\begin{eqnarray}
 \zeta^P&=&(12.8\pm1.2)\times 10^{-2},\;\;\; \zeta^P_J=(7.2\pm0.7)\times
 10^{-2},\nonumber\\
 \zeta^V&=&(12.4\pm1.8)\times 10^{-2},\;\;\; \zeta^V_J=(10.8\pm1.9)\times
 10^{-2},\nonumber\\
 \zeta_g&=&(-5.3\pm2.2)\times 10^{-2},\;\;\; \zeta_{Jg}=(-2.3\pm2.9)\times
 10^{-2},\nonumber\\
 |A_{cc}^{PP}|&=&(48.1\pm0.6)\times 10^{-4},\;\;\;
 arg[A_{cc}^{PP}]=(167.5\pm2.5)^\circ,\nonumber\\
 |A_{cc}^{VP}|&=&(40.6\pm0.9)\times 10^{-4},\;\;\;
 arg[A_{cc}^{VP}]=(10.7\pm4.3)^\circ,\nonumber\\
 |A_{cc}^{PV}|&=&(30.7\pm1.3)\times 10^{-4},\;\;\;
 arg[A_{cc}^{PV}]=(194.3\pm4.6)^\circ,\nonumber\\
 |A_{ccg}^{PP}|&=&(38.4\pm1.9)\times 10^{-4},\;\;\;
 arg[A_{ccg}^{PP}]=(83.0\pm3.8)^\circ,\nonumber\\
 |A_{ccg}^{VP}|&=&(23.0\pm2.4)\times 10^{-4},\;\;\;
 arg[A_{ccg}^{VP}]=(38.4\pm23.0)^\circ,
\end{eqnarray}
and one can obtain the predictions for $B\to P$ (here $P$ denotes a
pseudoscalar except $\eta$ and $\eta'$) and $B\to V$ form factors at
tree level:
\begin{eqnarray}
 F^{B\to P}= \zeta^P+\zeta^P_J=0.201\pm0.015,\;\;\;
 A_0^{B\to V}= \zeta^V+\zeta^V_J=0.232\pm0.037.
\end{eqnarray}
In the above equations (and also in the following), the
uncertainties are obtained through the $\chi^2$-fit program. After
including the chiraly enhanced penguin, the numerical results for
these inputs are (the charming penguins are given in units of GeV):
\begin{eqnarray}
 \zeta^P&=&(13.7\pm0.8)\times 10^{-2},\;\;\; \zeta^P_J=(6.9\pm0.7)\times
 10^{-2},\nonumber\\
 \zeta^V&=&(11.7\pm1.0)\times 10^{-2},\;\;\; \zeta^V_J=(11.6\pm0.9)\times
 10^{-2},\nonumber\\
 \zeta_g&=&(-4.9\pm2.4)\times 10^{-2},\;\;\; \zeta_{Jg}=(-2.7\pm3.2)\times
 10^{-2},\nonumber\\
 |A_{cc}^{PP}|&=&(40.0\pm0.6)\times 10^{-4},\;\;\;
 arg[A_{cc}^{PP}]=(165.2\pm2.8)^\circ,\nonumber\\
 |A_{cc}^{VP}|&=&(41.0\pm0.9)\times 10^{-4},\;\;\;
 arg[A_{cc}^{VP}]=(11.9\pm4.2)^\circ,\nonumber\\
 |A_{cc}^{PV}|&=&(39.9\pm1.0)\times 10^{-4},\;\;\;
 arg[A_{cc}^{PV}]=(191.5\pm3.6)^\circ,\nonumber\\
 |A_{ccg}^{PP}|&=&(37.7\pm1.8)\times 10^{-4},\;\;\;
 arg[A_{ccg}^{PP}]=(88.3\pm4.1)^\circ,\nonumber\\
 |A_{ccg}^{VP}|&=&(25.3\pm2.3)\times 10^{-4},\;\;\;
 arg[A_{ccg}^{VP}]=(-18.7\pm12.3)^\circ,\label{eq:solution1}
\end{eqnarray}
which gives the predictions for $B\to P$ and $B\to V$ form factors
at tree level:
\begin{eqnarray}
 F^{B\to P}=0.206\pm0.004,\;\;\; A_0^{B\to V}=0.233\pm0.017.
\end{eqnarray}

As shown in Fig.~\ref{diagram:chiralyenhancedpenguin}, chiraly
enhanced penguins have the same topology with the charming penguins.
The former two diagrams do not only contribute to decays without
iso-singlet mesons $\eta$ or $\eta'$ but also decays with these
mesons. The two diagrams in the lower line only contribute to decays
involving $\eta$ or $\eta'$, where $q=q'$. The inclusion of chirally
enhanced penguin will mainly change the size of three charming
penguins $A_{cc}^{PP}$, $A_{ccg}^{PP}$, $A_{cc}^{PV}$. Predictions
for branching fractions and CP asymmetries will not be changed
sizably. After including the chiraly enhanced penguins, the total
$\chi^2/d.o.f$ for observables $B\to PP$ and $B\to VP$ is
$301/(86-16)$. If only the 55 observables in $B\to VP$ decays are
concerned, the total $\chi^2$ is 112.

Besides the above results, there is another solution at leading
power:
\begin{eqnarray}
 \zeta^P&=&(13.4\pm0.3)\times 10^{-2},\;\;\; \zeta^P_J=(5.8\pm0.4)\times
 10^{-2},\nonumber\\
 \zeta^V&=&(22.9\pm1.3)\times 10^{-2},\;\;\; \zeta^V_J=(6.6\pm1.4)\times
 10^{-2},\nonumber\\
 \zeta_g&=&(-10.3\pm1.2)\times 10^{-2},\;\;\; \zeta_{Jg}=(5.8\pm1.5)\times
 10^{-2},\nonumber\\
 |A_{cc}^{PP}|&=&(48.4\pm0.4)\times 10^{-4},\;\;\;
 arg[A_{cc}^{PP}]=(167.1\pm2.6)^\circ,\nonumber\\
 |A_{cc}^{VP}|&=&(29.7\pm0.8)\times 10^{-4},\;\;\;
 arg[A_{cc}^{VP}]=(159.3\pm6.9)^\circ,\nonumber\\
 |A_{cc}^{PV}|&=&(44.9\pm1.1)\times 10^{-4},\;\;\;
 arg[A_{cc}^{PV}]=(-10.5\pm2.9)^\circ,\nonumber\\
 |A_{ccg}^{PP}|&=&(38.4\pm2.2)\times 10^{-4},\;\;\;
 arg[A_{ccg}^{PP}]=(83.8\pm4.5)^\circ,\nonumber\\
 |A_{ccg}^{VP}|&=&(18.6\pm2.3)\times 10^{-4},\;\;\;
 arg[A_{ccg}^{VP}]=(220.6\pm10.7)^\circ,
\end{eqnarray}
which gives:
\begin{eqnarray}
 F^{B\to P}=0.192\pm0.005,\;\;\; A_0^{B\to V}=0.295\pm0.009.
\end{eqnarray}
With the inclusion of chirally enhanced penguin, these inputs
become:
\begin{eqnarray}
 && \zeta^P=(14.1\pm0.8)\times 10^{-2},\;\;\;\;\;\;\;\; \zeta^P_J=(5.6\pm0.7)\times
 10^{-2},\nonumber\\
 && \zeta^V=(22.7\pm1.7)\times 10^{-2},\;\;\;\;\; \;\;\;\zeta^V_J=(6.5\pm1.8)\times
 10^{-2},\nonumber\\
 && \zeta_g=(-10.0\pm0.9)\times 10^{-2},\;\;\;\;\;\;\; \zeta_{Jg}=(5.1\pm1.1)\times
 10^{-2},\nonumber\\
 && |A_{cc}^{PP}|=(40.6\pm0.6)\times 10^{-4},\;\;\;
 arg[A_{cc}^{PP}]=(164.9\pm2.8)^\circ,\nonumber\\
 && |A_{cc}^{VP}|=(29.4\pm0.8)\times 10^{-4},\;\;\;
 arg[A_{cc}^{VP}]=(158.4\pm5.8)^\circ,\nonumber\\
 && |A_{cc}^{PV}|=(33.5\pm1.1)\times 10^{-4},\;\;\;
 arg[A_{cc}^{PV}]=(-14.3\pm3.8)^\circ,\nonumber\\
 && |A_{ccg}^{PP}|=(37.8\pm1.3)\times 10^{-4},\;\;\;
 arg[A_{ccg}^{PP}]=(87.5\pm2.1)^\circ,\nonumber\\
 && |A_{ccg}^{VP}|=(18.3\pm2.4)\times 10^{-4},\;\;\;
 arg[A_{ccg}^{VP}]=(225.6\pm10.0)^\circ,\label{eq:solution2}
\end{eqnarray}
with the form factors:
\begin{eqnarray}
 F^{B\to P}=0.198\pm0.003,\;\;\; A_0^{B\to V}=0.291\pm0.011.
\end{eqnarray}
The corresponding $\chi^2=271/(86-16)$ ($\chi^2$ for the $55$
observables in all $B\to VP$ decays is $69$). Comparing the results
in the leading order analysis and those with chiraly enhanced
penguins, we can see that the charming penguins $A_{cc}^{PP}$ and
$A_{cc}^{PV}$ are changed sizably. It is reasonable since chiraly
enhanced penguins and charming penguins have the same topology.  The
phase of $A_{ccg}^{VP}$ is also changed sizably. It implies that the
total statistical significance $\chi^2$ is not very sensitive to
$arg[A_{ccg}^{VP}]$. The large error in this parameter also confirms
this feature.

Using the two solutions for these non-perturbative inputs, we obtain
two different kinds of predictions (labeled as This work 1 and This
work 2) on branching fractions and CP asymmetries, where the chiraly
enhanced penguins are taken into account. As we have shown in the
above, the leading power results are not very different from these
results, as the inclusion of chiraly enhanced penguins only amounts
to a redefinition of charming penguins. Results for CP-averaged
branching fractions are summarized in table~\ref{tab:BRbtod},
table~\ref{tab:BRbtos} and table~\ref{tab:BRBs}, while predictions
on direct CP asymmetries are given in table~\ref{tab:Acpbtod},
table~\ref{tab:Acpbtos} and table~\ref{tab:AcpBs}. In $B^0/\bar
B^0\to\pi^\pm\rho^\mp$ decays, it is easy to identify the final
state mesons. Thus one can sum $B^0/\bar B^0\to\pi^-\rho^+$ up as
one channel, although the summed channels are not CP conjugates. The
$B^0/\bar B^0\to\pi^+\rho^-$ can be summed as another channel and it
is also similar for the branching ratios of $B^0/\bar B^0\to K^*K$
and $B_s^0/\bar B_s^0\to K^*K$ decays.  In table~\ref{tab:BRbtod}
and table~\ref{tab:BRBs}, we give our predictions on the summed
branching ratios in $B^0/\bar B^0\to\pi^\pm\rho^\mp, K^{*0} \bar
K^0(\bar K^{*0} K^0)$ and two $B_s\to K^* K$ decays. We also give
the predictions on the sum of the CP-averaged branching ratios of
$\bar B^0\to\pi^-\rho^+$ and $\bar B^0\to\pi^+\rho^-$ and the other
three $B_{(s)}$ decays  in table~\ref{tab:BRbtod} and
table~\ref{tab:AcpBs}. In order to compare with QCDF
approach~\cite{Beneke:2003zv,Du:2002up,Du:2002cf,Sun:2002rn,Dutta:2003hb,Li:2006jb}
and PQCD
approach~\cite{Lu:2000hj,Chen:2001pr,Chen:2001jx,Keum:2002qj,Mishima:2003wm,Li:2005kt,Liu:2005mm,
Xiao:2006hd,Guo:2006uq,Chen:2007qm,Guo:2007vw,Ali:2007ff,Akeroyd:2007fy},
we also collect their results in these tables, together with the
experimental data available at HFAG~\cite{Barberio:2006bi}.

Due to several approximations are made in this work, there are some
important possible corrections which we would like to address. First
of all, our results for the 16 inputs are obtained through the exact
flavor SU(3) symmetry for the form factors and charming penguins.
The amplitudes may receive sizable corrections from the SU(3)
symmetry breaking effect proportional to $m_s/\Lambda_{QCD}\sim
0.3$. Secondly, since we have concentrated on the leading order
analysis, the radiative corrections proportional to
$\alpha_s(\sqrt{m_b\Lambda_{QCD}})/\pi\sim0.1$ are also neglected.
Although we have included one of the most important power
corrections (chiraly enhanced penguins), the other parts of power
corrections proportional to
$\lambda=\sqrt{\Lambda_{QCD}/m_b}\sim0.3$ are not incorporated in
our analysis. At last, there are also   uncertainties from the input
parameters such as the $b$ quark mass, Wilson coefficients, etc. To
characterize these effects, we   vary the magnitudes of the
non-perturbative charming penguins   by $20\%$ and the phases by
$20^\circ$. We also assume that the gluonic form factors $\zeta_g$
and $\zeta_{Jg}$ have additional uncertainties ($\pm0.05$). In the
predictions for branching fractions and CP asymmetries collected in
tables~\ref{tab:BRbtod},~\ref{tab:BRbtos}
~\ref{tab:BRBs},~\ref{tab:Acpbtod},~\ref{tab:Acpbtos} and
~\ref{tab:AcpBs}, the first kinds of uncertainties are from these
hadronic uncertainties: charming penguins and gluonic form factors;
the second kinds of uncertainties are from those in the CKM matrix
elements.

\subsection{$b\to d$ transitions without $\eta(')$}

%

\begin{table}[htb]
\begin{center}{\tiny
\caption{Branching ratios (in units of $10^{-6}$) of $B\to VP$
decays induced by the $b\to d$ ($\Delta S =0$) transition: the first
solution (This work 1) and the second solution (This work 2). In
both cases, we have included the chiraly enhanced penguin in $B\to
VP$ decay amplitudes. The first kinds of uncertainties are from
uncertainties in charming penguins and gluonic form factors as
discussed in the text; the second kinds of uncertainties are from
those in the CKM matrix elements. We also cite the experimental data
and theoretical results given in QCDF~\cite{Beneke:2003zv} and
PQCD~\cite{Lu:2000hj,Liu:2005mm,Guo:2006uq,Guo:2007vw} approach to
make a comparison. }\label{tab:BRbtod}}
\begin{tabular} {|l|c|c|c|c|c|c}
\hline
   Channel                          &   Exp.                  & QCDF   &PQCD& This work 1 & This work 2\\\hline
 $ B^- \to \rho^- \pi^0$            & $10.9^{+1.4}_{-1.5}$    & $14.0^{+6.5+5.1+1.0+0.8}_{-5.5-4.3-0.6-0.7}$
                                    & $6$-$9$
                                    & $8.9_{-0.1-1.0}^{+0.3+1.0}$
                                    & $11.4_{-0.6-0.9}^{+0.6+1.1}$\\
 $ B^- \to \rho^0 \pi^-$            & $8.7^{+1.0}_{-1.1}$     & $11.9^{+6.3+3.6+2.5+1.3}_{-5.0-3.1-1.2-1.1}$
                                    & $10.4^{+3.3}_{-3.4}\pm2.1$
                                    & $10.7_{-0.7-0.9}^{+0.7+1.0}$
                                    & $7.9_{-0.1-0.8}^{+0.2+0.8}$\\
 $ B^- \to \omega \pi^- $           & $6.9\pm0.5$             & $8.8^{+4.4+2.6+1.8+0.8}_{-3.5-2.2-0.9-0.9}$
                                    & $11.3^{+3.3}_{-2.9}\pm1.4$
                                    & $6.7_{-0.3-0.6}^{+0.4+0.7}$
                                    & $8.5_{-0.3-0.8}^{+0.3+0.8}$\\
 $ B^- \to  K^{*0} K^- $            & $<1.1$                  & $0.30^{+0.11+0.12+0.09+0.57}_{-0.09-0.10-0.09-0.19}$
                                    & $0.31^{+0.12}_{-0.08}$
                                    & $0.49_{-0.20-0.08}^{+0.26+0.09}$
                                    & $0.51_{-0.16-0.06}^{+0.18+0.07}$\\
 $ B^- \to  K^{*-} K^0$             &                         & $0.30^{+0.08+0.41+0.08+0.58}_{-0.07-0.18-0.07-0.17}$
                                    & $1.83^{+0.68}_{-0.47}$
                                    & $0.54_{-0.21-0.08}^{+0.26+0.10}$
                                    & $0.51_{-0.17-0.07}^{+0.21+0.08}$\\
 $ B^- \to \phi \pi^- $             & $<0.24$                 & $\approx 0.005$
                                    &
                                    &$\approx0.003$
                                    &$\approx0.003$\\\hline
 \parbox[c]{2cm}{$\overline B^0 \to \rho^- \pi^+$ \\ $\bar B^0 \to \rho^+  \pi^-$}
  \Big\}& $24.0\pm2.5$    &  $36.5^{+18.2+10.3+2.0+3.9}_{-14.7-~8.6-3.5-2.9}$
                                    & $18$-$45$
                                    & $13.4_{-0.5-1.2}^{+0.6+1.2}$
                                    & $16.8_{-0.5-1.5}^{+0.5+1.6}$\\
 $ B^0/\bar B^0 \to \rho^+ \pi^- $  &           &
                                    & $24$-34
                                    & $12.0_{-1.6-1.1}^{+1.9+1.2}$
                                    & $14.8_{-1.5-1.4}^{+1.6+1.5}$\\
 $ B^0/\bar B^0 \to \rho^-  \pi^+ $ &             &
                                    & $24$-34
                                    & $14.9_{-1.9-1.3}^{+1.9+1.3}$
                                    & $18.7_{-1.6-1.6}^{+1.5+1.7}$\\
 $ \bar B^0 \to \rho^+  \pi^-$\footnote{We quote the branching ratios for $\bar B^0\to
 \rho^+\pi^-$ and $\bar B^0\to \rho^+\pi^-$ from
 Ref.~\cite{ref:Ntorhopi}.}
                                    & $8.9\pm2.5$
                                    & $15.4^{+8.0+5.5+0.7+1.9}_{-6.4-4.7-1.3-1.3}$
                                    &
                                    & $5.9_{-0.5-0.5}^{+0.5+0.5}$
                                    & $6.6_{-0.1-0.7}^{+0.2+0.7}$\\
 $ \bar B^0 \to \rho^-  \pi^+ $$^a$     & $13.9\pm2.7$            & $21.2^{+10.3+8.7+1.3+2.0}_{-~8.4-7.2-2.3-1.6}$
                                    &
                                    & $7.5_{-0.1-0.8}^{+0.3+0.8}$
                                    & $10.2_{-0.5-0.9}^{+0.4+0.9}$\\
 $ \bar  B^0 \to \rho^0 \pi^0$      & $1.8^{+0.6}_{-0.5}$      & $0.4^{+0.2+0.2+0.9+0.5}_{-0.2-0.1-0.3-0.3}$
                                    & $0.07$-$0.11$
                                    & $2.5_{-0.1-0.2}^{+0.2+0.2}$
                                    & $1.5_{-0.1-0.1}^{+0.1+0.1}$\\
 $\bar  B^0 \to \omega \pi^0$       & $<1.2$                   & $0.01^{+0.00+0.02+0.02+0.03}_{-0.00-0.00-0.00-0.00}$
                                    &  $0.10$-$0.28$
                                    & $0.0003_{-0.0000-0.0000}^{+0.0299+0.0000}$
                                    & $0.015_{-0.000-0.002}^{+0.024+0.002}$\\
 $\bar   B^0 \to K^{*0} \bar K^0$   &                          & $0.26^{+0.08+0.10+0.08+0.46}_{-0.07-0.09-0.08-0.15}$ &
                                    & $0.45_{-0.19-0.07}^{+0.24+0.09}$
                                    & $0.47_{-0.14-0.05}^{+0.17+0.06}$\\
 $\bar   B^0 \to \bar K^{*0} K^0$   &  $<1.9$                  & $0.29^{+0.10+0.39+0.08+0.60}_{-0.09-0.17-0.07-0.17}$ &
                                    & $0.51_{-0.20-0.08}^{+0.24+0.09}$
                                    & $0.48_{-0.16-0.06}^{+0.20+0.07}$\\
 \parbox[c]{2cm}{${\tiny\bar B^0\to K^{*0}\bar K^0}$ \\ ${\tiny\bar B^0\to\bar K^{*0} K^0} $} \Big\}
                                   &                           &
                                    & $\approx1.96$
                                    & $0.96_{-0.27-0.15}^{+0.34+0.18}$
                                    & $0.95_{-0.22-0.12}^{+0.26+0.14}$\\
 $B^0/\bar B^0 \to K^{*0} \bar K^0$ &                     &  &
                                    & $0.95_{-0.27-0.15}^{+0.34+0.18}$
                                    & $0.94_{-0.22-0.12}^{+0.26+0.14}$\\
 $B^0/\bar B^0 \to \bar K^{*0} K^0$   &                    & &
                                    & $0.97_{-0.27-0.15}^{+0.35+0.18}$
                                    & $0.97_{-0.22-0.12}^{+0.26+0.14}$\\
 $ \bar  B^0 \to \phi \pi^0$        & $<0.28$                  & $\approx 0.002$
                                    &  &$\approx0.001$
                                    &$\approx0.001$\\\hline
 $ B^- \to \rho^- \eta $            & $5.4\pm1.2$             & $9.4^{+4.6+3.6+0.7+0.7}_{-3.7-3.0-0.4-0.7}$
                                    & $8.5^{+3.0+0.8+0.4+1.2}_{-2.1-0.7-0.4-0.2}$\footnote{For
                                    $B\to\rho\eta$ decays, there are
                                    two different predictions given
                                    in Ref.~\cite{Liu:2005mm}
                                    according to the different mixing angles
                                    between $\eta$ and $\eta'$. We
                                    quote the results in which
                                    $\theta_P=-10^\circ$ is used.
                                    There are not too many changes for the other predictions
                                    as the value for the mixing angle $\theta_P=-17^\circ$ is very close to the first one. }
                                    & $3.9_{-1.7-0.4}^{+2.0+0.4}$
                                    & $3.3_{-1.6-0.3}^{+1.9+0.3}$\\
 $ B^- \to \rho^- \eta ^\prime$     & $9.1^{+3.7}_{-2.8}$     & $6.3^{+3.1+2.4+0.5+0.5}_{-2.5-2.0-0.3-0.5}$
                                    & $8.7^{+3.0+0.7+0.5+1.1}_{-2.2-0.9-0.7-0.3}$$^b$
                                    & $0.37_{-0.22-0.07}^{+2.46+0.08}$
                                    & $0.44_{-0.20-0.05}^{+3.18+0.06}$\\\hline
 $ \bar  B^0 \to \rho^0 \eta $      & $<1.5$                   & $0.03^{+0.02+0.16+0.02+0.05}_{-0.01-0.10-0.01-0.02}$
                                    & $0.024^{+0.012+0.004+0.002+0.102}_{-0.007-0.002-0.002-0.005}$$^b$
                                    & $0.04_{-0.01-0.00}^{+0.20+0.00}$
                                    & $0.14_{-0.13-0.01}^{+0.33+0.01}$\\
 $ \bar  B^0 \to \rho^0 \eta'$      & $<1.3$                   & $0.01^{+0.01+0.11+0.02+0.03}_{-0.00-0.06-0.00-0.01}$
                                    & $0.061^{+0.030+0.004+0.003+0.114}_{-0.018-0.003-0.003-0.009}$$^b$
                                    & $0.43_{-0.12-0.05}^{+2.51+0.05}$
                                    & $1.0_{-0.9-0.1}^{+3.5+0.1}$\\
 $ \bar  B^0\to \omega \eta $       & $<1.9$                   & $0.31^{+0.14+0.16+0.35+0.22}_{-0.12-0.11-0.14-0.16}$
                                    & $0.27^{+0.11}_{-0.10}$
                                    & $0.91_{-0.49-0.09}^{+0.66+0.09}$
                                    & $1.4_{-0.6-0.1}^{+0.8+0.1}$\\
 $ \bar  B^0 \to \omega \eta'$      & $<2.2$                   & $0.20^{+0.10+0.15+0.25+0.15}_{-0.08-0.05-0.10-0.11}$
                                    & $0.075^{+0.037}_{-0.033}$
                                    & $0.18_{-0.10-0.03}^{+1.31+0.04}$
                                    & $3.1_{-2.6-0.3}^{+4.9+0.3}$\\
 $ \bar  B^0 \to \phi \eta$         & $<0.6$                   & $\approx 0.001$
                                    & $0.0063^{+0.0033}_{-0.0019}$
                                    & $\approx0.0004$
                                    & $\approx0.0008$\\
 $ \bar  B^0 \to  \phi \eta'$       & $<0.5$                   & $\approx 0.001$
                                    & $0.0073^{+0.0035}_{-0.0026}$
                                    & $\approx0.0001$
                                    & $\approx0.0007$\\
\hline
\end{tabular}\end{center}
\end{table}


\begin{table}[htb]
\begin{center}
\caption{Direct CP asymmetries involving $b\to d$ ($\Delta S =0$)
transitions: the first solution (This work 1) and the second
solution (This work 2). In both solutions, we have included the
chiraly enhanced penguin in $B\to VP$ decay amplitudes. The first
kinds of uncertainties are from uncertainties in charming penguins
and gluonic form factors which are discussed in the text; the second
kinds of uncertainties are from those in the CKM matrix elements. We
also cite the experimental data and theoretical results given in
QCDF~\cite{Beneke:2003zv} and
PQCD~\cite{Lu:2000hj,Liu:2005mm,Guo:2006uq,Guo:2007vw} approach to
make a comparison.}\label{tab:Acpbtod}
\begin{tabular} {|l|c|c|c|c|c|c}
\hline
   Channel                          &   Exp.                  & QCDF   &PQCD& This work 1 & This work 2\\\hline
 $ B^- \to \rho^- \pi^0$            & $2\pm11$                & $-4.0^{+1.2+1.8+0.4+17.5}_{-1.2-2.2-0.4-17.7}$
                                    & $0$-$20$
                                    & $15.5_{-18.9-1.4}^{+16.9+1.6}$
                                    & $12.3_{-10.0-1.1}^{+9.4+0.9}$\\
 $ B^- \to \rho^0 \pi^-$            & $-7^{+12}_{-13}$ & $4.1^{+1.3+2.2+0.6+19.0}_{-0.9-2.0-0.7-18.8}$
                                    & $-20$-$0$
                                    & $-10.8_{-12.7-0.7}^{+13.1+0.9}$
                                    & $-19.2_{-13.4-1.9}^{+15.5+1.7}$\\
 $ B^- \to \omega \pi^- $           & $-4\pm6$          & $-1.8^{+0.5+2.7+0.8+2.1}_{-0.5-3.3-0.7-2.2}$
                                    & $\sim0$
                                    & $0.5_{-19.6-0.0}^{+19.1+0.1}$
                                    & $2.3_{-13.2-0.2}^{+13.4+0.2}$\\
 $ B^- \to  K^{*0} K^- $            & ...                     & $-23.5^{+6.9+7.8+5.5+25.2}_{-5.7-9.0-6.5-36.8}$
                                    & $-20\pm5\pm2$
                                    & $-3.6_{-5.3-0.4}^{+6.1+0.4}$
                                    & $-4.4_{-4.1-0.2}^{+4.1+0.2}$\\
 $ B^- \to  K^{*-} K^0$             & ...                     & $-13.4^{+3.7+7.8+4.2+27.4}_{-3.0-3.5-4.7-36.7}$
                                    & $-49^{+7+7}_{-3-7}$
                                    & $-1.5_{-2.3-0.1}^{+2.6+0.1}$
                                    & $-1.2_{-1.7-0.1}^{+1.7+0.1}$\\\hline
 $ \bar B^0 \to \rho^+  \pi^-$      & $-18\pm12$           & $0.6^{+0.2+1.3+0.1+11.5}_{-0.1-1.6-0.1-11.7}$
                                    &
                                    & $-9.9_{-16.7-0.7}^{+17.2+0.9}$
                                    & $-12.4_{-15.3-1.2}^{+17.6+1.1}$\\
 $ \bar B^0 \to \rho^-  \pi^+ $     & $11\pm6$           & $-1.5^{+0.4+1.2+0.2+8.5}_{-0.4-1.3-0.3-8.4}$
                                    &
                                    & $11.8_{-20.0-1.1}^{+17.5+1.2}$
                                    & $10.8_{-10.2-1.0}^{+9.4+0.9}$\\
 $ \bar  B^0 \to \rho^0 \pi^0$      & $-30\pm38$          & $-15.7^{+4.8+12.3+11.0+19.8}_{-4.7-14.0-12.9-25.8}$
                                    & $-75$-$0$
                                    & $-0.6_{-21.9-0.1}^{+21.4+0.1}$
                                    & $-3.5_{-20.3-0.3}^{+21.4+0.3}$\\
 $\bar  B^0 \to \omega \pi^0$       & ...                      & ...
                                    & $-20$-$75$
                                    & $-9.4_{-0.0-0.9}^{+24.0+1.1}$
                                    & $39.5_{-185.5-3.1}^{+79.1+3.4}$\\
 $\bar   B^0 \to K^{*0} \bar K^0$   & ...                      & $-26.7^{+7.4+7.2+5.7+10.9}_{-5.7-9.0-6.9-13.4}$&
                                    & $-3.6_{-5.3-0.4}^{+6.1+0.4}$
                                    & $-4.4_{-4.1-0.2}^{+4.1+0.2}$\\
 $\bar   B^0 \to \bar K^{*0} K^0$   & ...                      & $-13.1^{+3.8+5.4+4.5+5.8}_{-3.0-2.9-5.2-7.4}$ &
                                    & $-1.5_{-2.3-0.1}^{+2.6+0.1}$
                                    & $-1.2_{-1.7-0.1}^{+1.7+0.1}$\\\hline
 $ B^- \to \rho^- \eta $            & $1\pm16$           & $-2.4^{+0.7+6.3+0.4+0.2}_{-0.7-6.3-0.4-0.2}$
                                    & $-13^{+1.2+2}_{-0.5-14}$
                                    & $-6.6_{-21.3-0.7}^{+21.5+0.6}$
                                    & $-9.1_{-15.8-0.8}^{+16.7+0.9}$\\
 $ B^- \to \rho^- \eta ^\prime$     & $-4\pm28$          & $4.1^{+1.2+7.9+0.5+7.0}_{-1.1-6.9-0.8-7.0}$
                                    & $-18^{+3.0+1}_{-1.6-14}$
                                    & $-19.8_{-37.5-3.1}^{+66.5+2.8}$
                                    & $-21.7_{-24.3-1.7}^{+135.9+2.1}$\\\hline
 $ \bar  B^0 \to \rho^0 \eta $      & ...                      & ...
                                    & $-13^{+1.2+2}_{-0.5-14}$
                                    & $-46.7_{-74.3-3.7}^{+170.4+2.9}$
                                    & $33.3_{-62.4-2.8}^{+66.9+3.1}$\\
 $ \bar  B^0 \to \rho^0 \eta'$      & ...                      & ...
                                    & $-18^{+3.0+1}_{-1.6-14}$
                                    & $-51.7_{-42.9-3.9}^{+103.3+3.4}$
                                    & $52.2_{-80.6-4.1}^{+19.9+4.4}$\\
 $ \bar  B^0\to \omega \eta $       & ...                      & $-33.4^{+10.0+65.3+20.9+19.2}_{-~9.5-55.8-21.4-20.8}$
                                    & $-69.1^{+15.1}_{-13.4}$
                                    & $-9.4_{-30.2-1.0}^{+30.7+0.9}$
                                    & $-9.6_{-16.8-0.9}^{+17.8+0.9}$\\
 $ \bar  B^0 \to \omega \eta'$      & ...                      & $0.2^{+0.1+53.0+11.6+19.2}_{-0.1-76.5-11.5-20.1}$
                                    & $13.9^{+4.1}_{-3.5}$
                                    & $-43.0_{-38.8-5.1}^{+87.5+4.8}$
                                    & $-27.2_{-29.7-2.2}^{+18.1+2.4}$\\
\hline
\end{tabular}\end{center}
\end{table}

$b\to d$ transitions are induced by the operators whose CKM matrix
elements are $V_{ub}V_{id}^* (i=u,c,t)$.  To make it clear, we
decompose the decay amplitudes into three terms according to the CKM
matrix elements:
\begin{eqnarray}
 A(B\to M_1 M_2) =\frac{G_F}{\sqrt 2} m_B^2\left\{ V_{ub}V_{ud}^* A_u + V_{cb}V_{cd}^* A_c - V_{tb}V_{td}^*
 A_t\right\},
\end{eqnarray}
where $A_c$ is from the charming penguin term. The decomposition is
over complete since the unitarity property of CKM matrix can be used
to eliminate one of the three combinations of CKM matrix elements.
We keep all of them according to the different dynamics in the
corresponding amplitudes.  The values for CKM matrix elements:
\begin{eqnarray}
 |V_{ub}V_{ud}^*|= 3.48\times 10^{-3},\;\;\;
 |V_{cb}V_{cd}^*|= 9.17\times 10^{-3},\;\;\;
 |V_{tb}V_{td}^*|=8.60\times 10^{-3}\label{eq:CKMbtod}
\end{eqnarray}
will definitely character the branching fractions and CP
asymmetries.

$\bar B^0\to\pi^\pm\rho^\mp$ are dominated by tree operators which
has the CKM matrix elements: $V_{ub}V_{ud}^*$. To illustrate the
situation, we will use the second kind of inputs given in
Eq.~\eqref{eq:solution2} and take $\bar B^0\to\rho^+\pi^-$ as an
example (in units of GeV):
\begin{eqnarray}
 &&|A_u(\bar B^0\to \rho^+\pi^-)|=0.131\times (1.03\zeta^V
 +0.77\zeta^V_J)\sim260\times 10^{-4},\;\;\;\nonumber\\
 &&|A_c(\bar B^0\to \rho^+\pi^-)|= |A_{cc}^{PV}|\sim (30\sim 40)\times 10^{-4},\;\;\;\nonumber\\
 &&|A_t(\bar B^0\to \rho^+\pi^-)|=|0.131( -0.0015
 \zeta^V-0.007\zeta^V_J)|\sim5\times 10^{-4}.
\end{eqnarray}
Our predictions on branching fractions of $\bar
B^0\to\pi^\pm\rho^\mp$ decays are smaller than those in
QCDF~\cite{Beneke:2003zv}. Neglecting the small terms, the main
reason is our smaller $B\to P$ and $B\to V$ form factors: QCDF uses
much larger form factors $F^{B\to\pi}=0.28\pm0.05$ and
$A_0^{B\to\rho}=0.37\pm0.06$. In the present framework, ${\cal
BR}(\bar B^0\to\rho^+\pi^-)$ is smaller than ${\cal BR}(\bar
B^0\to\rho^-\pi^+)$. In the first solution, the fitted $B\to V$ form
factor $A_0=0.233$ is almost equal with the $B\to P$ form factor
$F=0.206$. Since the decay constant of $\rho$ meson is much larger
than that of $\pi$: $0.209/0.131\sim1.5$, we expect ${\cal BR}(\bar
B^0\to\rho^+\pi^-)$ is only one half of ${\cal BR}(\bar
B^0\to\rho^-\pi^+)$. Charming penguins $A_{cc}^{VP}$ and
$A_{cc}^{PV}$ can slightly change the ratio: the charming penguin
$A_{cc}^{PV}$ in $\bar B^0\to\rho^+\pi^-$ gives a destructive
contribution, while $A_{cc}^{VP}$ in $\bar B^0\to\rho^-\pi^+$ gives
a constructive contribution. In the second solution, contributions
proportional to form factors are almost equal with each other, as
the $B\to V$ form factor $A_0^{B\to V}=0.291$ is much larger than
$F^{B\to P}=0.198$ which can compensate differences caused by decay
constants. But unlike in the first solution, the role of charming
penguin totally changes: the charming penguin in $\bar
B^0\to\rho^+\pi^-$ gives a constructive contribution, while
$A_{cc}^{VP}$ in $\bar B^0\to\rho^-\pi^+$ can give a destructive
contribution. It is reasonable, since the charming penguins
$A_{cc}^{VP}$ and $A_{cc}^{PV}$ almost interchanges the phases.

Our predictions for branching ratios of $\bar B^0\to\pi^0\rho^0$ are
larger than that in QCDF especially the prediction utilizing the
inputs given in Eq.~\eqref{eq:solution1}. In this channel, two kinds
of charming penguin almost cancel with each other, since they have
similar magnitudes and but different signs as given in
Eq.~\eqref{eq:solution1} and Eq.~\eqref{eq:solution2}.  The tree
contribution proportional to the soft form factor $\zeta$ is
color-suppressed (the  Wilson coefficient $C_2+\frac{C_1}{N_c}\sim
0.12$ is small compared with that of $\bar B^0\to\rho^\pm\pi^\mp$
$C_1+\frac{C_2}{N_c}\sim 1.03$), thus the branching fractions of
$\bar B^0\to\pi^0\rho^0$ in QCDF approach and PQCD approach are much
smaller than  ${\cal BR}(\bar B^0\to\rho^\pm\pi^\mp)$. One important
feature of the  SCET framework is: the hard-scattering form factor
$\zeta_J$ is relatively large and comparable with the soft form
factor $\zeta$. Besides, this term has a large Wilson coefficient
$b_1^f$, since $C_2+\frac{1}{N_c}(1-\frac{m_b}{\omega_3})C_1\sim
1.23$ is large, it can give larger production rates which are
consistent with the present experimental data. The agreement is very
encouraging.

Branching ratios of $B\to K^* K$ are larger than those in QCDF for
the presence of charming penguins.  In $B^-\to K^{*-}K^0$ and $\bar
B^0\to\bar K^{*0}K^0$, both of penguin operators and charming
penguins can give contributions. The difference for these two
channels is: the spectator antiquark  in $B^-\to K^{*-}K^0$ is $\bar
u$ and it is $\bar d$ in $\bar B^0\to \bar K^{*0}K^0$. It does not
affect the contributions from either penguin operators or charming
penguins, thus we expect the relations ${\cal BR}(B^-\to
K^{*-}K^0)={\cal BR}(\bar B^0\to \bar K^{*0}K^0)$ and
${A_{CP}}(B^-\to K^{*-}K^0)={A_{CP}}(\bar B^0\to \bar K^{*0}K^0)$.
The small differences in branching fractions are induced by the
different lifetimes of $B^-$ and $\bar B^0$. The analysis is similar
for the other two $b\to d$ modes: $B^-\to K^{-}K^{*0}$ and $\bar
B^0\to \bar K^{0}K^{*0}$.

For the decays with sizable branching fractions, our predictions on
direct CP asymmetries are typically small and most of them have the
correct sign with experimental data. Predictions in QCDF approach on
these channels are also small in magnitude, but some of them have
different signs with our results and experimental data. In PQCD
approach, the strong phases mainly come from the $(S-P)(S+P)$
annihilation operators. These operators are chiraly enhanced and the
imaginary part are dominant. Thus the direct CP asymmetries in PQCD
approach are typically large in magnitude.

\subsection{ $b\to s$ transitions without $\eta$ and $\eta'$}

\begin{table}[tb]
\caption{Branching ratios (in units of $10^{-6}$) for $\Delta s=1$
processes: the first solution (This work 1) and the second solution
(This work 2). In both solutions, we have included the chiraly
enhanced penguin in $B\to VP$ decay amplitudes. The first kinds of
uncertainties are from uncertainties in charming penguins and
gluonic form factors which are discussed in the text; the second
kinds of uncertainties are from those in the CKM matrix elements. We
also cite the experimental data and theoretical results given in
QCDF~\cite{Beneke:2003zv} and PQCD~\cite{Li:2005kt,Akeroyd:2007fy}
to make a comparison. } \label{tab:BRbtos}
\begin{tabular}{|c|c|c|c|c|c|c}\hline
  Channel                           &   Exp.                     & QCDF   &  PQCD& This work 1 &  This work 2\\ \hline
  $B^-\to  K^{*-}\pi^0$             &  $6.9\pm2.3$               & $3.3^{+1.1+1.0+0.6+4.4}_{-1.0-0.9-0.6-1.4}$
                                    & $4.3^{+5.0}_{-2.2}$
                                    & $4.2_{-1.7-0.7}^{+2.2+0.8}$
                                    & $6.5_{-1.7-0.7}^{+1.9+0.7}$ \\
  $B^-\to \bar K^{*0}\pi^- $        & $10.7\pm0.8$               & $3.6^{+0.4+1.5+1.2+7.7}_{-0.3-1.4-1.2-2.3}$
                                    & $6.0^{+2.8}_{-1.5}$
                                    & $8.5_{-3.6-1.4}^{+4.7+1.7}$
                                    & $9.9_{-3.0-1.1}^{+3.5+1.3}$\\
  $B^-\to \rho^0  K^{-}$            & $4.25^{+0.55}_{-0.56}$     & $2.6^{+0.9+3.1+0.8+4.3}_{-0.9-1.4-0.6-1.2}$
                                    & $5.1^{+4.1}_{-2.8}$
                                    & $6.7_{-2.2-0.9}^{+2.7+1.0}$
                                    & $4.6_{-1.5-0.6}^{+1.8+0.7}$   \\
  $B^-\to \rho^-  \bar K^{0}$       & $8.0^{+1.5}_{-1.4}$        & $5.8^{+0.6+7.0+1.5+10.3}_{-0.6-3.3-1.3-~3.2}$
                                    & $8.7^{+6.8}_{-4.4}$
                                    & $9.3_{-3.7-1.4}^{+4.7+1.7}$
                                    & $10.1_{-3.3-1.3}^{+4.0+1.5}$\\
  $B^-\to \omega K^{-}$             & $6.7\pm0.5$                & $3.5^{+1.0+3.3+1.4+4.7}_{-1.0-1.6-0.9-1.6}$
                                    & $10.6^{+10.4}_{-5.8}$
                                    & $5.1_{-1.9-0.8}^{+2.4+0.9}$
                                    & $5.9_{-1.7-0.7}^{+2.1+0.8}$\\
  $B^-\to \phi K^{-}$               & $8.30\pm0.65$              & $4.5^{+0.5+1.8+1.9+11.8}_{-0.4-1.7-2.1-~3.3}$
                                    & $7.8^{+5.9}_{-1.8}$
                                    & $9.7_{-3.9-1.5}^{+4.9+1.8}$
                                    & $8.6_{-2.7-1.0}^{+3.2+1.2}$    \\\hline
  $\bar B^0\to  \bar K^{*0}\pi^0$   & $0.0^{+1.3}_{-0.1}$        & $0.7^{+0.1+0.5+0.3+2.6}_{-0.1-0.4-0.3-0.5}$
                                    & $2.0^{+1.2}_{-0.6}$
                                    & $4.6_{-1.8-0.7}^{+2.3+0.9}$
                                    & $3.7_{-1.2-0.5}^{+1.4+0.5}$\\
  $\bar B^0\to  \bar K^{*-}\pi^+$   & $9.8\pm1.1$                & $3.3^{+1.4+1.3+0.8+6.2}_{-1.1-1.2-0.8-1.6}$
                                    & $6.0^{+6.8}_{-2.6}$
                                    & $8.4_{-3.4-1.3}^{+4.4+1.6}$
                                    & $9.5_{-2.8-1.1}^{+3.2+1.2}$ \\
  $\bar B^0\to  \rho^0\bar K^{0}$   & $5.4^{+0.9}_{-1.0}$        & $4.6^{+0.5+4.0+0.7+6.1}_{-0.5-2.1-0.7-2.1}$
                                    & $4.8^{+4.3}_{-2.3}$
                                    & $3.5_{-1.5-0.6}^{+2.0+0.7}$
                                    & $5.8_{-1.8-0.7}^{+2.1+0.8}$\\
  $\bar B^0\to  \rho^+K^{-}$        & $15.3^{+3.7}_{-3.5}$       & $7.4^{+1.8+7.1+1.2+10.7}_{-1.9-3.6-1.1-~3.5}$
                                    & $8.8^{+6.8}_{-4.5}$
                                    & $9.8_{-3.7-1.4}^{+4.6+1.7}$
                                    & $10.2_{-3.2-1.2}^{+3.8+1.5}$   \\
  $\bar B^0\to \omega\bar K^{0}$    & $5.0\pm0.6$                & $2.3^{+0.3+2.8+1.3+4.3}_{-0.3-1.3-0.8-1.3}$
                                    & $9.8^{+8.6}_{-4.9}$
                                    & $4.1_{-1.7-0.7}^{+2.1+0.8}$
                                    & $4.9_{-1.6-0.6}^{+1.9+0.7}$\\
  $\bar B^0\to  \phi \bar K^{0}$    & $8.3^{+1.2}_{-1.0}$        & $4.1^{+0.4+1.7+1.8+10.6}_{-0.4-1.6-1.9-~3.0}$
                                    & $7.3^{+5.9}_{-1.8}$
                                    & $9.1_{-3.6-1.4}^{+4.6+1.7}$
                                    & $8.0_{-2.5-1.0}^{+3.0+1.1}$\\\hline
  $B^-\to K^{*-}\eta $              & $19.3\pm1.6$              & $10.8^{+1.9+8.1+1.8+16.5}_{-1.7-4.4-1.3-~5.5}$
                                    & $22.13^{+0.26}_{-0.27}$
                                    & $17.9_{-5.4-2.9}^{+5.5+3.5}$
                                    & $18.6_{-4.8-2.2}^{+4.5+2.5}$      \\
  $B^-\to K^{*-}\eta' $             & $4.9^{+2.1}_{-1.9}$       & $5.1^{+0.9+7.5+2.1+6.7}_{-1.0-3.8-3.0-3.3}$
                                    & $6.38\pm0.26$
                                    & $4.5_{-3.9-0.8}^{+6.6+0.9}$
                                    & $4.8_{-3.7-0.6}^{+5.3+0.8}$\\\hline
  $\bar B^0\to \bar K^{*0}\eta$     & $15.9\pm1.0$              & $10.7^{+1.1+7.8+1.4+16.2}_{-1.0-4.3-1.2-~5.5}$
                                    & $22.31^{+0.28}_{-0.29}$
                                    & $16.6_{-5.0-2.7}^{+5.1+3.2}$
                                    & $16.5_{-4.3-2.0}^{+4.1+2.3}$\\
  $\bar B^0\to \bar K^{*0}\eta'$    & $3.8\pm1.2$               & $3.9^{+0.4+6.6+1.8+6.2}_{-0.4-3.3-2.5-2.9}$
                                    & $3.35^{+0.29}_{-0.27}$
                                    & $4.1_{-3.6-0.7}^{+6.2+0.9}$
                                    & $4.0_{-3.4-0.6}^{+4.7+0.7}$\\\hline
 \hline
\end{tabular}
\end{table}

\begin{table}[tb]
\caption{Direct CP asymmetries (in $\%$) for $\Delta s=1$ processes:
the first solution (This work 1) and the second solution (This work
2).  In both solutions, we have included the chiraly enhanced
penguin in $B\to VP$ decay amplitudes. The first kinds of
uncertainties are from uncertainties in charming penguins and
gluonic form factors which are discussed in the text; the second
kinds of uncertainties are from those in the CKM matrix elements. We
also cite the experimental data and theoretical results given in
QCDF~\cite{Beneke:2003zv} and PQCD~\cite{Li:2005kt,Akeroyd:2007fy}
to make a comparison. } \label{tab:Acpbtos}
\begin{tabular}{|c|c|c|c|c|c|c}\hline
  Channel                           &   Exp.                     & QCDF   &  PQCD& This work 1 &  This work 2\\ \hline
  $B^-\to  K^{*-}\pi^0$             & $4\pm29$                   & $8.7^{+2.1+5.0+2.9+41.7}_{-2.6-4.3-3.4-44.2}$
                                    & $-32^{+21}_{-28}$
                                    & $-17.8_{-24.6-2.0}^{+30.3+2.2}$
                                    & $-12.9_{-12.2-0.8}^{+12.0+0.8}$ \\
  $B^-\to \bar K^{*0}\pi^- $        & $-8.5\pm5.7$               & $1.6^{+0.4+0.6+0.5+2.5}_{-0.5-0.5-0.4-1.0}$
                                    & $-1^{+1}_{-0}$
                                    &    $0$           & $0$ \\
  $B^-\to \rho^0  K^{-}$            & $31^{+11}_{-10}$           & $-13.6^{+4.5+6.9+3.7+62.7}_{-5.7-4.4-3.1-55.4}$
                                    & $71^{+25}_{-35}$
                                    & $9.2_{-16.1-0.7}^{+15.2+0.7}$
                                    & $16.0_{-22.4-1.6}^{+20.5+1.3}$     \\
  $B^-\to \rho^-  \bar K^{0}$       & $-12\pm17$                 & $0.3^{+0.1+0.3+0.2+1.6}_{-0.1-0.4-0.1-1.3}$
                                    & $1\pm1$
                                    &    $0$             &  $0$\\
  $B^-\to \omega K^{-}$             & $2\pm5$                    & $-7.8^{+2.6+5.9+2.4+39.8}_{-3.0-3.6-1.9-38.0}$
                                    & $32^{+15}_{-17}$
                                    & $11.6_{-20.4-1.1}^{+18.2+1.1}$
                                    & $12.3_{-17.3-1.1}^{+16.6+0.8}$ \\
  $B^-\to \phi K^{-}$               & $3.4\pm4.4$                & $1.6^{+0.4+0.6+0.5+3.0}_{-0.5-0.5-0.3-1.2}$
                                    & $1^{+0}_{-1}$              &   $0$            &  $0$      \\\hline\hline
  $\bar B^0\to  \bar K^{*0}\pi^0$   &  ...                       & $-12.8^{+4.0+4.7+2.7+31.7}_{-3.2-7.0-4.0-35.3}$
                                    & $-11^{+7}_{-5}$
                                    & $5.0_{-8.4-0.5}^{+7.5+0.5}$
                                    & $5.4_{-5.1-0.5}^{+4.8+0.4}$  \\
  $\bar B^0\to  \bar K^{*-}\pi^+$   & $-5\pm14$                   & $2.1^{+0.6+8.2+5.1+62.5}_{-0.7-7.9-5.8-64.2}$
                                    & $-60^{+32}_{-19}$
                                    & $-11.2_{-16.2-1.3}^{+19.0+1.3}$
                                    & $-12.2_{-11.3-0.8}^{+11.4+0.8}$     \\
  $\bar B^0\to  \rho^0\bar K^{0}$   & $-2\pm27\pm8\pm6$          & $7.5^{+1.7+2.3+0.7+8.8}_{-2.1-2.0-0.4-8.7}$
                                    & $7^{+8}_{-5}$
                                    &   $-6.6_{-9.7-0.9}^{+11.6+0.8}$
                                    & $-3.5_{-4.8-0.2}^{+4.8+0.3}$\\
  $\bar B^0\to  \rho^+K^{-}$        & $22\pm23$                  & $-3.8^{+1.3+4.4+1.9+34.5}_{-1.4-2.7-1.6-32.7}$
                                    & $64^{+24}_{-30}$
                                    & $7.1_{-12.4-0.7}^{+11.2+0.7}$
                                    &  $9.6_{-13.5-0.9}^{+13.0+0.7}$  \\
  $\bar B^0\to \omega\bar K^{0}$    & $21\pm19$                        & $-8.1^{+2.5+3.0+1.7+11.8}_{-2.0-3.3-1.4-12.9}$
                                    & $-3^{+2}_{-3}$
                                    & $5.2_{-9.2-0.6}^{+8.0+0.6}$
                                    & $3.8_{-5.4-0.3}^{+5.2+0.3}$\\
  $\bar B^0\to  \phi \bar K^{0}$    & $1\pm12$                   & $1.7^{+0.4+0.6+0.5+1.4}_{-0.5-0.5-0.3-0.8}$
                                    & $3^{+1}_{-2}$
                                    &  $0$              &$0$\\\hline
  $B^-\to K^{*-}\eta $              & $2\pm6$                    & $3.5^{+0.9+1.9+0.8+20.7}_{-0.9-2.7-0.8-20.5}$
                                    & $-24.57^{+0.72}_{-0.27}$
                                    & $-2.6_{-5.5-0.3}^{+5.4+0.3}$
                                    & $-1.9_{-3.6-0.1}^{+3.4+0.1}$    \\
  $B^-\to K^{*-}\eta' $             & $30^{+33}_{-37}$           & $-14.2^{+4.7+~8.5+~4.9+27.5}_{-4.2-13.8-14.6-26.1}$
                                    & $4.60^{+1.16}_{-1.32}$
                                    & $2.7_{-19.5-0.3}^{+27.4+0.4}$
                                    & $2.6_{-32.9-0.2}^{+26.7+0.2}$\\\hline
  $\bar B^0\to \bar K^{*0}\eta$     & $19\pm5$                   & $3.8^{+0.9+1.1+0.2+3.8}_{-1.1-0.8-0.2-3.5}$
                                    & $0.57\pm0.011$
                                    & $-1.1_{-2.4-0.1}^{+2.3+0.1}$
                                    & $-0.7_{-1.3-0.0}^{+1.2+0.1}$\\
  $\bar B^0\to \bar K^{*0}\eta'$    &  $-8\pm25$                 & $-5.5^{+1.6+3.1+1.8+6.2}_{-1.3-5.1-5.9-7.0}$
                                    & $-1.30\pm0.08$
                                    & $9.6_{-11.0-1.2}^{+8.9+1.3}$
                                    & $9.9_{-4.3-0.9}^{+6.2+0.9}$    \\\hline
 \hline
\end{tabular}
\end{table}

Like $b\to d$ processes, $b\to s$ decay amplitudes can also be
decomposed into three different parts according to the CKM matrix
elements. The values of the CKM matrix elements are given by:
\begin{eqnarray}
 |V_{ub}V_{us}^*|= 0.81\times 10^{-3},\;\;\;
 |V_{cb}V_{cs}^*|= 39.41\times 10^{-3},\;\;\;
 |V_{tb}V_{ts}^*|=40.66\times 10^{-3}.
\end{eqnarray}
Tree operators are highly CKM-suppressed, but the CKM matrix
elements for the rest two kinds of contributions $A_c$ and $A_t$ are
in similar size. Together with the hierarchy in Wilson coefficients:
$C_{1,2}\gg C_{3-10}$, charming penguins will provide a dominant
contribution. For example, the penguin operators in $B^-\to\pi^-
\bar K^0$ decay process is proportional to $a_4+r_\chi a_6$,
$B^-\to\pi^- \bar K^{*0}$ is proportional to $a_4$ while
$B^-\to\rho^- \bar K^0$ is proportional to $a_4-r_\chi a_6$, where
$a_{4,6}=C_{4,6}+C_{3,5}/N_c$ and $r_\chi=2\mu_P/m_b$. Thus if we
only consider the emission diagrams, ${\cal BR}(B^-\to\pi^- \bar
K^0)>{\cal BR}(B^-\to\pi^- \bar K^{*0})>{\cal BR}(B^-\to\rho^- \bar
K^0)$ holds, since $a_4\sim a_6$ and $r_\chi\sim1$. But in the
present framework, contributions from penguin operators proportional
to $V_{tb}V_{ts}^*$ do not play the most important role:
\begin{eqnarray}
 &&|A_t(B^-\to\pi^- \bar K^0)|=|0.16\times (-0.044\zeta^P-0.036\zeta_J^P)|\sim 15\times 10^{-4},\nonumber\\
 &&|A_t(B^-\to\rho^- \bar K^0)|= |0.16\times (0.0004\zeta^V+0.004\zeta^V_J)|\sim 1\times 10^{-4},\nonumber\\
 &&|A_t(B^-\to\pi^- \bar K^{*0})|=|0.217\times (-0.022\zeta^P-0.015\zeta^P_J)|\sim 10\times
 10^{-4}.\label{eq:penguinsbtos}
\end{eqnarray}
Compared with the results given in Eq.~\eqref{eq:solution1} and
Eq.~\eqref{eq:solution2}, we find penguin operators are smaller than
charming penguins. According to the size of charming penguins, we
expect the relation ${\cal BR}(B^-\to\rho^- \bar K^0)\sim {\cal
BR}(B^-\to\pi^- \bar K^{*0})$. This is well consistent with the
experimental data.

%

From table~\ref{tab:Acpbtos}, we can see the direct CP asymmetries
of $B^-\to \bar K^{*0}\pi^-$, $B^-\to \bar K^{0}\rho^-$, $B^-\to
K^{-}\phi$ and $B^-\to \bar K^{0}\phi$ are zero. In these channels,
tree operators do not contribute. The weak phases for penguin
operators and charming penguins are  equal to each other, which can
not induce any direct CP violations. CP asymmetries in other
channels are not large, because the strong phases of charming
penguins are either close to $0^\circ$ or $180^\circ$ and imaginary
parts are accordingly small. The PQCD results for most $B\to K^*\pi$
and $B\to \rho K$ channels are much larger than ours, since they
have more large imaginary part from annihilation diagrams. The QCDF
results are small and comparable with ours but with a relative minus
sign. We have to wait for the experiment data to resolve this
disagreements.

\subsection{$B$ Decays involving $\eta$ or $\eta'$}

As we can see from table~\ref{tab:BRbtod}, there is about
$3.1\sigma$ deviation for our prediction on the branching ratio of
$B^-\to \rho^-\eta'$ from the experimental data. Contributions from
penguin operators are suppressed by the CKM matrix elements as given
in Eq.~\eqref{eq:CKMbtod} and the dominant contribution is from the
tree operator. This kind of contribution is either proportional to
$B\to \eta_q$ or $B\to\eta_s$ form factor. Utilizing results given
in Eq.~\eqref{eq:solution1} and Eq.~\eqref{eq:solution2}, we obtain
$B\to\eta_q$ and $B\to\eta_s$ form factors as follows:
\begin{eqnarray}
 &&F^{B\to\eta_q} =(\zeta^P+\zeta^P_J+2\zeta_g +2\zeta_{Jg})=(0.053\pm0.068)\left[(0.100\pm0.021)\right],\nonumber\\
 &&F^{B\to\eta_s} =(\zeta_g +\zeta_{Jg})=(-0.076\pm0.055)[(-0.049\pm0.011)]
 ,\label{eq:Btoetaformfactors}
\end{eqnarray}
where the results in (out) the square brackets are predictions using
the second (first) kind of inputs. In
equation~\eqref{eq:Btoetaformfactors}, we can see: after taking the
gluonic form factors into account, the $F^{B\to\eta_q}$ and
$F^{B\to\eta_s}$ form factors are in the similar size but with
different signs in both kinds of inputs. In $B^-\to\rho^-\eta_q$,
another tree operator contributes in which $\eta_q$ is emitted.
Although this contribution is color-suppressed, terms proportional
to $\zeta_J^V$ give a sizable contribution. It can be estimated by
using a larger effective $B\to\eta_q$ form factor. Recalling that
physical states $\eta$ and $\eta'$ are mixtures of $\eta_q$ and
$\eta_s$ as in Eq.~\eqref{eq:etamix}, one obtains the expressions
for $B\to \eta(')$ form factors:
\begin{eqnarray}
 && F^{B\to\eta}=\frac{F^{B\to\eta_q}}{\sqrt 2} \cos(\theta)-
 F^{B\to\eta_s}\sin(\theta),\nonumber\\
 && F^{B\to\eta'}=\frac{F^{B\to\eta_q}}{\sqrt 2} \sin(\theta)+
 F^{B\to\eta_s}\cos(\theta).
\end{eqnarray}
The mixing angle between $\eta_q$ and $\eta_s$ has been determined
as
$\theta=(39.3\pm1.0)^\circ$~\cite{Feldmann:1998vh,Feldmann:1998sh,Feldmann:1999uf}
which is very close to $45^\circ$, thus we can obtain very small
$B\to\eta'$ form factors and relatively large $B\to\eta$ form
factors. Thus the branching fraction of $B^-\to\rho^-\eta'$ is
relatively suppressed for this flavor structure. In QCDF and PQCD
approaches, the form factors are different: $F^{B\to\eta_q}\gg
F^{B\to\eta_s}$. Thus the predicted branching ratio of $B^-\to
\rho^-\eta$ is comparable with ${\cal BR}(B^-\to\rho^-\eta')$ in
these two approaches.

As in $\bar B^0\to\pi^0\rho^0$ process, our predictions on branching
fractions of $\bar B^0\to \rho^0\eta(')$ and $\bar B^0\to
\omega\eta(')$ are much larger than the results evaluated in QCDF
and PQCD approach. These channels are the so-called color-suppressed
decays, as the contributions from terms proportional to $\zeta$ and
$\zeta_g$ are small due to the small Wilson coefficients. But in the
present framework, the hard-spectating form factors $\zeta_J$ and
$\zeta_{Jg}$ are comparable with $\zeta$ and $\zeta_g$.  Moreover,
the Wilson coefficients for these form factors are large. Thus
branching ratios of $\bar B^0\to \rho^0\eta(')$ and $\bar B^0\to
\omega\eta(')$  are much larger. 

Similar with $B\to K^*\pi$ and $B\to\rho K$ decays, $B\to
K^{*}\eta(\eta')$ are also induced by $b\to s$ transitions in which
charming penguins provide most important contributions. But compared
with $B\to K^{*}\pi$ and $B\to\rho K$ decays, there are something
new in these channels. In $B\to K^{*}\eta(\eta')$, there exist three
kinds of charming penguins:
\begin{eqnarray}
 A_{cc}^{K^*\eta_q}=\frac{1}{\sqrt 2} (A_{cc}^{VP}+2 A_{ccg}^{VP}),\;\;\;
 A_{cc}^{K^*\eta_s}= A_{ccg}^{VP}+A_{cc}^{PV}.
\end{eqnarray}
Substituting the values given in Eq.~\eqref{eq:solution1} and
Eq.~\eqref{eq:solution2}, we obtain ratios of charming penguins:
\begin{eqnarray}
\frac{|\frac{\cos(\theta)}{\sqrt 2} (A_{cc}^{VP}+2 A_{ccg}^{VP})-\sin(\theta)( A_{ccg}^{VP}+A_{cc}^{PV})|^2}
 {|\frac{\cos(\theta)}{\sqrt 2} (A_{cc}^{VP}+2 A_{ccg}^{VP})+\sin(\theta)(
 A_{ccg}^{VP}+A_{cc}^{PV})|^2}\sim 2.0.\nonumber
\end{eqnarray}
The branching fraction of $\bar B^0\to \bar K^{*0}\eta$ is about 4
times larger than that of $\bar B^0\to \bar K^{*0}\eta'$ for both
solutions. The main reason for the difference is:
$A_{cc}^{K^*\eta_s}$ is very small due to the cancelations between
$A_{cc}^{PV}$ and $A_{ccg}^{VP}$; the penguin operators play the
dominant role in the $B\to K^{*}\eta_s$ decay amplitudes. Our
results for these channels have a better agreement with experiments
than QCDF and PQCD.

\subsection{ $B_s\to VP$ Decays}

%

Since we have assumed the SU(3) symmetry for form factors and
charming penguins, branching fractions and direct CP asymmetries of
the $B_s$ decays are related to the corresponding $B$ decays:
\begin{eqnarray}
  &&{\cal BR}(\bar B_s^0\to K^{*+}
K^-)={{\cal BR}(\bar B^0\to\rho^+ K^-)},\;\;\;  {{\cal BR}(\bar
B_s^0\to K^{+} K^{*-})}={{\cal BR}(\bar B^0\to\pi^+ K^{*-})},\\
 &&
  A_{CP}(\bar B_s^0\to K^{*+}
K^-)={A_{CP}(\bar B^0\to\rho^+ K^-)},\;\;\;  {A_{CP}(\bar B_s^0\to
K^{+} K^{*-})}={A_{CP}(\bar B^0\to\pi^+ K^{*-})}.
\end{eqnarray}
These relations can also be applied to the following channels:
\begin{eqnarray}
 &&{{\cal BR}(\bar B_s^0\to K^{*+} \pi^{-})}={{\cal BR}(\bar
B^0\to\rho^+ \pi^-)},\;\;\;  {{\cal BR}(\bar B_s^0\to
K^{+}\rho^{-})}={{\cal BR}(\bar B^0\to \pi^+\rho^-)},\\
 &&
 {A_{CP}(\bar B_s^0\to K^{*+} \pi^{-})}={A_{CP}(\bar
B^0\to\rho^+ \pi^-)},\;\;\;  {A_{CP}(\bar B_s^0\to
K^{+}\rho^{-})}={A_{CP}(\bar B^0\to \pi^+\rho^-)}.
\end{eqnarray}

In tree-operator-dominated processes $\bar B_s^0\to\rho^- K^+$, we
obtain branching ratios which are much smaller than predictions in
the other two approaches: because  PQCD predicts $F^{B_s\to
K}=0.24^{+0.05+0.00}_{-0.04-0.01}$ and  QCDF use an even larger form
factor $F^{B_s\to K}=0.31\pm0.05$. ${\cal BR}(\bar B_s^0\to\pi^-
K^{*+})$ is consistent with results in QCDF and PQCD approach as the
$B\to K^*$ form factors are consistent. As in $B$ decays, we also
predict larger branching ratios for color-suppressed $B_s$ decays
than QCDF and PQCD which can be tested on the future experiments.

Our predictions on $b\to s$ processes $\bar B_s^0\to K^{*} K$ are
consistent with the other two approaches. But there are huge
differences in our predictions of ${\cal BR}(\bar
B_s\to\phi\eta(\eta'))$ with those in QCDF and PQCD. In PQCD
approach, contributions from gluonic components of $\eta$ and
$\eta'$ in $B\to\eta(')$ form factors are very small and can be
neglected~\cite{Charng:2006zj}. As shown in Ref.~\cite{Ali:2007ff},
decay amplitudes of $B_s\to\phi\eta_q$ are dynamically enhanced
sizably, as the Wilson coefficients $a_3-a_5$ strongly depend on the
factorization scale. In $B_s\to\phi\eta_s$, dominant penguin
operators are either proportional to $a_4-2r_\chi a_6$ or $a_4$. The
former Wilson coefficient is very small as $a_4\sim a_6$ and
$2r_\chi\sim1$. The total decay amplitudes of $B_s\to\phi\eta_q$ and
$B_s\to\phi\eta_s$ are in similar size but with different signs.
Thus branching ratio of $B_s\to\phi\eta$ predicted in PQCD approach
is relatively large while branching ratio of $B_s\to\phi\eta'$ is
small due to cancelations between the two
amplitudes~\cite{Ali:2007ff}. In the SCET framework, charming
penguins play the most important role: the charming penguin
$A_{cc}^{VP}$ almost cancels with $A_{cc}^{PV}$. Thus the dominant
contributions to $B_s\to\phi\eta(\eta')$ are from the gluonic
charming penguin and the penguin operators which are proportional to
$V_{tb}V_{ts}^*$. Neglecting the latter term, we have:
\begin{eqnarray}
 &&A_{cc}^{B_s\to\phi\eta}=\cos (\theta) \sqrt2A_{ccg}^{VP}-\sin(\theta)  A_{ccg}^{VP} \sim (\sqrt 2-1) A_{ccg}^{VP} ,\nonumber\\
 &&A_{cc}^{B_s\to\phi\eta'}=\sin (\theta)  \sqrt2A_{ccg}^{VP}+\cos(\theta) (  A_{ccg}^{VP}) \sim (\sqrt 2+1) A_{ccg}^{VP}.
\end{eqnarray}
These two equations can explain the small branching fraction for
$B_s\to\phi\eta$ together with the large one for $B_s\to\phi\eta'$.
The QCD penguin contributions do not change the ratios too much, but
sizable differences appear in the two solutions. The large
differences in two kinds of predictions on direct CP asymmetries
also confirm this feature.

In $B_s$ decays, there are 7 decays in which the direct CP
asymmetries are zero: $\bar B_s\to K^{*0}\bar K^0$, $\bar B_s\to\bar
K^{*0} K^0$, $\bar B_s\to \pi^0\phi$ and $\bar B_s\to
\rho^0(\omega)\eta(')$. As we know, in order to give a non-vanishing
direct CP violation, at least two decay amplitudes with different
weak phases and different strong phases are required. In the first
two decays, contributions from tree operators vanish at leading
order. The non-zero contribution is either proportional to the CKM
matrix elements $V_{tb}V_{ts}^*$ or $V_{cb}V_{cs}^*$ and both of
them are taken real in our calculation. Thus in these two channels,
there are only one weak phase and direct CP asymmetry is 0 in the
present framework. The latter 5 channels are induced by $b\to s$
transitions and one of the final state mesons is neither open nor
hidden strange. There is no contribution from charming penguins in
these modes. The direct CP asymmetries are zero for lack of
necessary strong phases.

\begin{table}[tb]
\caption{$CP$-averaged branching ratios ($\times 10^{-6}$) of
$B_s\to PV$ decays: the first solution (This work 1) and the second
solution (This work 2).  In both solutions, we have included the
chiraly enhanced penguin in $B\to VP$ decay amplitudes. The first
kinds of uncertainties are from uncertainties in charming penguins
and gluonic form factors which are discussed in the text; the second
kinds of uncertainties are from those in the CKM matrix elements. We
also cite theoretical results evaluated in QCDF~\cite{Beneke:2003zv}
and PQCD~\cite{Ali:2007ff} to make a comparison. }\label{tab:BRBs}
\begin{center}
 \begin{tabular}{|c|c|c|c|c|}
 \hline\hline {Modes}                        &   QCDF   & PQCD & This work 1 & This work 2  \\  \hline
   $\bar B^0_s\to K^+ K^{*-}$                & $4.1^{+1.7+1.5+1.0+9.2}_{-1.5-1.3-0.9-2.3}$
                 \                           & $6.0^{+1.7+1.7+0.7}_{-1.5-1.2-0.3}$
                                             & $8.4_{-3.4-1.3}^{+4.4+1.6}$
                                             & $9.5_{-2.8-1.1}^{+3.2+1.2}$\\
   $\bar B^0_s\to K^{*+} K^-$                & $5.5^{+1.3+5.0+0.8+14.2}_{-1.4-2.6-0.7-~3.6}$
                 \                           & $4.7^{+1.1+2.5+0.0}_{-0.8-1.4-0.0}$
                                             & $9.8_{-3.7-1.4}^{+4.6+1.7}$
                                             & $10.2_{-3.2-1.2}^{+3.8+1.5}$\\
   $\bar B^0_s\to K^{0}\overline K^{*0}$     & $3.9^{+0.4+1.5+1.3+10.4}_{-0.4-1.4-1.4-~2.8}$
                 \                           & $7.3^{+2.5+2.1+0.0}_{-1.7-1.3-0.0}$
                                             & $7.9_{-3.4-1.3}^{+4.4+1.6}$
                                             & $9.3_{-2.8-1.0}^{+3.2+1.2}$\\
   $\bar B^0_s\to K^{*0}\overline K^0$       & $4.2^{+0.4+4.6+1.1+13.2}_{-0.4-2.2-0.9-~3.2}$
                 \                           & $4.3^{+0.7+2.2+0.0}_{-0.7-1.4-0.0}$
                                             & $8.7_{-3.5-1.4}^{+4.4+1.6}$
                                             & $9.4_{-3.1-1.2}^{+3.7+1.4}$\\
   $B_s^0/\bar B^0_s\to K^+ K^{*-}$          &
                 \                           &
                                             & $16.5_{-4.9-2.6}^{+6.4+3.2}$
                                             & $17.5_{-4.4-2.1}^{+5.0+2.5}$\\
   $B_s^0/\bar B^0_s\to K^{*+} K^-$          &
                 \                           &
                                             & $19.8_{-5.6-2.9}^{+6.9+3.4}$
                                             & $21.8_{-4.7-2.4}^{+5.4+2.8}$\\
 \parbox[c]{2cm}{${\tiny\bar B_s^0\to K^{*+} K^-}$ \\ ${\tiny\bar B_s^0\to  K^{*-} K^+} $} \;\;\Big\}
                                             &
                                             &
                                             & $18.2_{-5.0-2.7}^{+6.3+3.3}$
                                             & $19.7_{-4.2-2.2}^{+5.0+2.6}$ \\%
 \parbox[c]{2cm}{${\tiny\bar B_s^0\to K^{*0}\bar K^0}$ \\ ${\tiny\bar B_s^0\to \bar K^{*0} K^0} $} \;\;\Big\}
                                                              &
                                       &
                                    & $16.6_{-4.9-2.7}^{+6.2+3.2}$
                                    & $18.7_{-4.2-2.2}^{+4.9+2.6}$ \\
   $\bar B^0_s\to\pi^0\phi$                  & $0.12^{+0.03+0.04+0.01+0.02}_{-0.02-0.04-0.01-0.01} $
                 \                           & $0.16^{+0.06+0.02+0.00}_{-0.05-0.02-0.00}$
                                             & $0.07_{-0.00-0.01}^{+0.00+0.01}$
                                             & $0.09_{-0.00-0.01}^{+0.00+0.01}$ \\\hline
   $\bar B^0_{s}{\to}\pi^-K^{*+}$            & $8.7^{+4.6+3.5+0.7+0.8}_{-3.7-2.9-1.0-0.7}$
                 \                           &  $7.6^{+2.9+0.4+0.5}_{-2.2-0.5-0.3}$
                                             & $5.9_{-0.5-0.5}^{+0.5+0.5}$
                                             & $6.6_{-0.1-0.7}^{+0.2+0.7}$\\
   $\bar B^0_{s}{\to}{\pi}^{0}K^{*0}$        & $0.25^{+0.08+0.10+0.32+0.30}_{-0.08-0.06-0.14-0.14}$
                 \                           & $0.07^{+0.02+0.04+0.01}_{-0.01-0.02-0.01}$
                                             & $0.90_{-0.01-0.11}^{+0.07+0.10}$
                                             & $1.07_{-0.15-0.09}^{+0.16+0.10}$\\
   $\bar B^0_{s}{\to}\rho^-K^{+}$            & $24.5^{+11.9+9.2+1.8+1.6}_{-9.7-7.8-3.0-1.6}$
                 \                           & $17.8^{+7.7+1.3+1.1}_{-5.6-1.6-0.9}$
                                             & $7.6_{-0.1-0.8}^{+0.3+0.8}$
                                             & $10.2_{-0.5-0.9}^{+0.4+0.9}$\\
   $\bar B^0_{s}{\to}{\rho}^{0}K^{0}$        & $0.61^{+0.33+0.21+1.06+0.56}_{-0.26-0.15-0.38-0.36}$
                 \                           & $0.08^{+0.02+0.07+0.01}_{-0.02-0.03-0.00}$
                                             & $2.0_{-0.2-0.2}^{+0.2+0.2}$
                                             & $0.81_{-0.02-0.09}^{+0.05+0.08}$\\
   $\bar B^0_s\to K^{0}\omega$               & $0.51^{+0.20+0.15+0.68+0.40}_{-0.18-0.11-0.23-0.25} $
                 \                           & $0.15^{+0.05+0.07+0.02}_{-0.04-0.03-0.01}$
                                             & $0.90_{-0.01-0.11}^{+0.08+0.10}$
                                             & $1.3_{-0.1-0.1}^{+0.1+0.1}$\\
   $\bar B^0_s\to K^0\phi$                   & $0.27^{+0.09+0.28+0.09+0.67}_{-0.08-0.14-0.06-0.18} $
                 \                           & $0.16^{+0.04+0.09+0.02}_{-0.03-0.04-0.01}$
                                             & $0.44_{-0.18-0.07}^{+0.23+0.08}$
                                             & $0.54_{-0.17-0.07}^{+0.21+0.08}$\\\hline
   $\bar B^0_s\to\rho^0\eta$                 & $0.17^{+0.03+0.07+0.02+0.02}_{-0.03-0.06-0.02-0.01} $
                 \                           & $0.06^{+0.03+0.01+0.00}_{-0.02-0.01-0.00}$
                                             & $0.08_{-0.03-0.01}^{+0.04+0.01}$
                                             & $0.06_{-0.02-0.00}^{+0.03+0.00}$\\
   $\bar B^0_s\to\rho^0\eta^\prime$          & $0.25^{+0.06+0.10+0.02+0.02}_{-0.05-0.08-0.02-0.02} $
                 \                           & $0.13^{+0.06+0.02+0.00}_{-0.04-0.02-0.01}$
                                             & $0.003_{-0.000-0.000}^{+0.082+0.000}$
                                             & $0.14_{-0.11-0.01}^{+0.24+0.01}$\\
   $\bar B^0_s\to\omega\eta$                 & $0.012^{+0.005+0.010+0.028+0.025}_{-0.004-0.003-0.006-0.006} $
                 \                           & $0.04^{+0.03+0.05+0.00}_{-0.01-0.02-0.00}$
                                             & $0.04_{-0.02-0.00}^{+0.04+0.00}$
                                             & $0.007_{-0.002-0.001}^{+0.011+0.001}$\\
   $\bar B^0_s\to\omega\eta^\prime$          & $0.024^{+0.011+0.028+0.077+0.042}_{-0.009-0.006-0.010-0.015}$
                 \                           & $0.44^{+0.18+0.15+0.00}_{-0.13-0.14-0.01}$
                                             & $0.001_{-0.000-0.000}^{+0.095+0.000}$
                                             & $0.20_{-0.17-0.02}^{+0.34+0.02}$\\
   $\bar B^0_s\to\phi\eta$                   & $0.12^{+0.02+0.95+0.54+0.32}_{-0.02-0.14-0.12-0.13} $
                 \                           & $3.6^{+1.5+0.8+0.0}_{-1.0-0.6-0.0}$
                                             & $0.59_{-0.59-0.10}^{+2.02+0.12}$
                                             & $0.94_{-0.97-0.13}^{+1.89+0.16}$\\
   $\bar B^0_s\to\phi\eta^\prime$            & $0.05^{+0.01+1.10+0.18+0.40}_{-0.01-0.17-0.08-0.04}$
                 \                           & $0.19^{+0.06+0.19+0.00}_{-0.01-0.13-0.00}$
                                             & $7.3_{-5.4-1.3}^{+7.7+1.6}$
                                             & $4.3_{-3.6-0.6}^{+5.2+0.7}$\\\hline
   $\bar B^0_s\to K^{*0}\eta$                & $0.26^{+0.15+0.49+0.15+0.57}_{-0.13-0.22-0.05-0.15}$
                 \                           & $0.17^{+0.04+0.10+0.03}_{-0.04-0.06-0.01}$
                                             & $1.7_{-0.3-0.1}^{+0.3+0.2}$
                                             & $0.62_{-0.14-0.08}^{+0.14+0.07}$\\
   $\bar B^0_s\to K^{*0}\eta^\prime$         & $0.28^{+0.04+0.46+0.23+0.29}_{-0.04-0.24-0.10-0.15} $
                 \                           & $0.09^{+0.02+0.03+0.01}_{-0.02-0.02-0.01}$
                                             & $0.64_{-0.26-0.11}^{+0.33+0.11}$
                                             & $0.87_{-0.32-0.08}^{+0.35+0.10}$\\
 \hline\hline\end{tabular}
\end{center}
 \end{table}

\begin{table}[tb]
\caption{Direct $CP$ asymmetries (in \%) in the $B_s\to PV$ decays:
the first solution (This work 1) and the second solution (This work
2).  In both solutions, the chiraly enhanced penguin has been taken
into account in $B\to VP$ decay amplitudes. The first kinds of
uncertainties are from uncertainties in charming penguins and
gluonic form factors which are discussed in the text; the second
kinds of uncertainties are from those in the CKM matrix elements. We
also cite theoretical results evaluated in QCDF~\cite{Beneke:2003zv}
and PQCD~\cite{Ali:2007ff} to make a comparison. } \label{tab:AcpBs}
\begin{center}
 \begin{tabular}{|c|c|c|c|c|c}
  \hline\hline {Modes}   &    QCDF   & PQCD  & This work 1& This work 2 \\  \hline
   $\overline B^0_s\to K^+ K^{*-}$                          & $2.2^{+0.6+8.4+5.1+68.6}_{-0.7-8.0-5.9-71.0}$
                                                            & $-36.6^{+2.3+2.8+1.3}_{-2.3-3.5-1.2}$
                                                            & $-11.2_{-16.2-1.3}^{+19.1+1.3}$
                                                            & $-12.3_{-11.3-0.8}^{+11.4+0.8}$  \\
   $\overline B^0_s\to K^{*+} K^-$                          & $-3.1^{+1.0+3.8+1.6+47.5}_{-1.1-2.6-1.3-45.0}$
                                                            &  $55.3^{+4.4+8.5+5.1}_{-4.9-9.8-2.5}$
                                                            & $7.1_{-12.4-0.7}^{+11.2+0.7}$
                                                            & $9.6_{-13.5-0.9}^{+13.0+0.7}$\\
   $\overline B^0_s\to K^{0}\overline K^{*0}$               & $1.7^{+0.4+0.6+0.5+1.4}_{-0.5-0.5-0.4-0.8}$
                                                            &  $0$ &$0$& $0$  \\
   $\overline B^0_s\to K^{*0}\overline K^0$                 & $0.2^{+0.0+0.2+0.1+0.2}_{-0.1-0.3-0.1-0.1}$
                                                            &  $0$  &$0$& $0$\\
   $\overline B^0_s\to\pi^0\phi$                            & $27.2^{+6.1+9.8+2.7+32.0}_{-6.8-5.6-2.4-37.1} $
                                                            & $13.3^{+0.3+2.1+1.5}_{-0.4-1.7-0.7}$
                                                            & $0$&$0$ \\\hline
   $\overline B^0_s\to\rho^0\eta$                           & $27.8^{+6.4+9.1+2.6+25.9}_{-6.7-5.7-2.2-28.4} $
                                                            &  $-9.2^{+1.0+2.8+0.4}_{-0.4-2.7-0.7}$
                                                            &$0$&$0$ \\
   $\overline B^0_s\to\rho^0\eta^\prime$                    & $28.9^{+6.1+10.3+1.5+24.8}_{-7.5-~6.3-1.8-27.5} $
                                                            &  $25.8^{+1.3+2.8+3.4}_{-2.0-3.6-1.5}$
                                                            & $0$&$0$\\
   $\overline B^0_s\to\omega\eta$                           &... &  $-16.7^{+5.8+15.4+0.8}_{-3.2-19.1-1.7}$
                                                            &$0$&$0$\\
   $\overline B^0_s\to\omega\eta^\prime$                    & ...&  $7.7^{+0.4+4.5+9.4}_{-0.1-4.2-0.4}$
                                                            &$0$&$0$\\
   $\overline B^0_s\to\phi\eta$                             &  $-8.4^{+2.0+30.1+14.6+36.3}_{-2.1-71.2-44.7-59.7} $
                                                            &$-1.8^{+0.0+0.6+0.1}_{-0.1-0.6-0.2}$
                                                            & $21.3_{-83.2-2.6}^{+53.5+2.5}$
                                                            & $16.9_{-18.3-1.6}^{+13.8+1.6}$\\
   $\overline B^0_s\to\phi\eta^\prime$                      & $-62.2^{+15.9+132.3+80.8+122.4}_{-10.2-~84.2-46.8-~49.9}$
                                                            &  $7.8^{+1.5+1.2+0.1}_{-0.5-8.6-0.4}$
                                                            & $4.4_{-7.1-0.6}^{+5.3+0.6}$
                                                            & $7.8_{-4.9-0.8}^{+5.0+0.8}$\\\hline\hline
   ${\overline{B}}^{0}_{s}{\to}\pi^-K^{*+}$                 & $0.6^{+0.2+1.4+0.1+19.9}_{-0.1-1.7-0.1-20.1}$
                                                            & $-19.0^{+2.5+2.7+0.9}_{-2.6-3.4-1.4}$
                                                            & $-9.9_{-16.7-0.7}^{+17.2+0.9}$
                                                            & $-12.4_{-15.3-1.2}^{+17.5+1.1}$\\
   ${\overline{B}}^{0}_{s}{\to}{\pi}^{0}K^{*0}$             & $-45.7^{+14.3+13.0+28.4+80.0}_{-16.0-11.6-28.0-59.7}$
                                                            & $-47.1^{+7.4+35.5+2.9}_{-8.7-29.8-7.0}$
                                                            & $22.9_{-40.2-1.9}^{+33.1+2.1}$
                                                            & $13.4_{-18.8-1.2}^{+18.6+0.8}$ \\
   ${\overline{B}}^{0}_{s}{\to}\rho^-K^{+}$                 & $-1.5^{+0.4+1.2+0.2+12.1}_{-0.4-1.4-0.3-12.1}$
                                                            & $14.2^{+2.4+2.3+1.2}_{-2.2-1.6-0.7}$
                                                            & $11.8_{-20.0-1.1}^{+17.5+1.2}$
                                                            & $10.8_{-10.2-1.0}^{+9.4+0.9}$ \\
   ${\overline{B}}^{0}_{s}{\to}{\rho}^{0}K^{0}$             & $24.7^{+7.1+14.0+22.8+51.3}_{-5.2-12.4-17.7-52.3}$
                                                            & $73.4^{+6.4+16.2+2.2}_{-11.7-47.8-3.9}$
                                                            & $-12.0_{-19.6-0.7}^{+20.1+1.0}$
                                                            & $-32.5_{-23.4-2.9}^{+30.7+2.7}$ \\
   $\overline B^0_s\to K^{0}\omega$                         & $-43.9^{+13.6+18.0+30.6+57.7}_{-13.4-18.2-30.2-49.3} $
                                                            & $-52.1^{+3.2+22.7+3.2}_{-0.0-15.1-2.0}$
                                                            & $24.4_{-41.4-2.0}^{+33.7+2.2}$
                                                            & $18.2_{-17.0-1.7}^{+16.4+1.2}$ \\
   $\overline B^0_s\to K^0\phi$                             &  $-10.3^{+3.0+4.7+3.7+5.0}_{-2.4-3.0-4.1-7.5} $
                                                            &  $0$
                                                            & $-3.0_{-4.7-0.3}^{+5.3+0.3}$
                                                            & $-2.2_{-2.9-0.1}^{+3.0+0.1}$\\\hline
   $\overline B^0_s\to K^{*0}\eta$                          & $40.2^{+17.0+24.6+~7.8+65.9}_{-11.5-30.8-14.0-96.3}$
                                                            & $51.2^{+6.2+14.1+2.0}_{-6.4-12.4-3.3}$
                                                            & $-25.7_{-22.0-1.3}^{+23.4+2.0}$
                                                            & $-62.7_{-22.5-3.9}^{+28.1+2.6}$\\
   $\overline B^0_s\to K^{*0}\eta^\prime$                   & $-58.6^{+16.9+41.4+19.9+44.9}_{-11.9-11.7-13.9-35.7} $
                                                            & $-51.1^{+4.6+15.0+3.2}_{-6.6-18.2-4.1}$
                                                            & $-35.2_{-49.4-3.8}^{+63.3+3.1}$
                                                            & $-32.1_{-23.2-1.7}^{+22.8+2.6}$\\
 \hline\hline\end{tabular}
\end{center}
 \end{table}


\subsection{Mixing-induced CP asymmetries}

In this subsection, we will discuss mixing-induced CP asymmetries
which can be studied via time-dependent measurements of decay
widths. The four decay amplitudes in $B^0/\bar B^0\to f(\bar f)$
decays are defined by:
\begin{eqnarray}
 A_f=\langle f| {\cal H}_{eff}|B^0\rangle,
 \bar A_f=\langle f|  {\cal H}_{eff}|\bar B^0\rangle,
 A_{\bar f}=\langle \bar f|  {\cal H}_{eff}|B^0\rangle,
 \bar A_{\bar f}=\langle\bar f|  {\cal H}_{eff}|\bar B^0\rangle.
\end{eqnarray}
Considering the width differences of the two mass eigenstates $B_H$
and $B_L$, the decay amplitudes squared at time $t$ of the state
that was a pure $B^0$ state at time $t=0$ can be parameterized by:
\begin{eqnarray}
 |A_f(t)|^2\equiv |\langle f|B(t)\rangle |^2= \frac{e^{-\Gamma
 t}}{2}(|A_f|^2+|\bar A_f|^2)
\Big[\cosh \Big(\frac{\Delta \Gamma t}{2}\Big)
+H_f \sinh \Big(\frac{\Delta \Gamma t}{2}\Big)\nonumber\\
+C_f\cos( \Delta m t)- S_f \sin (\Delta m t)
\Big],\label{eq:timedependent}
\end{eqnarray}
where $\Delta m=m_H-m_L>0$ and  $\Delta \Gamma=\Gamma_H-\Gamma_L$ is
the difference of decay widths for the heavier and lighter $B^0$
mass eigenstates. The time-dependent decay amplitudes squared of
another channel $\bar B^0\to f$ is obtained from the above
expression by flipping the signs of the $\cos(\Delta m t)$ and
$\sin(\Delta m t)$ terms. For decays to the CP-conjugate final
state, one replaces $f$ by $\bar f$.

Time-dependent decay amplitudes squared can be simplified in two
kinds of cases. In $B^0$-$\bar B^0$ system, the small width
difference $\Delta\Gamma$ can be safely neglected. Thus the first
two terms $\cosh \Big(\frac{\Delta \Gamma t}{2}\Big) $ and $ \sinh
\Big(\frac{\Delta \Gamma t}{2}\Big)$ in Eq~\eqref{eq:timedependent}
can be reduced to $1$ and $0$ and the decay amplitudes squared
becomes:
\begin{eqnarray}
 |A_f(t)|^2\equiv |\langle f|B(t)\rangle |^2= \frac{e^{-\Gamma
 t}}{2}(|A_f|^2+|\bar A_f|^2)
\Big[1 +C_f\cos( \Delta m t)- S_f \sin (\Delta m t) \Big],
\end{eqnarray}
In the following, we use the phase convention $CP|B^0\rangle =|\bar
B^0\rangle$ and define the following amplitudes ratios:
\begin{eqnarray}
 \lambda_f=\frac{q}{p}\frac{\bar A_f}{A_f},\;\;\; \lambda_{\bar f}=\frac{q}{p}\frac{\bar
 A_{\bar f}}{A_{\bar f}},
\end{eqnarray}
and $q$ and $p$ are the mixing parameters between $B^0$ and $\bar
B^0$. The definitions for $C_f$ and $S_f$ are given by:
\begin{eqnarray}
 C_f=\frac{1-|\lambda_f|^2}{1+|\lambda_f|^2}=\frac{| A_{ f}|^2-|\bar A_f|^2}{|A_{
 f}|^2+|\bar A_f|^2}
 ,\;\;\; S_f=2\frac{{\mbox
 {Im}}(\lambda_f)}{1+|\lambda_f|^2},\nonumber\\
 C_{\bar f}=\frac{1-|\lambda_{\bar f}|^2}{1+|\lambda_{\bar f}|^2}=\frac{| A_{\bar f}|^2-|\bar A_{\bar f}|^2}{|A_{
 \bar f}|^2+|\bar A_{\bar f}|^2}
 ,\;\;\; S_{\bar f}=2\frac{{\mbox
 {Im}}(\lambda_{\bar f})}{1+|\lambda_{\bar
 f}|^2}.\label{eq:mixingCS}
\end{eqnarray}
The system of four decay modes defines five asymmetry parameters,
$C_f$, $S_f$, $C_{\bar f}$, $S_{\bar f}$ together with the global
charge asymmetry related to the overall normalization:
\begin{eqnarray}
 A_{CP}=\frac{|A_f|^2+|\bar A_f|^2-|A_{\bar f}|^2-|\bar A_{\bar f}|^2}
 {|A_f|^2+|\bar A_f|^2+|A_{\bar f}|^2+|\bar A_{\bar f}|^2}.\label{eq:mixingAcp}
\end{eqnarray}
One can also use the parameters $C\equiv \frac{1}{2}(C_f+C_{\bar
f})$, $S\equiv \frac{1}{2}(S_f+S_{\bar f})$, $\Delta C\equiv
\frac{1}{2}(C_f-C_{\bar f})$, $\Delta S\equiv
\frac{1}{2}(S_f-S_{\bar f})$. If there is no direct CP violation,
only two independent decay amplitudes squared are left. Thus
$A_{CP}=0$, $C_f=-C_{\bar f}$ and $S_f=-S_{\bar f}$ which also
implies $C=0$ and $S=0$. If we recall that the CP invariance
conditions at the decay amplitudes level are $A_f=\bar A_{\bar f}$
and $A_{\bar f}=\bar A_f$, one can  study the following two
parameters:
\begin{eqnarray}
 A_{f\bar f}=\frac{|\bar A_{\bar f}|^2-|A_f|^2}{|\bar A_{\bar
 f}|^2+|A_f|^2} ,\;\;\; A_{\bar ff}=\frac{|\bar A_{ f}|^2-|A_{\bar f}|^2}{|\bar A_{ f}|^2+|A_{\bar
 f}|^2}.\label{eq:mixingAfbarf}
\end{eqnarray}
Sometimes, they are considered as more physically intuitive
parameters since they characterize direct CP violations. In
$B^0\to\rho^\pm\pi^\mp$ decays, (choosing $f=\rho^+\pi^-$ and $\bar
f=\rho^-\pi^+$), we use  $A_{\rho\pi}^{+-}$ parameterizes the direct
CP violation in decays in which the produced $\rho$ meson does not
contain the spectator quark, while $A^{-+}_{\rho\pi}$ parameterizes
the direct CP violation in decays in which it does. Of course, these
two parameters are not independent of the other sets of parameters
given above, and can be written by:
\begin{eqnarray}
 A_{\rho\pi}^{+-}=-\frac{A_{CP}+C_{f\bar f}+A_{CP}\Delta C_{f\bar f}}{1+\Delta C_{f\bar f} +A_{CP} C_{f\bar
 f}},\;\;\;
 A_{\rho\pi}^{-+}=-\frac{A_{CP}+C_{f\bar f}+A_{CP}\Delta C_{f\bar f}}{-1+\Delta C_{f\bar f} +A_{CP} C_{f\bar
 f}}.
\end{eqnarray}
Predictions on these parameters are given in
table~\ref{tab:mixingpirho}.  Most of them are consistent with the
data except $\Delta C$ and $\Delta S$.

\begin{table}[tb]
\caption{Mixing-induced CP asymmetries in $B\to \pi^\pm\rho^\mp$
decay processes: the first solution (This work 1) and the second
solution (This work 2).  In both cases, the chiraly enhanced penguin
has been taken into account. The first kinds of uncertainties are
from uncertainties in charming penguins which are discussed in the
text; the second kinds of uncertainties are from those in the CKM
matrix elements. We also cite theoretical results evaluated in QCDF
approach \cite{Beneke:2003zv} to make a comparison. }
\label{tab:mixingpirho}
\begin{tabular}{|c|c|c|c|c|c|c}\hline
  Parameter  &   Exp.                     & QCDF  & This work 1 &  This work 2\\ \hline \
  $A_{CP}$   &  $-0.13 \pm 0.04$          & $0.00^{+0.00+0.01+0.00+0.10}_{-0.00-0.01-0.00-0.10}$
                                          & $-0.12_{-0.05-0.03}^{+0.04+0.04}$
                                          & $-0.21_{-0.02-0.03}^{+0.03+0.02}$ \\
  $C$        &  $0.01\pm0.07$             & $0.00^{+0.00+0.01+0.00+0.02}_{-0.00-0.01-0.00-0.02}$
                                          & $-0.01_{-0.12-0.00}^{+0.13+0.00}$
                                          & $0.01_{-0.10-0.00}^{+0.09+0.00}$   \\
  $S$        &  $0.01\pm0.09$             & $0.13^{+0.60+0.04+0.02+0.02}_{-0.65-0.03-0.01-0.01}$
                                          & $-0.11_{-0.08-0.13}^{+0.07+0.08}$
                                          & $-0.01_{-0.07-0.14}^{+0.06+0.08}$ \\
  $\Delta C$ &  $0.37\pm0.08$             & $0.16^{+0.06+0.23+0.01+0.01}_{-0.07-0.26-0.02-0.02}$
                                          & $0.11_{-0.13-0.01}^{+0.12+0.01}$
                                          & $0.12_{-0.10-0.01}^{+0.09+0.01}$\\
  $\Delta S$ &  $-0.04\pm0.10$            & $-0.02^{+0.01+0.00+0.00+0.01}_{-0.00-0.01-0.00-0.01}$
                                          & $-0.47_{-0.06-0.04}^{+0.08+0.05}$
                                          & $0.43_{-0.07-0.03}^{+0.05+0.03}$\\%
                                          \hline
 \hline
\end{tabular}
\end{table}

If the final state $f$ is a CP eigenstate, there are only two
different amplitudes since $|f\rangle=\pm|\bar f\rangle$ and the
time-dependent decay amplitudes squared can also be simplified.
Restricting the final state $f$ to have definite CP-parity, the
time-dependent decay width for the $B \to f$ decay is:
\begin{eqnarray}
\Gamma(B^0(t)\to f)= e^{-\Gamma t}\; \overline \Gamma(B\to f)
\Big[\cosh \Big(\frac{\Delta \Gamma t}{2}\Big)
+H_f \sinh \Big(\frac{\Delta \Gamma t}{2}\Big)\nonumber\\
-{\cal A}_{\mathrm{CP}}^{dir}\cos( \Delta m t)- S_f \sin (\Delta m
t) \Big].
\end{eqnarray}
The time dependent decay width $\Gamma(\bar B(t)\to f)$ is obtained
from the above expression by flipping the signs of the $\cos(\Delta
m t)$ and $\sin(\Delta m t)$ terms. In the $B_d$ system, the width
differences are small which can be safely neglected, but in the
$B_s$ system, we expect a much larger decay width difference
$(\Delta\Gamma/\Gamma)_{B_s}$. This is estimated within the standard
model to have a value $(\Delta\Gamma/\Gamma)_{B_s}=-0.147\pm 0.060$~
\cite{Beneke:1998sy}, while experimentally $(\Delta
\Gamma/\Gamma)_{B_s}=-0.33^{+0.09}_{-0.11}$ \cite{Barberio:2006bi},
so that both  $S_f$ and $H_f$,  can be extracted from the time
dependent decays of  $B_s$ mesons. The definition of the various
quantities in the above equation are as follows:
\begin{eqnarray}
S_f =\frac{2{\mbox {Im}}[\lambda]}{1+|\lambda|^2}, ~~~ H_f
=\frac{2\mbox {Re}[\lambda]}{1+|\lambda|^2},
\end{eqnarray}
with
\begin{eqnarray}
\lambda=\eta_f \frac{q}{p}\frac{A(\bar B\to f)}{A(B\to\bar
f)},\label{eq:mixinglambda}
\end{eqnarray}
where  $\eta_f$ is $+1(-1)$ for a CP-even (CP-odd) final state $f$.
$q/p=e^{-2i\beta}$ for the $B_d$ system while $q/p=e^{+2i\epsilon}$
for the $B_s$ system where $\epsilon=\mbox{arg}[
-V_{cb}V_{ts}V^*_{cs} V^* _{tb}]$. With the convention
$\mbox{arg}[V_{cb}]=\mbox{arg}[V_{cs}]=0$, the parameter can be
reduced to $\epsilon=\mbox{arg}[-V_{ts}V_{tb}^*]$. For $b\to s$
transition induced $\bar B^0$ decays,  the ratios of decay
amplitudes $\frac{A(\bar B\to f)}{A(B\to\bar f)}$ are almost real
and thus $S_f\sim \sin(2\beta)$. These channels provide a good way
to measure $\sin (2\beta)$.
Experimentalists often use the following parameters in $b\to s$
transitions:
\begin{eqnarray}
 -\eta_f S_f=-2 \frac{{\mbox {Im}}[\frac{q}{p}\frac{A(\bar B_s\to f)}{A(B_s\to\bar
 f)}]}{1+|\lambda|^2}, \;\;\;
 -\eta_f H_f=-2 \frac{{\mbox {Re}}[\frac{q}{p}\frac{A(\bar B_s\to f)}{A(B_s\to\bar
 f)}]}{1+|\lambda|^2},
\end{eqnarray}
while the latter parameter is only defined for the $B_s^0-\bar
B_s^0$ system. Although the $K^{*0}$ meson is not a CP eigenstate,
its daughter-mesons $K_S\pi^0$ behave as CP eigenstates. Thus we
also give the predictions on mixing-induced CP asymmetries in the
decays involving a $K^{*0}$ meson and other related decays. Results
for these parameters are collected in table~\ref{tab:mixingbtos} and
table~\ref{tab:mixingBs}, where predictions on decays with branching
ratios smaller than $10^{-7}$ are omitted.

\begin{table}[tb]
\caption{Mixing-induced CP asymmetries $S_f$ in $B\to VP$ decay
processes:the first solution (This work 1) and the second solution
(This work 2). In both cases, the chiraly enhanced penguin has been
taken into account. The first kinds of uncertainties are from
uncertainties in charming penguins and gluonic form factors which
are discussed in the text; the second kinds of uncertainties are
from those in the CKM matrix elements. We also quote the
experimental results to make a comparison.  } \label{tab:mixingbtos}
\begin{tabular}{|c|c|c|c|c|c|c}\hline
  Channel                           &   Exp.                       &  This work 1 &  This work 2\\ \hline \
  $\bar B^0\to  \rho^0 K_S$         &  $0.61^{+0.22}_{-0.24}\pm0.09\pm0.08$
                                    & $0.85_{-0.05-0.01}^{+0.04+0.01}$
                                    & $0.56^{+0.02+0.01}_{-0.03-0.01}$  \\
  $\bar B^0\to \omega K_S$          &  $0.48\pm0.24$
                                    &  $0.51_{-0.06-0.02}^{+0.05+0.02}$
                                    & $0.80^{+0.02+0.01}_{-0.02-0.01}$\\
  $\bar B^0\to  \phi  K_S$    &  $0.39\pm0.17$  & $0.69$  &$0.69$\\\hline
  $\bar B^0\to  K^{*-}  \pi^+\to K_S\pi^-\pi^+$    &  ...
                                    &  $0.93_{-0.07-0.02}^{+0.04+0.01}$
                                    & $0.34^{+0.06+0.03}_{-0.07-0.03}$\\
  $\bar B^0\to  K^{*0}  \pi^0\to K_S\pi^0\pi^0$    & ...     &$0.52_{-0.05-0.02}^{+0.04+0.02}$
                                    & $0.79^{+0.02+0.01}_{-0.02-0.01}$\\
  $\bar B^0\to  K^{*0}  \eta\to K_S\pi^0\eta$    &   ...     & $0.75_{-0.01-0.01}^{+0.01+0.01}$
                                    & $0.64^{+0.01+0.00}_{-0.01-0.00}$\\
  $\bar B^0\to  K^{*0}  \eta'\to K_S\pi^0\eta'$    &  ...
                                    & $0.76_{-0.06-0.01}^{+0.07+0.01}$
                                    & $0.66^{+0.04+0.00}_{-0.05-0.00}$
  \\\hline\hline
  $S(\bar B^0\to\pi^0\rho^0)$& $0.12\pm0.38$     &$-0.11_{-0.14-0.15}^{+0.14+0.10}$
                                          & $-0.19_{-0.14-0.15}^{+0.14+0.10}$     \\
  $S(\bar B^0\to\pi^0\omega)$&  ...              & $-0.87_{-0.00-0.01}^{+0.44+0.02}$
                                          & $0.72_{-1.54-0.11}^{+0.36+0.07}$  \\\hline
  $\bar B^0\to  \rho^0\eta$         &  ...
                                    & $0.86_{-2.03-0.07}^{+0.15+0.03}$
                                    & $0.29_{-0.44-0.15}^{+0.36+0.09}$ \\
  $\bar B^0\to  \rho^0\eta'$        &  ...
                                    & $0.79_{-1.73-0.09}^{+0.20+0.05}$
                                    & $0.38_{-1.24-0.14}^{+0.22+0.09}$\\
  $\bar B^0\to  \omega\eta$         &   ...
                                    & $0.12_{-0.20-0.17}^{+0.19+0.10}$
                                    & $-0.16_{-0.15-0.15}^{+0.14+0.10}$\\
  $\bar B^0\to  \omega\eta'$        &  ...
                                    & $0.23_{-1.10-0.10}^{+0.59+0.10}$
                                    & $-0.27_{-0.33-0.14}^{+0.17+0.09}$\\
  \hline
 \hline
\end{tabular}
\end{table}

\begin{table}[tb]
\caption{Mixing-induced $CP$ asymmetries $(S_f)_{B_s}$ and
$(H_f)_{B_s}$ in $B_s\to PV$ decays. Results obtained in the PQCD
approach~\cite{Ali:2007ff} are also collected here; the errors for
these entries correspond to the uncertainties in the input hadronic
quantities (charming penguins and the two form factors $\zeta_g$ and
$\zeta_{Jg}$ ), and the CKM matrix elements, respectively. }
\label{tab:mixingBs}
\begin{center}
 \begin{tabular}{|c|c|c|c|}
  \hline\hline {Modes}   &     PQCD   &  This work 1& This work 2   \\  \hline
 $\overline B^0_s\to\pi^0\phi$        & $-0.07^{+0.01+0.08+0.02}_{-0.01-0.09-0.03}$
                                      & $0.89_{-0.00-0.05}^{+0.00+0.04}$
                                      & $0.90_{-0.00-0.03}^{+0.00+0.02}$\\
                                      & $0.98^{+0.00+0.01+0.01}_{-0.00-0.03-0.00}$
                                      & $-0.45_{-0.00-0.10}^{+0.00+0.09}$
                                      & $0.44_{-0.00-0.05}^{+0.00+0.05}$ \\
 $\overline B^0_s\to\rho^0\eta$       & $0.15^{+0.06+0.14+0.01}_{-0.06-0.16-0.01}$
                                      & $1.00_{-0.06-0.01}^{+0.00+0.00}$
                                      & $0.60_{-0.53-0.03}^{+0.30+0.03}$\\
                                      & $0.98^{+0.01+0.01+0.00}_{-0.01-0.03-0.00}$
                                      & $-0.04_{-0.32-0.08}^{+0.41+0.08}$
                                      & $0.80_{-0.36-0.02}^{+0.20+0.02}$ \\
 $\overline B^0_s\to\rho^0\eta^\prime$&$-0.16^{+0.00+0.10+0.04}_{-0.00-0.12-0.05}$
                                      & $0.95_{-1.60-0.02}^{+0.00+0.02}$
                                      & $-0.41_{-0.75-0.15}^{+0.75+0.10}$\\
                                      & $0.95^{+0.01+0.01+0.01}_{-0.00-0.02-0.02}$
                                      & $0.32_{-1.29-0.06}^{+0.67+0.06}$
                                      & $-0.91_{-0.08-0.04}^{+0.82+0.08}$\\
 $\overline B^0_s\to\omega\eta$       & $-0.02^{+0.01+0.02+0.00}_{-0.03-0.08-0.00}$
                                      & $-0.62_{-0.18-0.12}^{+0.41+0.08}$
                                      & $0.93_{-0.98-0.04}^{+0.04+0.03}$\\
                                      & $0.99^{+0.01+0.01+0.00}_{-0.01-0.06-0.00}$
                                      & $-0.79_{-0.20-0.06}^{+0.16+0.11}$
                                      & $-0.37_{-0.65-0.10}^{+1.37+0.09}$\\
 $\overline B^0_s\to\omega\eta^\prime$& $-0.11^{+0.01+0.04+0.02}_{-0.00-0.04-0.03}$
                                      & $-0.25_{-0.74-0.16}^{+1.23+0.10}$
                                      & $-1.00_{-0.00-0.00}^{+0.04+0.01}$\\
                                      & $0.99^{+0.00+0.00+0.00}_{-0.00-0.00-0.00}$
                                      & $-0.97_{-0.00-0.02}^{+2.12+0.05}$
                                      & $-0.09_{-0.22-0.08}^{+0.32+0.12}$\\
 $\overline B^0_s\to\phi\eta$         & $-0.03^{+0.02+0.07+0.01}_{-0.01-0.20-0.02}$
                                      & $-0.39_{-0.15-0.04}^{+0.43+0.04}$
                                      & $0.23_{-0.16-0.02}^{+0.35+0.02}$\\
                                      & $1.00^{+0.00+0.00+0.00}_{-0.00-0.01-0.00}$
                                      & $0.90_{-0.24-0.02}^{+0.14+0.02}$
                                      & $0.96_{-0.12-0.01}^{+0.04+0.01}$\\
 $\overline B^0_s\to\phi\eta^\prime$  & $0.00^{+0.00+0.02+0.00}_{-0.00-0.02-0.00}$
                                      & $-0.07_{-0.06-0.01}^{+0.06+0.01}$
                                      & $0.10_{-0.05-0.01}^{+0.07+0.01}$\\
                                      & $1.00^{+0.00+0.00+0.00}_{-0.00-0.00-0.02}$
                                      & $1.00_{-0.01-0.00}^{+0.00+0.00}$
                                      & $0.99_{-0.01-0.00}^{+0.01+0.00}$\\\hline\hline
 $\overline B^0_s\to K_S\phi$         & $-0.72$
                                      & $0.09^{+0.04+0.01}_{-0.03-0.01}$
                                      & $-0.13^{+0.02+0.01}_{-0.02-0.01}$\\
                                      & $-0.69$
                                      & $-1.00^{+0.00+0.00}_{-0.00-0.00}$
                                      & $-0.99^{+0.00+0.00}_{-0.00-0.00}$\\
 ${\overline{B}}^{0}_{s}{\to}{\rho}^{0}K_S$&$-0.57^{+0.22+0.51+0.02}_{-0.17-0.39-0.05}$
                                      & $0.99_{-0.05-0.01}^{+0.00+0.00}$
                                      & $-0.03^{+0.22+0.17}_{-0.17-0.12}$\\
                                      & $-0.36^{+0.10+0.46+0.04}_{-0.13-0.15-0.04}$
                                      & $0.04^{+0.13+0.09}_{-0.11-0.13}$
                                      & $0.95^{+0.05+0.01}_{-0.13-0.02}$ \\
 $\overline B^0_s\to K_S\omega$       & $-0.63^{+0.09+0.28+0.01}_{-0.09-0.11-0.02} $
                                      & $-0.11^{+0.28+0.18}_{-0.22-0.14}$
                                      & $0.98^{+0.02+0.00}_{-0.04-0.01}$\\
                                      & $-0.57^{+0.11+0.31+0.02}_{-0.13-0.38-0.02}$
                                      & $0.96^{+0.02+0.01}_{-0.16-0.03}$
                                      & $-0.07^{+0.11+0.08}_{-0.09-0.12}$\\\hline
  $\bar B^0_s\to  K^{*+}  \pi^-\to K_S\pi^+\pi^-$ & ...
                                      & $0.98^{+0.01+0.01}_{-0.04-0.02}$
                                      & $0.35^{+0.11+0.15}_{-0.09-0.11}$ \\
                                      &...
                                      & $0.16^{+0.11+0.09}_{-0.09-0.13}$
                                      & $0.93^{+0.03+0.03}_{-0.07-0.07}$\\
  $\bar B^0_s\to  K^{*0}  \pi^0\to K_S\pi^0\pi^0$    & ...
                                      & $-0.07^{+0.26+0.18}_{-0.22-0.14}$
                                      & $0.94^{+0.03+0.02}_{-0.05-0.04}$\\
                                      & ...
                                      & $0.97^{+0.01+0.01}_{-0.15-0.02}$
                                      & $-0.30^{+0.12+0.07}_{-0.09-0.10}$  \\
  $\bar B^0_s\to  K^{*0}  \eta\to K_S\pi^0\eta$    & ...
                                      & $0.94^{+0.06+0.02}_{-0.09-0.03}$
                                      & $-0.77^{+0.23+0.04}_{-0.16-0.03}$\\
                                      & ...
                                      & $-0.22^{+0.15+0.08}_{-0.14-0.11}$
                                      & $0.10^{+0.26+0.11}_{-0.22-0.11}$\\
  $\bar B^0_s\to  K^{*0}  \eta'\to K_S\pi^0\eta'$   & ...
                                      & $-0.94^{+0.33+0.03}_{-0.09-0.01}$
                                      & $0.72^{+0.15+0.04}_{-0.16-0.05}$\\
                                      & ...
                                      & $0.01^{+0.45+0.16}_{-0.39-0.16}$
                                      & $-0.62^{+0.20+0.05}_{-0.16-0.06}$ \\
 \hline\hline\end{tabular}
\end{center}
 \end{table}

After studying the two simplified cases, we come to the
time-dependent CP asymmetries in $\bar B^0_s\to K^{*+}K^-$, where
the final state are not CP eigenstate and the width difference of
$B_s^0-\bar B_s^0$ can not be neglected either. In the following, we
choose $f=K^{*+}K^-$ and $\bar f=K^{+}K^{*-}$. One needs to consider
two additional CP asymmetries:
\begin{eqnarray}
 H_f=2\frac{{\mbox
 {Re}}(\lambda_f)}{1+|\lambda_f|^2},\;\;\;\;
 H_{\bar f}=2\frac{{\mbox {Re}}(\lambda_{\bar f})}{1+|\lambda_{\bar
 f}|^2},
\end{eqnarray}
which can be redefined as: $H= \frac{H_f+ H_{\bar f}}{2}$ and
$\Delta H=\frac{H_f- H_{\bar f}}{2}$. Our predictions for these
parameters are given in table~\ref{tab:mixingBstoKstarK}, but we
have not considered the global charge asymmetries because of the
presence of $\Delta\Gamma$. These predictions will be tested at the
forthcoming LHCb experiments

\begin{table}[tb]
\caption{Mixing-induced CP asymmetries in $\bar B_s^0\to K^{*+} K^-$
decay processes:  the first solution (This work 1) and the second
solution (This work 2). In both  predictions, we have included the
chiraly enhanced penguin and chosen $f=K^{*+}K^-$. The first kinds
of uncertainties are from uncertainties in charming penguins which
are discussed in the text; the second kinds of uncertainties are
from those in the CKM matrix elements.} \label{tab:mixingBstoKstarK}
\begin{tabular}{|c|c|c|c|c|c|c}\hline
  Parameter  &      This work 1                               & This work 2\\ \hline \
  $C$        &  $0.02_{-0.11-0.00}^{+0.10+0.00}$             & $0.01_{-0.09-0.00}^{+0.09+0.00}$        \\
  $S$        &  $-0.02_{-0.07-0.01}^{+0.07+0.01}$             & $0.02_{-0.05-0.00}^{+0.05+0.01}$   \\
  $H$        &  $0.92_{-0.04-0.02}^{+0.02+0.02}$              & $0.91_{-0.03-0.02}^{+0.02+0.02}$   \\
  $\Delta C$ &  $-0.09_{-0.10-0.01}^{+0.11+0.01}$             & $-0.11_{-0.09-0.01}^{+0.09+0.01}$  \\
  $\Delta S$ &  $0.38_{-0.07-0.04}^{+0.07+0.04}$              & $-0.41_{-0.05-0.03}^{+0.05+0.03}$   \\
  $\Delta H$ &  $0.01_{-0.02-0.00}^{+0.04+0.00}$              & $0.01_{-0.02-0.00}^{+0.02+0.00}$   \\%
  \hline
 \hline
\end{tabular}
\end{table}

\subsection{Isospin asymmetries and U-spin asymmetries}

Currently, there are many experimental methods to measure CKM
angles: $\alpha$, $\beta$ and $\gamma$. But in order to reduce the
uncertainties, a good way is to use SU(3) symmetry although this
will induce the errors from SU(3) symmetry breaking effect. Here we
will present some tests on this kind of symmetry breaking, although
the flavor SU(3) symmetry for $B\to P$, $B\to V$ form factors and
various charming penguins are used.

In the $B\to\pi\pi$ and $B\to\pi\rho$ system, one often uses the
following ratios~\cite{Beneke:2003zv}:
\begin{eqnarray}
   &&R_1 \equiv \frac{\Gamma(\bar B^0\to \pi^+\rho^-)}
                    {\Gamma(\bar B^0\to \pi^+\pi^-)} ,\;\;\;
   R_2 \equiv \frac{\Gamma(\bar B^0\to \pi^+\rho^-)+
                     \Gamma(\bar B^0\to \pi^-\rho^+)}
                    {2\Gamma(\bar B^0\to \pi^+\pi^-)} , \nonumber\\
   &&R_3 \equiv \frac{\Gamma(\bar B^0\to \pi^+\rho^-)}
               {\Gamma(\bar B^0\to \pi^-\rho^+)} ,  \;\;\;
   R_4 \equiv \frac{2\,\Gamma(B^-\to \pi^-\rho^0)}
                    {\Gamma(\bar B^0\to \pi^-\rho^+)} - 1 ,\;\;\;
   R_5 \equiv \frac{2\,\Gamma(B^-\to \pi^0\rho^-)}
                    {\Gamma(\bar B^0\to \pi^+\rho^-)} - 1 \,,\label{eq:rhopiratios}
\end{eqnarray}
where the partial decay widths are CP averaged. Our predictions are
given in table~\ref{tab:rhopiratios}, where we have used the
experimental results on branching ratios to evaluate the ratios and
these values are collected as experimental results. The predictions
in QCDF approach are also collected in this table. In $\bar
B^0\to\pi^+\pi^-$ and $\bar B^0\to\pi^+\rho^-$, tree operators
dominate. If we only consider the tree operators, $R_1$ becomes
ratios of decay constants: $R_1=(f_\rho/f_\pi)^2\sim 2$. Our
predictions are smaller than 2 for both solutions. In the first
solution, the ratio is much smaller which is mainly caused by
charming penguin terms: $A_{cc}^{PP}$ gives a constructive
contribution to the decay width of $\bar B^0\to\pi^+\pi^-$ while
$A_{cc}^{VP}$ gives a destructive contribution to $\Gamma(\bar
B^0\to\pi^+\rho^-)$. In the second solution, the deviation of $R_1$
from 2 is not too large as the phase of $A_{cc}^{PP}$ is almost the
same as $A_{cc}^{VP}$. $R_4$ and $R_5$ are larger than predictions
in QCDF approach and the present experimental data. $B^-\to
\pi^-\rho^0$ contains two different contributions from tree
operators: color-allowed contribution with $\rho^-$  emitted;
color-suppressed contribution with $\pi^-$  emitted. In QCDF
approach, the second contribution is small and the first
contribution is related to tree operators in $B^-\to \pi^-\rho^+$.
Neglecting the color-suppressed contribution and contributions from
penguin operators, $R_4$ is equal to zero. In SCET, color-suppressed
tree operators can give sizable contributions as we have discussed.
Thus the branching ratio of $B^-\to \pi^-\rho^0$ is enhanced which
can give a large value for $R_4$. The analysis is also similar for
the ratio $R_5$.

\begin{table}
{\caption{ \label{tab:rhopiratios}Two kinds of results for the
ratios $R_{1-5}$ in $B\to\pi\pi$ and $B\to\pi\rho$ decays, together
with the predictions in QCDF~\cite{Beneke:2003zv} and experimental
data evaluated using the results of branching fractions. The first
kinds of uncertainties are from uncertainties in charming penguins
as discussed in the text; the second kinds of uncertainties are from
those in the CKM matrix elements.}}
\begin{tabular}{|c|c|c|c|c|c|c|c}
 \hline & Exp.   & QCDF                                            & This work 1 & This work  2  \\
 \hline\hline
 $R_1$          & $2.69^{+0.54}_{-0.53}$
                & $2.39^{+0.31+0.04+0.15+0.05}_{-0.25-0.08-0.12-0.11}$
                & $1.32_{-0.12-0.12}^{+0.15+0.10}$
                & $1.84_{-0.19-0.06}^{+0.22+0.05}$\\
 $R_2$          & $2.21^{+0.37}_{-0.37}$
                & $2.06^{+0.40+0.53+0.12+0.03}_{-0.30-0.36-0.09-0.06}$
                & $1.17_{-0.11-0.07}^{+0.13+0.06}$
                & $1.52_{-0.15-0.09}^{+0.17+0.07}$ \\
 $R_3$          & $1.56^{+0.68}_{-0.46}$
                & $1.38^{+0.18+0.82+0.03+0.02}_{-0.17-0.59-0.04-0.05}$
                & $1.28_{-0.10-0.10}^{+0.12+0.08}$
                & $1.54_{-0.09-0.07}^{+0.07+0.11}$ \\
 $R_4$          & $0.96^{+0.80}_{-0.49}$
                & $0.42_{-0.04-0.11-0.21-0.20}^{+0.04+0.15+0.45+0.23}$
                & $2.38_{-0.20-0.02}^{+0.21+0.02}$
                & $1.23_{-0.04-0.02}^{+0.04+0.02}$ \\
 $R_5$          & $0.57^{+0.43}_{-0.33}$
                & $0.22_{-0.08-0.06-0.12-0.12}^{+0.07+0.08+0.23+0.14}$
                & $1.21_{-0.05-0.03}^{+0.05+0.02}$
                & $1.09_{-0.08-0.03}^{+0.08+0.04}$\\
 \hline\hline
\end{tabular}
\end{table}

In $\overline{B_d^0}\to K^{*-}\pi^+$, $\overline{B_s^0}\to
K^+\rho^-$, $\overline{B_d^0}\to K^{-}\rho^+$ and
$\overline{B_s^0}\to K^{*+}\pi^-$, the branching ratios are very
different from each other due to the differing strong and weak
phases entering in the tree and penguin amplitudes. However, as
shown by Gronau~\cite{Gronau:2000zy}, the two relevant products of
the CKM matrix elements entering in the expressions for the direct
CP asymmetries in these decays are equal, and, as stressed by
Lipkin~\cite{Lipkin:2005pb} subsequently, the final states in these
decays are charge conjugates, and the strong interactions being
charge-conjugation invariant, the direct CP asymmetry in
$\overline{B_s^0}\to K^+\pi^-$ can be related to the well-measured
CP asymmetry in the decay $\overline{B_d^0}\to K^-\pi^+$ using
U-spin symmetry. In this symmetry limit, we
have~\cite{Gronau:2000zy,Lipkin:2005pb}:
\begin{eqnarray}
&&|A(B_s^0\to\pi^+K^{*-})|^2-|A(\bar B_s^0\to\pi^-K^{*+})|^2=|A(\bar
B_d\to\rho^+K^{-})|^2-|A(B_d\to\rho^-K^{+})|^2,\\
&&A^{dir}_{CP}(\bar B_d\to\rho^+K^{-})=-A^{dir}_{CP}(\bar
B_s^0\to\pi^-K^{*+})\cdot\frac{BR(\bar B_s^0\to\pi^-K^{*+})}{BR(\bar
B_d^0\to\rho^+K^{-})}\cdot\frac{\tau(B_d)}{\tau(B_s)}.
\end{eqnarray}
Following the suggestions in the literature, we can test these
equations and search for possible new physics effects which would
likely violate these relations. Accordingly, one can define the
following parameters:
\begin{eqnarray}
 R_{6}&\equiv&\frac{|A(B_s\to\pi^+K^{*-})|^2-|A(\bar
 B_s\to\pi^-K^{*+})|^2}{|A(B_d\to\rho^-K^{+})|^2-|A(\bar
 B_d\to\rho^+K^{-})|^2}=\frac{{\cal BR}(\bar B_s\to\pi^-K^{*+})
 A_{CP}^{dir}(\bar B_s\to\pi^-K^{*+})\tau(B_d)}{{\cal BR}(\bar
 B\to K^-\rho^{+}) A_{CP}^{dir}(\bar B\to K^-\rho^{+})\tau(B_s)},\\
 \Delta_1&=&\frac{A^{dir}_{CP}(\bar B_d\to\rho^+K^{-})}{A^{dir}_{CP}(\bar
 B_s\to\pi^-K^{*+})}+\frac{BR(B_s\to\pi^+K^{*-})}{BR(\bar
 B_d\to\rho^+K^{-})}\cdot\frac{\tau(B_d)}{\tau(B_s)},\\
 R_{7}&\equiv&\frac{|A(B_s\to\rho^+K^-)|^2-|A(\bar
 B_s\to\rho^-K^+)|^2}{|A(B_d\to\rho^-K^+)|^2-|A(\bar
 B_d\to\rho^+K^-)|^2}=\frac{{\cal BR}(\bar B_s\to\rho^-K^{+})
 A_{CP}^{dir}(\bar B_s\to\rho^-K^{+})\tau(B_d)}{{\cal BR}(\bar
 B\to K^{*-}\pi^+) A_{CP}^{dir}(\bar B\to K^{*-}\rho^+)\tau(B_s)},\\
 \Delta_2&=&\frac{A^{dir}_{CP}(\bar
 B_d\to\pi^+K^{*-})}{A^{dir}_{CP}(\bar
 B_s\to\rho^-K^+)}+\frac{BR(\bar B_s\to\rho^-K^+)}{BR(\bar
 B_d\to\pi^+K^{*-})}\cdot\frac{\tau(B_d)}{\tau(B_s)}.
\end{eqnarray}
We also consider $\bar B^0\to \pi^+\rho^-$, $\bar B_s^0\to K^+
K^{*-}$,  $\bar B^0\to \pi^-\rho^+$ and $\bar B_s^0\to K^{*+} K^{-}$
which are related by U-spin transformation and define the following
ratios:
\begin{eqnarray}
 R_{8}&\equiv&\frac{|A(B_s\to K^{*+} K^-)|^2-|A(\bar B_s\to
 K^{*-}K^{+})|^2}{|A(B_d\to\rho^+\pi^-)|^2-|A(\bar
 B_d\to\rho^-\pi^+)|^2}=\frac{{\cal BR}(\bar B_s\to K^+K^{*-})
 A_{CP}^{dir}(\bar B_s\to K^+K^{*-})\tau(B_d)}{{\cal BR}(\bar
 B\to \pi^+\rho^{-}) A_{CP}^{dir}(\bar B\to \pi^+\rho^{-})\tau(B_s)},\\
 \Delta_3&=&\frac{A^{dir}_{CP}(\bar
 B_d\to\rho^-\pi^+)}{A^{dir}_{CP}(\bar B_s\to
 K^{*-}K^+)}+\frac{BR(\bar B_s\to K^{*-}K^+)}{BR(\bar
 B_d\to\rho^-\pi^+)}\cdot\frac{\tau(B_d)}{\tau(B_s)},\\
 R_{9}&\equiv&\frac{|A(B_s\to K^{+} K^{*-})|^2-|A(\bar B_s\to
 K^{-}K^{*+})|^2}{|A(B_d\to\pi^+\rho^-)|^2-|A(\bar
 B_d\to\pi^-\rho^+)|^2}=\frac{{\cal BR}(\bar B_s\to K^-K^{*+})
 A_{CP}^{dir}(\bar B_s\to K^-K^{*+})\tau(B_d)}{{\cal BR}(\bar
 B\to \pi^-\rho^{+}) A_{CP}^{dir}(\bar B\to \pi^-\rho^{+})\tau(B_s)},\\
 \Delta_4&=&\frac{A^{dir}_{CP}(\bar
 B_d\to\pi^-\rho^+)}{A^{dir}_{CP}(\bar B_s\to
 K^{-}K^{*+})}+\frac{BR(\bar B_s\to K^{-}K^{*+})}{BR(\bar
 B_d\to\pi^-\rho^+)}\cdot\frac{\tau(B_d)}{\tau(B_s)}.
\end{eqnarray}
In the flavor SU(3) symmetry limit, the ratios are $ R=-1$ and
$\Delta$ is zero. Using the first solution for the 16 inputs, we
obtain the following values:
\begin{eqnarray}
 R_{6}&=&-0.89,\;\;\; \Delta_1=-0.08_{-0.04-0.01}^{+0.02+0.01},\nonumber\\
 R_{7}&=&-0.99,\;\;\; \Delta_2=-0.01_{-0.01-0.00}^{+0.00+0.00},\nonumber\\
 R_{8}&=&-1.11,\;\;\; \Delta_3=0.11_{-0.05-0.02}^{+0.06+0.03},\nonumber\\
 R_{9}&=&-1.24,\;\;\; \Delta_4=0.33_{-0.11-0.05}^{+0.12+0.06},
\end{eqnarray}
where the tiny uncertainties of $R_{6-8}$ are omitted here.  Our
predictions using the second kind of inputs are given by:
\begin{eqnarray}
 R_{6}&=&-0.87,\;\;\; \Delta_1=-0.10_{-0.05-0.02}^{+0.03+0.02},\nonumber\\
 R_{7}&=&-0.99,\;\;\; \Delta_2=-0.01_{-0.00-0.00}^{+0.00+0.00},\nonumber\\
 R_{8}&=&-1.10,\;\;\; \Delta_3=0.09_{-0.02-0.01}^{+0.03+0.01},\nonumber\\
 R_{9}&=&-1.25,\;\;\; \Delta_4=0.33_{-0.11-0.05}^{+0.13+0.06}.
\end{eqnarray}
Since the form factors and charming penguins are assumed to the
respect flavor SU(3) symmetry, the small deviations for the ratios
$R$ and $\Delta$ are reasonable.

\section{Comparisons with the PQCD approach}
\label{sec:differences}

\begin{figure}[thb]
\begin{center}
\includegraphics[scale=0.7]{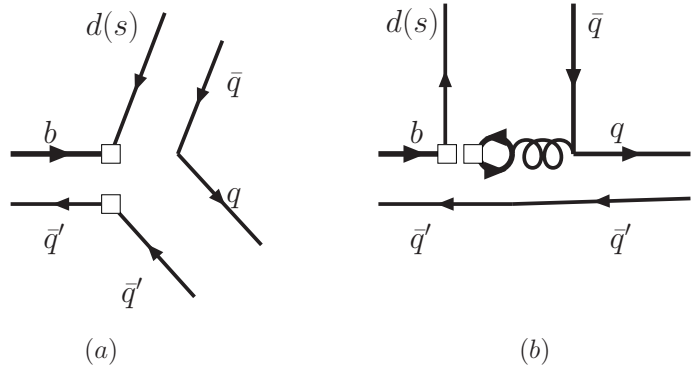}
\caption{Feynman diagrams for the $(S-P)(S+P)$ annihilation
operators in PQCD approach and charming penguins in SCET. }
\label{diagram:chiralyenhancedpenguinannihilation}
\end{center}
\end{figure}

PQCD approach is based on  $k_T$ factorization, where one keeps the
intrinsic transverse momentum of quark degrees of freedom. The
intrinsic transverse momentum can smear the end-point singularities
which often appear in collinear factorization. Resummation of double
logarithms results in the Sudakov factor which suppresses
contributions from the end-point region to make the PQCD approach
more self-consistent. This approach can explain many problems to
achieve great successes. Currently, radiative corrections
~\cite{Li:2005kt,Li:2006cva,Li:2006jv,Nandi:2007qx} and power
corrections in $1/m_b$ \cite{Huang:2004hw,Wu:2007rt} in this
approach are under studies. In PQCD approach, annihilation diagrams
can be directly calculated. Among them, the $(S-P)(S+P)$
annihilation penguin operators (from the Fierz transformation of
$(V-A)(V+A)$ operators) are the most important one. According to the
power counting in PQCD approach, annihilation diagrams are
suppressed by $\Lambda_{QCD}/m_b$ but the suppression for
$(S-P)(S+P)$ annihilation penguin operators is $2r_\chi$. This
factor is comparable with 1. Thus annihilations play a very
important role in PQCD approach. Phenomenologically, the large
annihilations can explain the correct branching ratios and direct CP
asymmetries of $B^0\to\pi^+\pi^-$ and $\bar B^0\to
K^-\pi^+$~\cite{Hong:2005wj}, the polarization problem of $B\to \phi
K^*$~\cite{Li:2004mp}, etc. In
Fig.~\ref{diagram:chiralyenhancedpenguinannihilation}(a), we draw
the Feynman diagrams for this term. Comparing with charming
penguins, we can see they have the same topologies in flavor space.
So generally speaking, charming penguins in SCET as shown in
Fig.~\ref{diagram:chiralyenhancedpenguinannihilation}(b) have the
same role with $(S-P)(S+P)$ annihilation penguin operators in PQCD.
Both of them are essential to explain the branching ratios in these
two different approaches. But there are indeed some differences in
predictions on other parameters such as direct CP asymmetries and
mixing-induced CP asymmetries.

First of all, the CKM matrix elements associated with charming
penguins and $(S-P)(S+P)$ annihilation penguin operators are
different. If we consider $\bar B$ decays in which a $b$ quark
annihilates, the $(S-P)(S+P)$ annihilation penguin operators are
proportional to $V_{tb}V_{tD}^*$, while charming penguins are
proportional to $V_{cb}V_{cD}^*$. The differences in the CKM matrix
elements will affect direct CP asymmetries and mixing-induced CP
asymmetries sizably. For example, in $\bar B_s^0\to\phi K_S$ decay,
the mixing-induced CP asymmetries in SCET are dramatically different
from predictions in PQCD approach. In the SCET framework, there is
no contributions from tree operators to $B_s\to\phi K_S$ at tree
level and penguin operators are much smaller than charming penguins.
As the CKM matrix element $V_{cb}V_{cD}^*$ for the charming penguin
is real, the parameter $\lambda$ defined in
Eq.~\eqref{eq:mixinglambda} becomes $\lambda=-e^{+2i\epsilon}$,
where we have neglected contributions from penguin operators. Thus
in SCET the two parameters $S_f$ and $H_f$ are given by:
\begin{eqnarray}
 S_f=-\sin(2\epsilon)=-0.03,\;\;\; H_f=-\cos(2\epsilon)= -1.00.
\end{eqnarray}
In PQCD approach, the CKM matrix element for the $(S-P)(S+P)$
annihilation penguin operators is $V_{tb}V_{td}^*$ which gives
$\lambda= -e^{+2i\epsilon+2i\beta}$:
\begin{eqnarray}
 S_f=-\sin(2\epsilon+2\beta)=-0.72,\;\;\; H_f=-\cos(2\epsilon+2\beta)= -0.69.
\end{eqnarray}
The differences in the mixing-induced CP asymmetries between SCET
and PQCD will be tested at the future experiments.

In PQCD approach, contributions from the $(S-P)(S+P)$ annihilation
penguin operators can be calculated using perturbation theory. These
contributions are expressed as the convolution of light-cone
distribution amplitudes and a hard kernel. We can also include SU(3)
symmetry breaking effects in the calculation in PQCD approach.  In
SCET, charming penguins are from the charm quark loops. Since the
charm quark is heavy, one can not factorize charming penguins (see
Ref.~\cite{Beneke:1999br,Beneke:2000ry,Beneke:2003zv,Beneke:2004bn}
for another point of view). Thus charming penguins are
non-perturbative in nature which is similar with the final state
interactions~\cite{Cheng:2004ru,Lu:2005mx}.  In the present work
based on SCET, we have assumed SU(3) symmetries for the
contributions from charming penguins. The magnitudes and strong
phases of charming penguins can not be calculated using perturbation
theory which obtained by fitting the experimental data.

The third difference is the magnitudes of charming penguins in SCET
and contributions from the $(S-P)(S+P)$ annihilation penguin
operators in PQCD approach. This difference arises from the
different power counting in the two approaches.  We take $b\to s$
transitions to illustrate the difference. In PQCD approach, the
$(S-P)(S+P)$ annihilation penguins are enhanced to be of the same
order with penguins in emission diagrams. In SCET, charming penguins
are more important. Comparing the values given in
Eq.~\eqref{eq:solution1}, Eq.~\eqref{eq:solution2} and
Eq.~\eqref{eq:penguinsbtos}, we can see charming penguins in SCET
always larger than contributions from emission penguin diagrams.

In PQCD approach, the $(S-P)(S+P)$ annihilation penguin operators
are chiraly enhanced and the dominant contribution is from the
imaginary part. The main strong phases in PQCD approach which are
essential to explain the large CP asymmetries in many channels are
also produced through from these operators.  But in SCET, as we have
shown in Eq.~\eqref{eq:solution1} and Eq.~\eqref{eq:solution2},
strong phases of charming penguins are not too large. Accordingly,
our predictions on direct CP asymmetries are small compared with
predictions in PQCD approach.

\section{conclusions}\label{sec:conclusions}

We provide the analysis of charmless two-body $B\to VP$ decays under
the framework of soft-collinear-effective theory. Besides the
leading power contributions, we also take some power corrections
(chiraly enhanced penguins) into account.  In the present framework,
decay amplitudes of $B \to PP$ and $B\to VP$ decay channels can be
expressed as functions of 16 non-perturbative inputs: 6 form factors
and 5 complex (10 real) charming penguins. Using the $B\to PP$ and
$B\to VP$ experimental data on branching fractions and CP asymmetry
variables, we find two kinds of solutions in $\chi^2$ fit for these
16 non-perturbative inputs. Chiraly enhanced penguin could change
some charming penguins sizably, since they have the same topology
with each other. However, most of other non-perturbative inputs and
predictions on branching ratios and CP asymmetries are not changed
too much. With the two sets of inputs, we predict branching
fractions and CP asymmetries. Agreements and differences with
results in QCD factorization and perturbative QCD approach are also
analyzed. Our conclusions are as follows:

\begin{itemize}
\item In color-allowed processes such as $\bar B^0\to \pi^\pm\rho^\mp$ decays,
tree operators provide the dominant contributions. Our predictions
on branching fractions are smaller than the ones calculated in the
QCDF approach and PQCD approach. The main reason is that: both $B\to
P$ and $B\to V$ form factors in SCET are smaller.
$B^0\to\pi^0\rho^0$ and other color-suppressed channels are
predicted with a larger branching ratios in SCET, because the hard
scattering form factors $\zeta_J^{P,V}$ are comparable with
$\zeta^{P,V}$ who also have a large Wilson coefficients. The large
branching ratios for $B^0\to\pi^0\rho^0$ are consistent with the
experimental data.

\item $b\to s$ decay processes such as $B\to \pi K^*$, $B\to \rho K$  and
the corresponding $B_s$ decays are dominated by contributions from
charming penguins. Since we have assumed flavor SU(3) symmetry for
charming penguins, branching fractions of $b\to s$ transition decays
can be estimated by analyzing the corresponding charming penguin
terms. Decays with iso-singlet mesons $\eta$ and $\eta'$ are
slightly different since there exists cancelations between different
charming penguins.

\item In the PQCD approach, annihilation diagrams do not suffer from the endpoint singularity
problem, which can be directly calculated. Among the three kinds of
penguin operators, the $(S-P)(S+P)$ operators are most important
which provide the main strong phase in the PQCD approach. In the
SCET framework, charming penguins play an important role especially
in $b\to s$ transitions. The $(S-P)(S+P)$ annihilations have the
same topology with charming penguin. Besides the commons, there
exists many differences in these two objects including weak phases,
magnitudes, strong phases, SU(3) symmetry property and factorization
property. These differences will mainly affect the direct CP
asymmetries and time-dependent CP asymmetry variables.

%

\end{itemize}

\section*{Acknowledgements}
This work is partly supported by National Nature Science Foundation
of China under the Grant Numbers 10735080, 10625525 and 10705050. We
would like to thank H.Y. Cheng, T. Huang, Y. Jia, M.Z. Yang, Y.D.
Yang and Q. Zhao for valuable discussions and comments. W. Wang
would like to acknowledge G.F. Cao and G. Li for the great help on
the $\chi^2$-fit program.

\appendix

\section{Expressions for Hard kernels}\label{sec:Hardkernels}

For explicit decay channels, the hard kernels depend the Lorentz
structure and flavor structures. They can be evaluated using the
Wilson coefficients given in Eq.~\eqref{eq:wcleadingpower} and
Eq.~\eqref{eq:wcsubleadingpower}. In this appendix, we intend to
write the decay amplitudes in a compact form. In doing it, the
following meson matrices are required:
\begin{eqnarray}
&& B^-=(1,0,0),\;\;\; \bar B^0=(0,1,0),\;\;\; \bar
 B^0_s=(0,0,1),\nonumber\\
 &&M_{\pi^+}=M_{\rho^+}= \left(%
\begin{array}{ccc}
  0  & 0    &0 \\
  1  & 0    &0 \\
  0  & 0    &0 \\
\end{array}%
\right),\;\;\;
 M_{K^+}=M_{K^{*+}}= \left(%
\begin{array}{ccc}
  0  & 0    &0 \\
  0  & 0    &0 \\
  1  & 0    &0 \\
\end{array}%
\right),\;\;\;
 M_{K^0}=M_{K^{*0}}= \left(%
\begin{array}{ccc}
  0  & 0    &0 \\
  0  & 0    &0 \\
  0  & 1    &0 \\
\end{array}%
\right),\nonumber\\
&&\sqrt 2 M_{\pi^0}= \sqrt 2M_{\rho^{0}}= \left(%
\begin{array}{ccc}
    1 & 0    &0 \\
  0  &  -1    &0 \\
  0  & 0    &0 \\
\end{array}%
\right),\;\;\;
\sqrt 2 M_{\eta_q}= \sqrt 2M_{\omega}= \left(%
\begin{array}{ccc}
   1  & 0    &0 \\
  0  &   1    &0 \\
  0  & 0    &0 \\
\end{array}%
\right),\;\;\;
 M_{\eta_s}=M_{\phi}= \left(%
\begin{array}{ccc}
  0  & 0    &0 \\
  0  & 0    &0 \\
  0  & 0    &1 \\
\end{array}%
\right),\nonumber\\
 &&M_{\pi^-}=M_{\rho^-}= M_{\pi^+}^T,\;\;\;
 M_{K^-}=M_{K^{*-}}= M_{K^+}^T,\;\;\;
 M_{\bar K^0}=M_{\bar K^{*0}}= M_{K^0}^T,
\end{eqnarray}
we also need the following matrices:
\begin{eqnarray}
 \delta_u= \left(%
\begin{array}{ccc}
  1  & 0    &0 \\
  0  & 0    &0 \\
  0  & 0    &0 \\
\end{array}%
\right),\;\;\;
\Lambda^d= \left(%
\begin{array}{c}
  0   \\
  1   \\
  0   \\
\end{array}%
\right),\;\;\;
\Lambda^s= \left(%
\begin{array}{c}
  0   \\
  0   \\
  1   \\
\end{array}%
\right),
\end{eqnarray}

Using the meson matrices, one can write the hard kernels appearing
in $B\to M_1 M_2$ decays as:
\begin{eqnarray}
 T_{1} &=& c^{f}_1 B M_2\delta_u M_1 \Lambda^f + (c^{f}_2\pm c_3^{f}) B M_2 \Lambda^f \mbox {Tr} [\delta_u M_1]\nonumber\\
 &&+
 c^{f}_4 B M_2  M_1 \Lambda^f + (c^{f}_5\pm c_6^{f}) B M_2 \Lambda^f \mbox {Tr} [M_1],\nonumber\\
 T_{1g} &=& c_1^f B \delta_u M_1 \Lambda^f \mbox{Tr}[M_2] + (c^{f}_2\pm c_3^{f}) B\Lambda^f \mbox {Tr} [\delta_u M_1] \mbox {Tr}[M_2]\nonumber\\
 &&+ c_4^f B M_1 \Lambda^f \mbox{Tr}[M_2] + (c^{f}_5\pm c_6^{f}) B\Lambda^f \mbox {Tr} [M_1] \mbox {Tr} [M_2],\nonumber\\
 T^g_{1} &=& c^{f}_g B M_2 \Lambda^f \mbox{Tr}[M_1],\;\;\;T^g_{1g} =  c^{f}_g B\Lambda^f \mbox {Tr} [M_1] \mbox {Tr} [M_2],\nonumber\\
  T_{1J} &=& T_1(c_i^f\to b_i^f),\;\;\;  T_{1Jg} = T_{1J}(c_i^f\to b_i^f),\;\;\;
    T^g_{1J} = T_{1}^g(c_i^f\to b_i^f),\;\;\;  T^g_{1Jg}=  T_{1g}^g(c_i^f\to
    b_i^f).
\end{eqnarray}
If the emitted meson $M_2$ is a pseudoscalar, $c^f_2-c_3^f$ and
$c^f_5-c_6^f$ in $T_i$ are used. But for vector meson emission, we
use plus signs in the combinations.

Using meson matrices, the charming penguins responsible for $B\to
M_1M_2$ decays can be determined in the same way. If the charming
penguins in $B\to PP$ decays are considered, the master equation is:
\begin{eqnarray}
 A_{cc}^{M_1M_2}= B M_2 M_1 \Lambda^f A^{PP}_{cc} +BM_1\Lambda^f {\mbox
 {Tr}}[M_2]A^{PP}_{ccg} ,
\end{eqnarray}
where the $A_{ccg}$ term is only responsible for the iso-singlet
mesons $\eta_q$ and $\eta_s$. In $B\to VP$ decays, the charming
penguins are:
\begin{eqnarray}
 A_{cc}^{M_1M_2}= B M_2 M_1 \Lambda^f A^{VP}_{cc} +B M_1 M_2 \Lambda^f A^{PV}_{cc} +BM_1\Lambda^f {\mbox
 {Tr}}[M_2]A^{VP}_{ccg},
\end{eqnarray}
where we take $M_1$ as a vector meson and $M_2$ as a pseudo-scalar
meson.

The master equations for hard kernels for chiraly enhanced penguins
are given by:
\begin{eqnarray}
 T_1^{\chi}&=& c_{1(qfq)}^{1\chi} B M_2 M_1 \Lambda^f + c_{2(qfq)}^{1\chi} B M_2 Q M_1
 \Lambda^f,\nonumber\\
 T_{1g}^{\chi }&=& c_{1(qfq)}^{1\chi} B  M_1 \Lambda^f {\rm Tr}[M_2]+ c_{2(qfq)}^{1\chi} B   Q M_1
 \Lambda^f {\rm Tr}[M_2],\nonumber\\
  T_{1J}^\chi&=&T_1^{\chi}(c_{1(qfq)}^{1\chi}\to c_{3(qfq)}^{2\chi},
 c_{2(qfq)}^{1\chi}\to c_{4(qfq)}^{2\chi} ),\nonumber\\
  T_{1Jg}^\chi&=&T_{1g}^{\chi}(c_{1(qfq)}^{1\chi}\to c_{3(qfq)}^{2\chi},
 c_{2(qfq)}^{1\chi}\to c_{4(qfq)}^{2\chi} ).
\end{eqnarray}

\end{document}